\documentclass[preprintnumbers,floatfix,superscriptaddressins,prd,nofootinbib]{revtex4-1}

\usepackage{enumerate}
\usepackage{amsmath,amssymb}  
\usepackage{bm}               
\usepackage{amscd}            
\usepackage{graphicx}         
\usepackage{epsfig}
\usepackage{hhline,multirow}  
\usepackage{dcolumn}          
\usepackage{wrapfig}
\usepackage[dvipsnames]{xcolor}
\usepackage[normalem]{ulem}
\usepackage{mathtools}

\usepackage{slashed}
\usepackage{hyperref}   

\newcommand{\psl}{p\!\!\!\slash} 
\newcommand{\ksl}{k\!\!\!\slash} 
\newcommand{\qsl}{q\!\!\!\slash} 
\newcommand{\kpsl}{k\!\!\!\slash'} 

\newcommand{\kTsl}{k\!\!\slash\!_T} 
\newcommand{\kpTsl}{k\!\!\!\slash'_T}

\newcommand{\hatpsl}{\hat{p}\!\!\!\slash}

\newcommand{\xbmax}{x_{B,\text{max}}}
\newcommand{\xBbreak}{x_{B,\text{break}}}
\newcommand{\Wsqbreak}{W^2_\text{break}}
\newcommand{\Rbreak}{R_\text{break}}

\newcommand{\ktmax}{\bm{k_{T,\text{max}}}}
\newcommand{\ktmaxsq}{\ktmax^2}

\newcommand{\Lambdatw}{\Lambda_\text{QCD}}

\newcommand{\ie}{\textit{i.e.}}

\newcommand{\nbar}{{\overline n}}
\newcommand{\nslash}{n\hspace*{-0.22cm}\slash\hspace*{0.022cm}}
\newcommand{\nbslash}{\nbar\hspace*{-0.22cm}\slash\hspace*{0.022cm}}
\newcommand{\Tr}{{\rm Tr}}
\newcommand{\vect}[1]{\boldsymbol{#1}}

\newcommand{\vpsq}{v^{\,\prime\,2}}

\newcommand{\vsqbar}{\bar{v}^{\,2}}
\newcommand{\vpsqbar}{\bar{v}^{\,\prime\,2}}
\newcommand{\vsqbarnew}{\bar{v}_*^{\,2}}
\newcommand{\vsqbarnewT}{\bar{v}_{*T}^{\,2}}
\newcommand{\xiqnew}{\xi^*_q}
\newcommand{\xiqTnew}{\xi^*_{qT}}
\newcommand{\dfourk}{d^4 k \,}

\newcommand{\xipinew}{\xi^*_\pi}

\newcommand{\dtwokT}{d^2 \boldsymbol{k_T} \,}
\newcommand{\dkTtwo}{d \boldsymbol{k_T}^2 \,}

\newcommand{\dtwokTp}{d^2 \boldsymbol{k_T'} \,}

\newcommand{\kTsq}{\bm{k_T}^{\,2}}
\newcommand{\avekTsq}{\langle\kTsq\rangle}
\newcommand{\kpTsq}{\bm{k_T'}^{\,2}}

\newcommand{\CF}{\text{CF}}
\newcommand{\QCD}{\text{QCD}}
\newcommand{\DIS}{\text{DIS}}
\newcommand{\INT}{\text{INT}}
\newcommand{\RES}{\text{RES}}

\newcommand{\id}{{\mathbb{I}}}

\newcommand{\Jfull}{J_{x,k^2,\kTsq}}

\begin{document}

\preprint{\vbox{\hbox{JLAB-THY-20-3254} }}

\title{Collinear Factorization at sub-asymptotic kinematics \\ and validation in a diquark spectator model}

\author{Juan V. Guerrero}\email{e-mail: juanvg@jlab.org}
\affiliation{
	Jefferson Lab, Newport News, VA 23606, USA }
\affiliation{ Department of Physics, Old Dominion University, Norfolk, Virginia 23529, USA}
\author{Alberto Accardi}\email{e-mail: accardi@jlab.org, ORCID: 0000-0002-2077-6557}
\affiliation{
	Hampton University, Hampton, VA 23668, USA}
\affiliation{
	Jefferson Lab, Newport News, VA 23606, USA \\ }

\date{\today}

\begin{abstract}

We revisit the derivation of collinear factorization for Deep Inelastic Scattering at sub-asymptotic values of the four momentum transfer squared, where the masses of the particles participating in the interaction cannot be neglected. By using an inclusive jet function to describe the scattered quark final state, we can restrict the needed parton kinematic approximations just to the four-momentum conservation of the hard scattering process, and explicitly expand the rest of the diagram in powers of the unobserved parton transverse momenta rather than neglecting those. This procedure provides one with more flexibility in fixing the virtuality of the scattered and recoiling partons than in the standard derivation, and naturally leads to scaling variables that more faithfully represent the partonic kinematic at sub-asymptotic energy than the Bjorken's $x_B$ variable.

We then verify the validity of the obtained factorization formula by considering a diquark spectator model designed to reproduce the main features of electron-proton scattering at large $x_B$ in Quantum Chromo-Dynamics. In the model, the Deep Inelastic Scattering contribution to the cross section can be explicitly isolated and analytically calculated, then compared to the factorized approximation. Limiting ourselves to the leading twist contribution, we then show that use of the new scaling variables maximizes the kinematic range of validity of collinear factorization, and highlight the intrinsic limitations of this approach due to the unavoidably approximate treatment of four momentum conservation in factorized diagrams. Finally, we briefly discuss how these limitations may be overcome by 
including higher-twist corrections to the factorized calculation.

\end{abstract}

\maketitle

\newpage
\tableofcontents

\newpage


\section{Introduction}

\subsection{Motivation}

Unraveling the quark and gluon structure of the nucleon still remains a major challenge in hadronic and particle physics, notwithstanding the significant experimental and theoretical advances made in this area throughout the last decade~\cite{Jimenez-Delgado:2013sma,Gao:2017yyd,Lin:2017snn,Ethier:2020way,Lin:2020rut}. 

The Large Hadron Collider can measure a large variety of observables, especially at the energy frontier, and access the proton structure at the smallest spatial scales. However utilizing its data remains challenging due to tensions between various observables, and its impact on the determination of unpolarized Parton Distribution Functions (PDFs) is so far somewhat statistically limited \cite{Hou:2019efy,Bailey:2019yze,AbdulKhalek:2020jut}.
Proton-proton collisions at the Relativistic Heavy Ion Collider (RHIC) provide complementary access to PDFs at lower energy scale and higher parton fractional momentum, most notably in polarized collisions \cite{deFlorian:2014yva,Nocera:2014gqa,DeFlorian:2019xxt}. Use of RHIC data in unpolarized PDF fits has however not received much attention until very recently, despite its potential for flavor separation of sea quarks via weak boson production data \cite{Park:2021kgf,Cocuzza:2021cbi} and gluon PDF determination through jet observables. 
Lowering the collision energy and changing reaction to electron-proton collisions, recent data from the Jefferson Lab 6 GeV program and those being collected at its 12 GeV upgrade \cite{Dudek:2012vr,Burkert:2018nvj}, as well as those expected from the future Electron Ion Collider \cite{Accardi:2012qut,Aidala:2020mzt}
will enable us to access quarks and gluons in unprecedented ways, and to build an accurate, 3-dimensional picture of the inner structure of the proton.

In order to use high-energy scattering data to describe the proton's structure in terms of quark and gluon PDFs one relies on QCD factorization theorems, such as Collinear Factorization (CF) \cite{Collins:1989gx}. These theorems allow one to write the cross sections of large momentum transfer scattering processes such as Deep Inelastic Scattering (DIS) as a convolution of a short distance matrix element, which can be computed perturbatively and describes the quark and gluon ``hard''
interaction with a probe, and long distance non perturbative matrix elements -- the PDFs -- that describe the quark and gluon momentum distribution within the proton.

In this paper we are interested in assessing the viability of Collinear Factorization in describing DIS events with large enough 4-momentum transfer to justify a perturbative QCD analysis of the cross sections in terms of quark and gluon interactions, but not large enough to neglect any other mass or dynamical momentum scale characterizing the process. For example, experiments at Jefferson Lab with a 6~GeV energy beam involve low photon virtualities $Q^2$  that require control of $1/Q^2$ power corrections to the calculation of cross sections. With the 12~GeV beam the accessible $Q^2$ increases, without, however, reaching asymptotic values where other scales can be neglected. In this sub-asymptotic regime, the mass of the proton target and the mass of an observed hadron, collectively denoted  by $\mu$, induce finite-$Q^2$ corrections of order $\mu^2/ Q^2$ which we call ``Hadron Mass Corrections'' (HMCs). These can compete with experimental uncertainties at Jefferson Lab energy and can also affect higher-energy experiments such as HERMES and COMPASS, see Refs.~\cite{Accardi:2009md,Guerrero:2015wha,Guerrero:2017yvf,Scimemi:2019cmh}. These papers take into account into account the mass of the target and of the observed hadron through a rescaling of the Bjorken variable $x_B$, as already discussed for example in \cite{Aivazis:1993kh,Kretzer:2003iu,Albino:2008fy}\footnote{And many other papers starting from the seminal inclusive DIS analysis by Nachtmann \cite{Nachtmann:1973mr} and Georgi, Politzer and De Rujula in the operator product expansion (OPE) formalism \cite{Georgi:1976ve,DeRujula:1976baf}, and by Ellis, Furmanski and Petronzio in collinear factorization \cite{Ellis:1982cd}; see \cite{Schienbein:2007gr} for a review, and \cite{Brady:2011hb} for a comparison of mass corrections methods. See also Refs~\cite{Accardi:2008ne,Steffens:2012jx} for recent proposals to address the ``threshold'' problem within collinear and OPE approaches to target mass corrections.}. Refs.~\cite{Accardi:2009md,Guerrero:2015wha,Guerrero:2017yvf} go a step further, and argue that one also needs to take into account the fact that the (unobserved) scattered parton needs to have a virtuality substantially different from 0 in order to fragment into a massive hadron, and show that this requirement can be implemented in a gauge invariant way through a modified scaling variable. Numerical estimates at JLab kinematics suggest large effects for semi-inclusive pion production \cite{Accardi:2009md}, and even more for kaons or heavier hadrons \cite{Guerrero:2015wha}. 
In fact, HMCs implemented in this way may even explain \cite{Guerrero:2017yvf} the apparent large discrepancy between the measurements of transverse momentum integrated kaon multiplicities performed at HERMES \cite{Airapetian:2012ki} and COMPASS \cite{Adolph:2016bwc,Seder:2015sdw}.

Two subsequent papers from the COMPASS collaboration have furthermore analyzed kaons and protons produced at even larger hadron momentum fractions than reported before, highlighting  strong departures from pQCD calculations \cite{Akhunzyanov:2018ysf,Alexeev:2020jia}.
This discrepancy between theory and experiment seems too large to be only due to the phase space limitations induced by finite mass effects (which, as argued in Refs.~\cite{Guerrero:2015wha,Guerrero:2017yvf}, can be treated as a correction to the usual collinear pQCD calculations) and may indicate that the factorization formalism is being applied in a kinematic region where this is not a good approximation to the semi-inclusive cross section. 
If the correct treatment of the partonic kinematics and the very validity of QCD factorization are under question at high-energy experiments such as HERMES and COMPASS -- and already for transverse momentum integrated observables! -- investigating these issues becomes essential 
for a correct interpretation of the upcoming semi-inclusive measurements at the JLab 12 GeV upgraded facility \cite{Dudek:2012vr,Burkert:2018nvj}, that are largely focused on the 3D imaging of nucleons and nuclei. Indeed, the transverse momentum dependent cross sections are naturally more sensitive to HMCs than their integrated counterparts, and factorization encounters novel challenges of its own~\cite{Boglione:2016bph,Gonzalez-Hernandez:2018ipj,Boglione:2019nwk,Scimemi:2019cmh}. 

Motivated by these considerations, in this paper we revisit the ``standard'' derivation of collinear factorization in DIS processes \cite{Collins:2011zzd,Bacchetta-lectures-2012} with the goal of identifying under what conditions this can be extended to the sub-asymptotic kinematic region, and how one can maximize its regime of applicability. As a first step, we will discuss inclusive DIS scattering, where we can avoid purely technical complications due to the interplay of initial and final state kinematics \cite{Guerrero:2015wha}, and limit ourselves to Leading-Order (LO) perturbative calculations, that do not require renormalization of the quark fields and limit the final state to 2 particles. Nonetheless, we will be able to address in full the need for, and means of, an improved kinematic approximation. The use of an inclusive jet function \cite{Accardi:2008ne,Accardi:2017pmi,Accardi:2019luo,Accardi:2020iqn} to describe the scattered quark final state will prove essential to our goals. We will then validate the obtained factorization formula in the framework of a QCD-like idealized field-theoretical model describing a spin 1/2 idealized nucleon, which contains an active quark as well as a scalar diquark spectator that does not participate in the interaction~\cite{Bacchetta:2008af,Moffat:2017sha}, 
and complete the analysis presented in Refs.~\cite{Guerrero:2019qbn,Guerrero:PhD_thesis_2019}.
In the chosen ``diquark spectator model'', one can perform fully analytic calculations of the DIS cross section, as well as collinear factorization with or without HMCs. The model is designed to reproduce the main feature of the QCD process at large $x_B$, and a comparison of the full and collinearly factorized cross sections will determine the validity of the proposed HMC scheme, as well as test the limits of collinear factorization itself. Furthermore, working within an explicit model we will be able to investigate the role of the parton's transverse momentum, that is by necessity neglected in the Leading-Twist calculation of the inclusive DIS cross section, but contributes to Higher-twist (HT) $O(\avekTsq/Q^2)$ corrections. As we will very briefly discuss in the closing section, we believe that an extension of our HMC scheme to Next-to-Leading Order (NLO) -- and, in fact, also to Semi-Inclusive DIS (SIDIS) processes -- should not encounter essential difficulties.

\subsection{Paper organization and overview of results}

As this paper is quite long, it is worthwhile to provide the readers with an overview of its structure, the philosophy behind our approach and the novelties compared to the standard derivation, and the main results of each Section, before delving into the details of calculations and derivations.

In Section~\ref{sec:CF}, we discuss our proposal for performing Collinear Factorization of the DIS structure functions at sub-asymptotic hard scales $Q^2$, and how one can account for hadron masses and non-zero parton virtualities in the treatment of partonic kinematics. Our central philosophy, adopted from Ref.~\cite{Collins:2007ph}, is to minimize the number of uncontrolled approximations needed to achieve the desired factorization formula. In particular, we confine the needed ``pure'' kinematic approximations just to the external legs of the partonic hard-scattering, and perform a controlled ``twist'' expansion of the rest of the diagram. Gauge invariance is guaranteed by the use of an inclusive quark jet function \cite{Accardi:2008ne,Accardi:2017pmi,Accardi:2019luo,Accardi:2020iqn} to describe the scattered quark in the DIS handbag diagram, rather than utilizing an on-shell quark propagator as in standard derivations \cite{Collins:2011zzd,Bacchetta-lectures-2012}. In fact, our calculation parallels the analogous one for single inclusive hadron production in SIDIS \cite{Bacchetta:2006tn,Bacchetta-lectures-2012}, and highlights how the parton's transverse momentum needs not be altogether neglected, but can be instead dynamically included in higher-twist terms \cite{Ellis:1982cd,Qiu:1988dn}.

The end result is quite simple: at leading twist (LT), the factorized formula for the hadronic tensor, and therefore for the cross section and its structure functions, are given by their asymptotic (or massless) counterparts evaluated at a suitably defined scaling variable $\bar x$ instead of $x_B$, see Eqs.~\eqref{eq:factorized_Wmunu_LT}-\eqref{eq:Jacobian}. The choice of the $\bar x$ scaling variable is not prescribed by the factorization procedure itself, but can be guided by kinematic consideration at the parton level and by respect of momentum and baryon number conservation laws, see Section~\ref{sec:kin_approx_QCD} and in particular Eq.~\eqref{eq:xbar_thr_QCD}. 

The use of a scaling variable is not a new concept, as it has been proposed in a similar context, for example, in Refs.~\cite{Aivazis:1993kh,Kretzer:2003iu,Moffat:2017sha,Tung:2001mv,Nadolsky:2009ge} and even earlier in Refs.~\cite{Ellis:1982cd,Georgi:1976ve,Frampton:1976mu,Nachtmann:1973mr}. Here we attempt, however, at a more systematic treatment that avoids \textit{a priori} parton model considerations. In fact our end result cannot be interpreted in parton model terms except in a well defined limit, but, conversely, gives one enhanced freedom in devising realistic kinematic approximations in the sub-asymptotic regime. In particular, it turns out that the light-cone virtualities $v^2$ and $\vpsq$ of the partons participating in the initial and final state of hard-scattering process need not be approximated to zero, and can be chosen differently for the incoming and scattered quarks without breaking gauge invariance. This added flexibility may also facilitate the study of the transition from perturbative to non-perturbative degrees of freedom in data beyond the deep inelastic regime, where the virtual photon excites proton resonances \cite{Melnitchouk:2005zr} and may be sensitive to multi-parton nucleon substructures \cite{Moffat:2019qll}.

The theoretical results outlined above are compelling, but call for a benchmark validation. To this end, in Section~\ref{sec:model} we present the diquark spectator model adopted for our validation study, and use this to analytically calculate the inelastic lepton-proton cross section at LO. We then show how the cross section can be decomposed in a gauge invariant way into DIS, proton resonance, and interference contributions, and study in detail the proton's $F_T$ and $F_L$ transverse and longitudinal structure functions, as well as their scaling properties with respect to the photon virtuality $Q^2$. The low-$x_B$ behavior of the model also turns out to be interesting, even if the model is not designed to provide one with a realistic description of experimental measurements in that regime. Indeed, our explicit calculation will highlight a quite different $Q^2$ scaling of the DIS component of $F_L$ compared to simple dimensional arguments and to what happens for $F_T$. As we will explain, this is however a general consequence of gauge invariance rather than a model artifact. It may also explain the need for phenomenological $O(1/Q^2)$ corrections in order for CF calculation of $F_L$ to agree with recent HERA data at small values of the Bjorken invariant $x_B$ \cite{Abt:2016vjh}.

Sections~\eqref{sec:testing_kin} and \eqref{sec:fact_limits} are devoted to the validation of the sub-asymptotic factorization formulae \ref{eq:factorized_Wmunu_LT}-\ref{eq:Jacobian} for the DIS transverse structure function $F_T^{\DIS}$. The goal is to verify: (i) to what extent the ''internal'' (\ie, unobserved) partonic variables can be replaced in the factorized cross section by the proposed sub-asymptotic kinematic approximations and the use of a scaling variable; (ii) how large transverse momentum corrections of order $O(\avekTsq/Q^2)$, with $\kTsq$ the parton's transverse momentum, are since these cannot be kinematically included in the scaling variables and dynamically contribute to the factorized cross section starting only at next-to-leading twist; and (iii) explore the intrinsic limitations of collinear factorization. The conclusions we will reach are rather robust versus variations of model parameters, as demonstrated in the  Appendix, and therefore indicative of what we may expect to happen in QCD. 

In Section~\ref{sec:testing_kin}, the accuracy to which the $\bar x$ scaling variable describes the internal partonic kinematics is investigated, and this and the corresponding light-cone virtuality $\bar v$ are compared to the average parton momentum fraction and virtuality values calculated in the full model. 
In particular, we find that one can approximate at the 90\% level the average parton fractional momentum by including all external mass scales in the quark-mass-corrected Nachtmann scaling variable $\xi_q= \xi (1 + m_q^2/Q^2)$, where $m_q$ is the quark mass and the  Nachtmann variable $\xi$ \cite{Nachtmann:1973mr} accounts for the proton mass; small corrections of order $O(\avekTsq/Q^2)$ account for the rest.

These conclusions are confirmed in Section~\ref{sec:fact_limits}, where we compare the full and factorized transverse $F_T$ structure functions and show that only corrections of order $O(\avekTsq/Q^2)$ are needed to describe the full structure functions after removing all mass corrections by using of the $\xi_q$ in our sub-asymptotic CF formula.  These additional corrections are not experimentally controllable in inclusive measurements, but given their small size one can hope to theoretically treat them in the twist expansion \cite{Ellis:1982cd,Qiu:1988dn} without resorting to the Transverse-Momentum Dependent (TMD) factorization formalism \cite{Bacchetta:2006tn,Collins:2011zzd}, or phenomenologically by adding a power suppressed term to a PDF fit analysis. A brief discussion of these issues is offered in Section~\ref{sec:beyond_LT}, and a detailed analysis is left for future work. 

In Section~\ref{sec:fact_limits} we also demonstrate the inherent limits of collinear factorization, that breaks down at very large $x_B$ because the kinematical approximations needed to factorize the PDFs and the partonic hard scattering coefficient do not respect four momentum conservation in transverse momentum, as already argued in Ref.~\cite{Moffat:2017sha}. Fortunately, this effect can be circumscribed by simple kinematic cuts on the invariant mass $W$ of the final state, and our model estimate indicates that factorization breaks down only in the resonance region at $W^2 \lesssim 4$ GeV$^2$.

Finally, in Section~\ref{sec:conclusions} we summarize the many results of our paper and their implications, and in the Appendices we included a study of the dependence (or rather independence) of our conclusions on the model parameters. In the Appendix section, we also provide: a complete discussion of the structure function projectors and their small-$x_B$ limit; details of the sub-asymptotic kinematic limits; an analytic calculation of the small-$x_B$ scaling behaviour of the model structure functions; and an explicit illustration of resonant electron-nucleon scattering when the masses of the constituents are smaller than the mass of the target itself.


\section{Collinear factorization at sub-asymptotic momentum transfer}
\label{sec:CF}

\begin{figure}[bt]
	\centering
	\includegraphics[width=15cm]{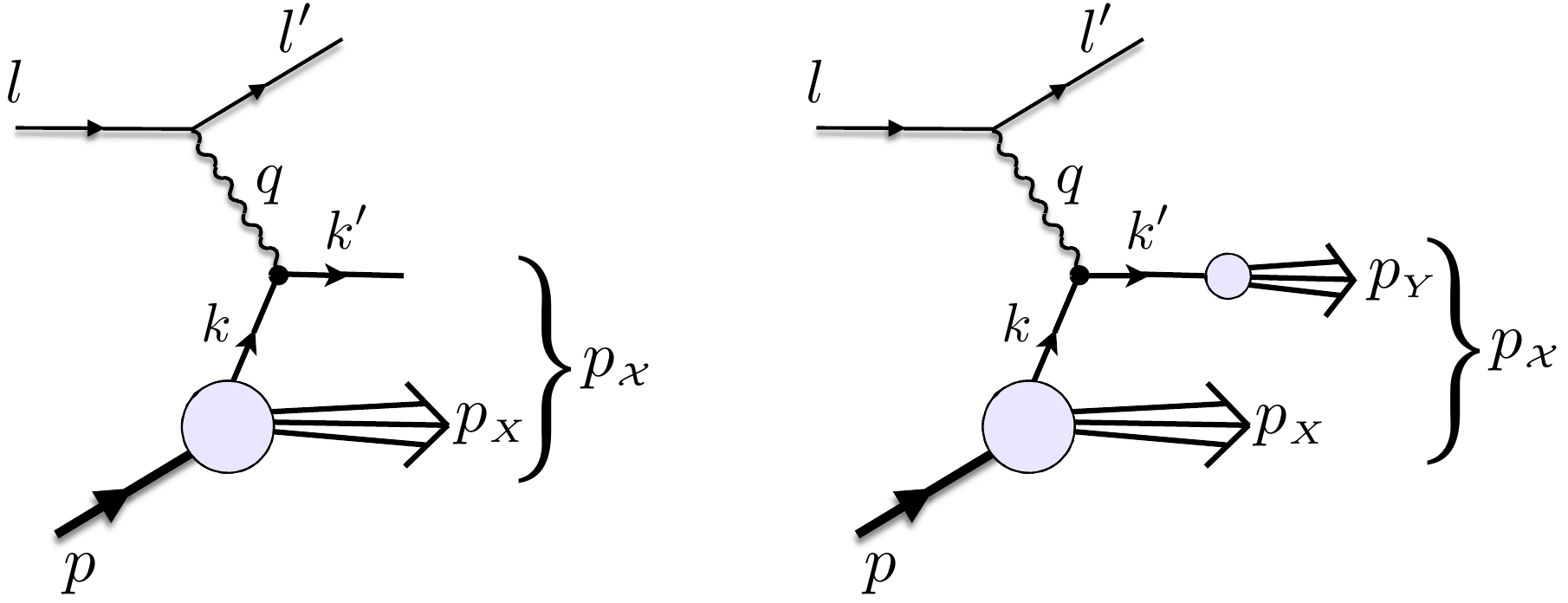}	
	\caption{Electron-proton DIS at leading order in the strong coupling constant. {\it Left}: in the impulse approximation, far from kinematic thresholds. {\it Right}: in the impulse approximation, with a quark remnant $Y$ accounting for the hadronization of the final state quark.}   
	\label{fig:DIS_diagram}
\end{figure}

Deeply inclusive lepton-nucleon scattering on a proton or neutron target is illustrated in Figure~\ref{fig:DIS_diagram}, where the incident lepton (with four-momentum momentum $l$) interacts with a nucleon ($p$) through the exchange of a virtual photon ($q$). 
At large values of the virtuality $Q^2= - q^\mu q_\mu$, the virtual photon scatters, on a short time scale, on a quark of four momentum $k$ belonging to the nucleon. 
In the final state, one measures the recoil lepton momentum $l'$, while the recoiled quark with four momentum $k'$, as well as the remnant $X$ of the proton are unobserved. The remnant is a system of many particles produced by the fragmentation of the target after the photon extracted one of its quarks. In fact, the colored quark and remnant are subject to QCD confinement, and, on a much longer time scale compared to the photon-quark scattering process, hadronize into a system of color neutral hadrons.
Far from kinematic thresholds, unitarity arguments show that color neutralization can be ignored in an inclusive measurement such as we are discussing, and the process calculated as if quarks where asymptotic states, see the left panel of Figure~\ref{fig:DIS_diagram}. 
However, closer to the pion production threshold the final state phase space shrinks, and one needs to take into account the fact that on-shell quarks cannot be present in the final state. In this regime, it is possible to consider the diagram on the right panel of Figure~\ref{fig:DIS_diagram} where one includes a quark remnant $Y$ to account for quark hadronization. Note that we are considering diagrams in which the final state in the current direction, $Y$, does not interact with the target remnant $X$. This assumption is indeed justified at large enough values of the Bjorken invariant $x_B = Q^2/2(p^\mu q_\mu)$, which is the focus of this paper, because the finite value final state invariant mass $W^2=(1-1/x_B)Q^2 + M^2$ kinematically limits the transverse momentum of particles produced in the quark's direction, squeezing these in a jet-like configuration aligned with the quark's momentum~\cite{Manohar:2005az,Accardi:2018gmh}.

\subsection{Kinematics}
\label{sec:kinematics_QCD}

We parametrize the four-momenta of the proton, photon and incoming quark in Figure~\ref{fig:DIS_diagram} in terms of light-cone unit vectors $n$ and $\nbar$, which satisfy $n^2 = \nbar^2 = 0$ and $n \cdot \nbar=1$ \cite{Ellis:1982cd}.
The ``plus'' and ``minus'' components of a four-vector $a^\mu$ are defined by
$a^+ = a \cdot n     = (a^0 + a^z)/\sqrt{2}$ and
$a^- = a \cdot \nbar = (a^0 - a^z)/\sqrt{2}$. 
Then, one can decompose
\begin{equation}
  a^\mu = a^+ \nbar^\mu
    + a^- n^\mu +a_{T}^{\,\mu} \equiv (a^+, a^-, \boldsymbol{a_T}),
\label{eq:vmu}
\end{equation}
where $a_T^\mu$ is the vector's transverse four-momentum, which satisfies $a_{T} \cdot n = a_{T} \cdot \nbar = 0$, with norm $a_{T}^2 = -\bm{a_T}^{\,2}$, and $\bm{a_T} \equiv (a_{Tx},a_{Ty})$ is the 2D Euclidean transverse momentum. 


We will work in the ``$(p,q)$ frames'' class \cite{Accardi:2009md}, in which the 
initial proton momentum and virtual photon are collinear in 3-dimensional space and oriented along the $z-$direction. We can thus decompose them as
\begin{align}
    p^\mu &= p^+\, \nbar^\mu
    + \frac{M^2}{2 p^+}\, n^\mu,		\label{eq:p} \\
    q^\mu &= - \xi p^+\, \nbar^\mu 
    + \frac{Q^2}{2\xi p^+}\, n^\mu,		\label{eq:q} 
\end{align}
where $\xi$ is the so-called Nachtmann variable and is defined as
\begin{align}
\label{eq:xi}
\xi & \equiv -\frac{q^+}{p^+} =
\frac{2 x_B}{1 + \sqrt{1 + 4 x_B^2 M^2/Q^2}} \ .
\end{align}
The $p^+$ component of the nucleon's momentum parametrizes the Lorentz boosts in the $z$ direction, and interpolates between the nucleon rest frame ($p^+=M/\sqrt2$) and the infinite momentum frame ($p^+\rightarrow\infty$).
The use of a light-cone reference frame is justified for hard scattering at large $Q^2$, where the proton and scattered quark momenta are dominated by their light-cone plus and minus components, respectively. For the same reason, this is also the frame used to perform collinear factorization, as discussed more extensively in the rest of this section.

Both the target momentum $p$ and the photon momentum $q$ are ``external'' variables, namely they are experimentally measured in the process of interest. On the contrary, the incoming and outgoing quark momenta are not even in principle measurable and therefore we will consider them ``internal'' variables. The parton momenta can be decomposed as 
\begin{align}
k^\mu &= xp^+\, \bar{n}^\mu
+ \frac{v^2}{2 x p^+}\, n^\mu
+ k^\mu_T	\ ,			\label{eq:k} \\
k'^\mu &= \frac{\vpsq}{2 k'^-}\, \bar{n}^\mu
+ k'^-\, n^\mu
+ k'^{\mu}_T	\ \label{eq:kp}
\end{align}
where $x \equiv \frac{k^+}{p^+}$ is the light cone momentum fraction carried by the parton. 
In Eq.~\eqref{eq:k}, the struck quark's ``light-cone virtuality''
\begin{align}
    v^2 = k^2+\kTsq
    \label{eq:v2_def}
\end{align}
is a mass scale that will be relevant to our derivation  of sub-asymptotic collinear factorization. The name is justified by noticing that $k^-=v^2/(2k^+)$, so that $v^2$ quantifies how far the quark momentum is from the light-cone plus direction. 
As we will discuss, it is this scale, rather than the quark's virtuality $k^2$ alone, that controls the partonic kinematics in the diagram and determines the applicability of collinear factorization assumptions. Similarly, the outgoing quark's light-cone virtuality, $\vpsq$, is defined as
\begin{align}
   \vpsq = k'^2+ \kpTsq \, .
\end{align}


\subsection{The DIS hadronic tensor at LO}

The differential cross section for the inelastic scattering of an unpolarized lepton from an unpolarized nucleon target can be written in the Born approximation as 
\begin{equation}
    \frac{d \sigma}{dx_B dQ^2} 
    = \frac{\pi \alpha^2 y^2}{Q^6} L_{\mu \nu} \, 2M W^{\mu\nu}\, ,
\label{eq:cross_section_DIS}
\end{equation}
where $\alpha = \frac{e^2}{4 \pi}$ is the fine structure constant, $Q^2= -q \cdot q$ is the photon's virtuality, $x_B=Q^2/(2p\cdot q)$ is the Bjorken variable, and the Lorentz invariant $y$ is defined as $y=\frac{p\cdot l}{p \cdot q}$ (here and in the following we use the shorthand $a \cdot b \equiv a^\mu b_\mu$ for the Lorentz contraction of 2 four-vectors). The leptonic $L_{\mu\nu}$ tensor for unpolarized leptons can be directly computed from QED and reads,
\begin{align}
	L_{\mu\nu}(l,l') &=  2 (l_\mu l'_\nu + l_\nu l'_\mu - \, l \cdot l' g_{\mu \nu})\, .
	\label{eq:Lmunu_unp}
\end{align}
The hadronic tensor $W^{\mu \nu}$, on the other hand, is an inclusive tensor containing all the information on the structure of the nucleon target. It is defined by summing the transition matrix elements of the electromagnetic current operator $J^\mu$ between the initial state nucleon and all possible unobserved final states $\mathcal{X}$,
\begin{align}
	2 M W^{\mu\nu}(p,q)	=
	\frac{1}{2\pi}\sum_\mathcal{X} \int \frac{ d^3\bm{p}_\mathcal{X} } 
	{(2 \pi )^3 2 E_\mathcal{X} } \, 
	(2\pi)^4 \delta^{(4)} \big( p + q - p_\mathcal{X}   \big)	\, 
	\langle N  | J^\mu(0) |  \mathcal{X} \rangle
	\langle  \mathcal{X} | J^\nu(0) | N \rangle \, ,
	\label{eq:Wmunu_DIS} 
\end{align}
\noindent where we have used the shorthand notation
$d^3\bm{p}_\mathcal{X}  /(2 \pi )^3 2E_\mathcal{X}  = \prod_{i\in \mathcal{X} } d^3\bm{p}_i /(2 \pi )^3 2E_i$,
with $p_\mathcal{X}$ the total momentum of the unobserved hadrons. For a derivation of these formulae, see for example Ref.~\cite{Bacchetta-lectures-2012} and the works cited therein.
%


We now wish to factorize the hadronic tensor in terms of a non-perturbative quark distribution function $q(x)$, and a perturbatively calculable photon-quark hard scattering term $\mathcal{H}^{\mu \nu}$, without relying on the assumption that $Q^2$ is asymptotically large - as done in most derivations, see for example \cite{Collins:2011zzd} - but still assuming that this is large enough to resolve individual quarks within the target.

\begin{figure}
	\centering
	\includegraphics[width=7cm]{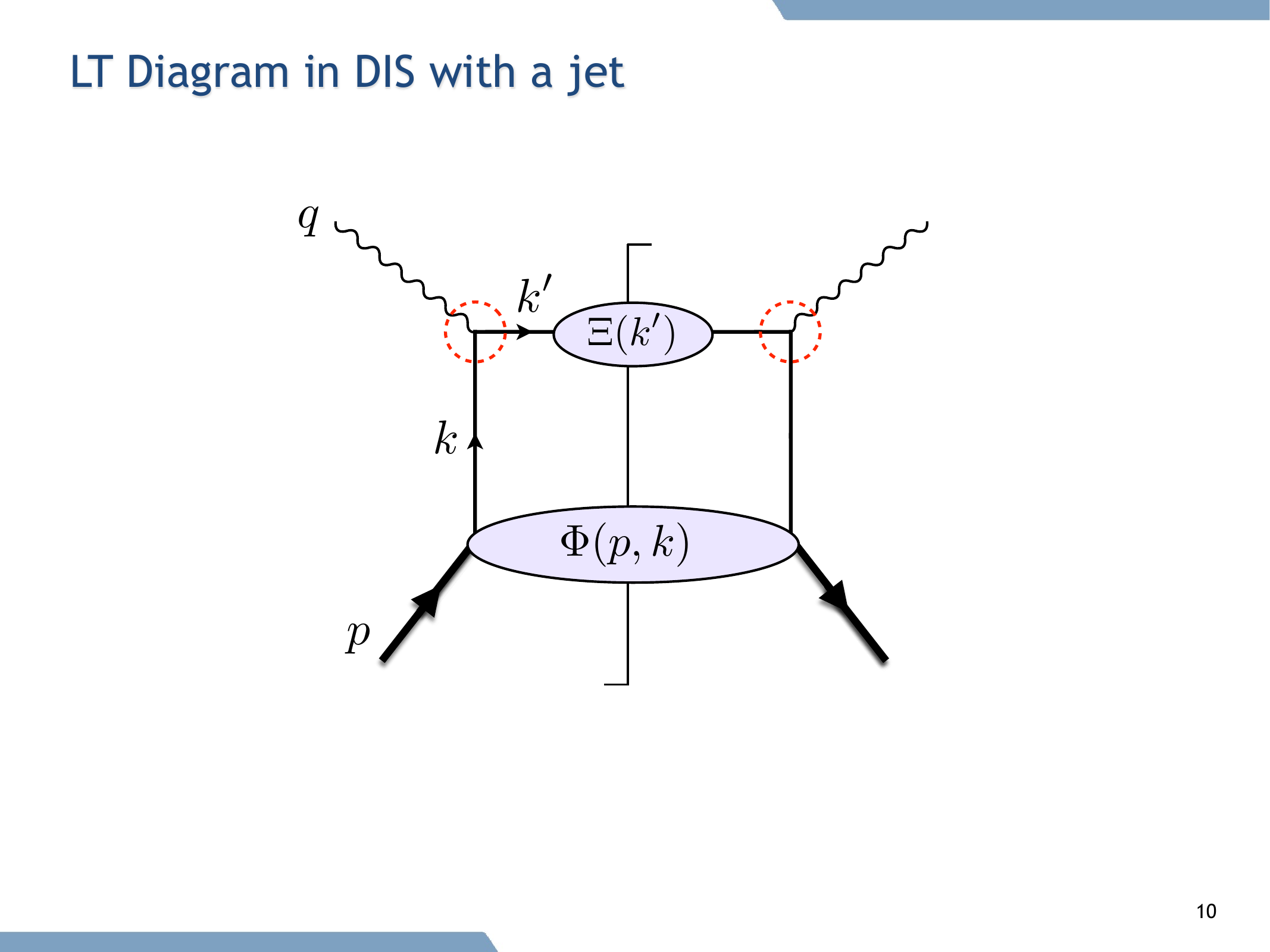}	
	\caption{Leading order inclusive DIS scattering handbag diagram, including a jet correlator in the top part. 
	The red circles indicate the hard scattering vertex, where kinematic approximations will be performed (see Section~\ref{sec:Derivation} for details).
	} 
	\label{fig:DIS_jet}
\end{figure}

Working at LO in the coupling constant, we consider the DIS handbag diagram shown in Figure~\ref{fig:DIS_jet}, where we have included the customary quark correlator $\Phi(p,k)$ in the bottom part \cite{Collins:2011zzd}, and an ``inclusive quark jet correlator'' $\Xi(k')$ in the top part \cite{Collins:2011zzd,Accardi:2019luo,Accardi:2020iqn,Procura:2009vm}. In the context of collinear factorization, the quark jet correlator was already used in Ref~\cite{Collins:2007ph,Accardi:2008ne,Moffat:2017sha} in order to correctly handle the external, hadron-level kinematic constraints in the DIS endpoint region, while allowing one to perform the parton-level momentum approximations needed to prove the factorizability of the DIS hadronic tensor. As a field theoretical object in its own right, the quark jet correlator has also been recently studied in Refs.~\cite{Accardi:2019luo,Accardi:2020iqn}, where it was used to derive a complete set  of fragmentation function sum rules and to provide a new way to study the dynamical breaking of chiral symmetry in QCD. In our derivation of factorization, we will incorporate insights from that analysis. In fact, using the jet correlator, we will be able to weaken the needed approximations on the quark's transverse momentum compared to other collinear factorization derivations \cite{Collins:2011zzd,Moffat:2017sha}. 

One can then write the hadronic tensor Eq.~\eqref{eq:Wmunu_DIS} as 
\begin{equation}
    2MW^{\mu\nu} 
    = 
    (2\pi)^3 \int d^4k\ d^4k'\
    \Tr \big[ \Phi(p,k)\, \gamma^\mu\, \Xi(k') \gamma^\nu
    \big] \, \, \delta^{(4)}(k + q - k') \, ,
\label{eq:Wmunufact}
\end{equation}
where the $\delta$-function encodes 4-momentum conservation in the photon quark hard-scattering vertex, indicated by red circles in Figure~\ref{fig:DIS_jet}, 
and the factor $(2\pi)^3$ in front of the integral comes from the phase space over the momentum of the $Y$ blob.
Following Ref.~\cite{Accardi:2020iqn}, the quark-distribution correlator $\Phi$ is defined as
\begin{equation}
    \Phi(p,k)
    = 
    \text{Disc} \int \frac{d^4 \xi}{(2\pi)^4}\, e^{i k \cdot \xi}\,
    \langle N(p) |\, \bar{\psi}(0)\, \psi(\xi)\, | N(p) \rangle \ ,
\label{eq:Phi_def}
\end{equation}
where $\psi$ is the quark field operator and $|N(p)\rangle$ single nucleon state with momentum $p$. The quark-to-jet correlator $\Xi(k')$ is analogously defined as 
\begin{align}
    \Xi(k') 
    = 
    \text{Disc} \int \frac{d^4 \eta}{(2\pi)^4}\, e^{i k' \cdot \eta}\,
    \langle\Omega|\, \psi(\eta)\, {\overline\psi}(0)\,  |\Omega\rangle \ ,
\label{eq:xi_def}
\end{align}
where $|\Omega\rangle$ is the interacting vacuum state, and can be interpreted as the discontinuity of the quark propagator~\cite{Accardi:2019luo,Accardi:2020iqn}. For simplicity, we work in light-cone gauge and therefore we can ignore the Wilson lines in the definition of either $\Phi$ or $\Xi$. Nonetheless, the sub-asymptotic kinematic assumptions we consider will only be made at the hard-scattering vertex, and will not change the derivation of the Wilson lines in QCD. Therefore, the results obtained in this work can be extended to any gauge.

\subsection{Factorization at asymptotic \boldmath{$Q^2$}}
\label{sec:asymptotic_CF}

We start the discussion on factorization, by reviewing the approximations taken in standard Collinear Factorization derivations, performed in the Bjorken limit at asymptotically high values of $Q^2$~\cite{Collins:2011zzd,Moffat:2017sha,Bacchetta-lectures-2012}. This will help us understanding where the assumption can be weakened if one wants to extend the procedure to sub-asymptotic values of the scale.

First of all, since the final state invariant mass is large, one can sum Eq.~\eqref{eq:Wmunu_DIS} over a complete set of states $\mathcal{X}$
replacing the jet correlator with a single quark line that passes the cut and can be considered a particle of zero mass\footnote{For light quarks, $m_q^2 \ll Q^2$ and quarks can indeed be considered massless. However, this is not strictly necessary, and one can treat the ``massless" $m_q^2/Q^2$ limit independently of the Bjorken limit. Hence the quark mass can be retained in the $\delta$ function, see for example Ref.~\cite{Moffat:2017sha}.}:
\begin{align}
  \Xi(k)  \longrightarrow \kpsl \delta(k'^2) \ .     
\end{align}
If one also immediately integrates over $d^4k'$, this turns into $(\ksl + \qsl)\, \delta\big((k+q)^2\big)$. We have thus \textit{arrived} at the starting point of most standard CF derivations, which is only valid if one assumes asymptotic values of $Q^2$ from the outset \cite{Collins:2011zzd}.

Next, in order to decouple the hard scattering from the soft quark dynamics in the target, one needs a suitable set of kinematic approximations on the subleading components of the unobserved initial state quark momentum $k$. For an inclusive DIS cross section the goal is to reduce the remaining 4-dimensional integral over $k$ to a 1-dimensional integral over the dominant $k^+ \sim Q$ component. This can be done in two independent steps. Firstly, one can neglect the quark's transverse momentum and assume that $\bm{k_T} \approx 0$. In other words, the incoming quark's 3-momentum is assumed to be parallel, or ``collinear'' to the 3-momentum of the target's nucleon . This ``collinear (kinematic) approximation'' is certainly a valid approximation at asymptotic $Q^2$ values because $|\bm{k_T}| \sim \Lambda_{QCD}$. Secondly, in order to preserve gauge invariance at the hard scattering vertex, one also needs to assume that $k^- \approx 0$, \textit{i.e.}, to also consider the scattering quark to be on-shell with $m_q=0$. This is a justified kinematic approximation, as well, since $k^- \sim \Lambda_{QCD}^2 / Q$, and moreover agrees with the intuitive DIS picture provided by Feynman's parton model~\cite{Feynman:1969ej,Feynman:1973xc}. We stress, nonetheless, that any assumption about the virtuality of the quark is in addition to its being or not collinear in 3-dimensional space to its parent hadron\footnote{In fact, collinear but virtuality-dependent quark distributions have been discussed in Ref.~\cite{Radyushkin:2015gpa}}.

Finally, applying the collinear approximation and setting $m_q=0$, one finds $\Phi(k) \approx k^+ \gamma^-$ and the hadronic tensor can be written as a 1-dimensional convolution: 
\begin{align}
    2MW^{\mu\nu} = \int \frac{dx}{x} \, q(x)  
    \left\{ \frac{1}{4} x
    \Tr\left[ \nbslash\,\gamma^\mu\,\nslash\,\gamma^\nu \right] \, 
    \delta(x-x_B) 
    \right\} \ ,
\label{eq:hadronic_tensor_asymptotic_CF}
\end{align}
where
\begin{align}
    q(x) = \frac{1}{2} \int dk^- \dtwokT \Tr\left[\Phi(k) \gamma^+ \right]_{k^+=xp^+} 
\end{align}
is the quark's light-cone plus momentum distribution, usually called ``collinear" quark PDF. As discussed above, this name involves a mild abuse of language, the essential feature being that it provides a 1-dimensional momentum distribution, integrated over the transverse and light-cone minus components.
The rest of the integrand, enclosed in curly brackets, can be interpreted as the tensor describing the scattering of a virtual photon with a massless quark traveling in the direction of its parent hadron (\ie, a ``collinear'' parton) and Eq.~\eqref{eq:hadronic_tensor_asymptotic_CF} provides one with a field-theoretical realization of Feynman's parton model -- or, more accurately, it shows how the parton model emerges in the large $Q^2$ limit of QCD calculations.

\subsection{Factorization at finite \boldmath{$Q^2$}}
\label{sec:Derivation}

At sub-asymptotic values of $Q^2$, we need to be more careful with the kinematic approximations since now have to deal with a set of hadrons in the final state's current region rather than a single on-shell quark. We then go back to the starting point, Eq.~\eqref{eq:Wmunufact}, \textit{before} integrating over $k'$ at variance with what did in the previous subsection. In this, we proceed similarly to derivations of factorization in SIDIS processes \cite{Bacchetta-lectures-2012}, which we take as a template for our derivation.

Since neither the $k$ nor $k'$ momenta are directly measurable, see Fig.~\ref{fig:DIS_jet}, we treat them as internal variables. We then need to approximate both the scattering and recoiled $k$ and $k'$ quark momenta appearing in the 4-momentum conservation $\delta$-function. Namely, we take
\begin{equation}
\delta^{(4)}(k + q - k') \simeq \delta^{(4)}(\widetilde{k} + q - \widetilde{k'})\, ,
\label{eq:four_delta_approx}
\end{equation}
with $\widetilde{k}$ and $\widetilde{k'}$ defined as,
\begin{align}
    k^{\mu} \approx \widetilde k^{\mu}
        & = \Big(xp^+,\frac{\vsqbar}{2xp^+},\bm{k_T}\Big) \label{eq:ktilde} \\
    k'^{\mu} \approx \widetilde k'^{\mu}
        & = \Bigg(\frac{\vpsqbar}{2k'^-},k'^-,\bm{k_{T}'}\Bigg) \ ,
\label{eq:kptilde}    
\end{align}
and the approximate light cone $\vsqbar$ and $\vpsqbar$ virtualities ideally chosen such that they approximate the respective averages, $\vsqbar \approx \langle v^2 \rangle= \langle k^2 +\kTsq \rangle$ and $\vpsqbar \approx \langle \vpsq  \rangle = \langle k'^2 + \kpTsq \rangle$. Note, that we are approximating only the sub-sub-leading momentum components of the scattering and recoiled partonic momenta, but we fully retain their individual transverse components. In this respect, we depart from the treatment of Refs~\cite{Collins:2011zzd,Moffat:2017sha}, and do not need further kinematic assumptions. 

Before carrying on with the derivation, it is important to remark that the approximation~\eqref{eq:four_delta_approx} is only made at the hard scattering vertex level, denoted with red circles in Figure~\ref{fig:DIS_jet}, and that this is the \textit{only} approximation we perform\footnote{As already stressed in Refs.~\cite{Collins:2007ph} and \cite{Collins:2011zzd}, working locally at the level of the hard scattering vertex instead of globally at the level of the whole Feynman diagram provides one with flexibility to adjust the kinematic approximation to the situation under discussion. In this paper we exploit this flexibility to address the factorization of DIS at sub-asymptotic $Q^2$ values, and we make our approximations very explicit as recommended by Ref.~\cite{Collins:2011zzd}.}.
Next, instead of neglecting the transverse momenta as in the standard derivation, we will perform a twist expansion of the nonperturbative correlators, and only then we integrate over the transverse transverse momenta. Finally (as in the standard derivation) we will obtain the hadronic tensor written as a 1-dimensional integral over the light-cone plus direction. In this sense, with a mild abuse of language as also discussed in the previous subsection, the factorized result can still be considered ``collinear''.

We can now carry on. In the approximated delta function~\eqref{eq:four_delta_approx}, the light-cone $k^+$ and $k'^-$ momentum components decouple from the transverse momenta,
\begin{align}
    \delta^{(4)}(\widetilde{k} + q - \widetilde{k'})
    =
    \delta^{(2)}(\boldsymbol{k_T}-\boldsymbol{k_T'}) \, \delta\Big(k^+ + q^+ -\frac{\vpsqbar}{2k'^-}\Big)  \delta\Big(\frac{\vsqbar}{2 k^+} + q^- -k'^-\Big) 
\end{align}
and the integrations over $dk^-$ and $dk'^{\,+}$ in Eq.~\eqref{eq:Wmunufact} can act directly on $\Phi(k)$ and $\Xi(k')$. 
Therefore, by defining the Transverse-Momentum Dependent (TMD) quark correlator as
\begin{align}
   \Phi(x,\bm{k_T}) \equiv \int dk^- \Phi(k) 
\label{eq:phi_int_minus}
\end{align}
and the TMD inclusive jet correlator \cite{Accardi:2019luo} as
\begin{align}
  J(k'^{-},\bm{k_T'}) 
  \equiv \frac{1}{2} \int dk'^{\, +} \Xi(k') ,
\label{eq:J_TMD}
\end{align}
we can write the hadronic tensor as
\begin{align}
    2MW^{\mu\nu} 
    & = 2 (2\pi)^3  \int dx \, dk'^- 
        \delta\Big(x - \xi -\frac{\vpsqbar}{2 p^+ k'^- }\Big)
        \delta\Big(\frac{\vsqbar}{2 x p^+} + q^- -k'^-\Big)
        \int \dtwokT\, \Tr \big[ \Phi(x,\bm{k_T}) \, \gamma^\mu\, J(k'^{-},\bm{k_T}) \gamma^\nu \big] \ .
\label{eq:Wmunu_II}
\end{align}
Note that we have written the integral over $dk^+$ in terms of $dx$, and that the integration over $\dtwokTp$ has set $\bm{k'_T}=\bm{k_T}$.
The remaining $\dtwokT$ transverse momentum integration acts only over the trace term, and the plus- and minus-direction delta functions fix the values of the light cone fraction $x$ and of the dominant $k'^-$ component of the recoiled quark momentum, respectively. One can here appreciate the importance of approximating the quark light-cone virtualities rather than their mass in Eqs.~\eqref{eq:ktilde}-\eqref{eq:kptilde}: in that case, we would be able to achieve delta function transverse decoupling only if we also approximated to zero the quark transverse momenta. As it will become clear by the end of this subsection, this additional approximation on $\bm{k_T}$ is not necessary to achieve factorization.

We still need to decouple the quark and jet correlators in the trace appearing in Eq.~\eqref{eq:Wmunu_II}. To this end, we introduce the ``operational'' twist expansion~\cite{Jaffe:1996zw,Bacchetta-lectures-2012} for the TMD correlators, $\Phi(x,\bm{k_T})$ and $J(k'^{-},\bm{k_T'})$. As with the kinematic approximations discussed above, this dynamical expansion is predicated on the existence of a hard scale determining a large boost in the light cone direction such that the scattering quark momentum component satisfy $k^+ \gg |\boldsymbol{k_T}| \gg k^-$, and for the recoiled momentum $k'^- \gg |\boldsymbol{k'_T}| \gg k'^+$. In DIS, such a scale is provided by the photon's virtuality $Q^2$, and one can consider $p^+ \sim k'^- \sim Q$.
The quark correlator can then be expressed as a power expansion in $M/p^+$, where the power counting scale $M$ can be identified with the proton mass ~\cite{Bacchetta:2006tn}. Limiting ourselves to the unpolarized sector, we write
\begin{align}
    \Phi(x,\bm{k_T}) = \frac{1}{2} q(x,\kTsq) \, \nbslash + \frac{M}{2 p^+}\Big[e(x,\kTsq)\, \id  + q^\perp(x,\kTsq) \frac{\kTsl}{M} \Big] + O\bigg(\frac{M^2}{(p^+)^2}\bigg)
\label{eq:phi_int_minus_II}
\end{align}
where $q(x,\kTsq)$ is the unintegrated unpolarized parton distribution function, while $e(x,\kTsq)$ and $q^\perp(x,\kTsq)$ are twist-3 level parton distributions. The latter describes the non-perturbative dynamics of the quark's intrinsic transverse momentum.
For the TMD inclusive jet correlator, the power counting scale can be identified with the $\Lambdatw$ confinement scale and the correlator expanded in powers of $\Lambdatw/k'^-$~\cite{Accardi:2018gmh,Accardi:2019luo}:
\begin{align}
J(k'^-,\bm{k'_T}) & = 
\frac{1}{2} \alpha(k'^-)\nslash + \frac{\Lambdatw}{2 k'^-} \Big[\zeta(k'^-)\, \id  +  \alpha(k'^-) \frac{\kpTsl}{\Lambdatw} \Big] +  O\bigg(\frac{\Lambdatw^2}{(k'^-)^2}\bigg) \ .
\label{eq:J_TMD_II}
\end{align}
The leading twist coefficient $\alpha(k'^-) = \frac{\theta(k^-)}{2 (2\pi)^3}$, independent of $\bm{k'_T}$, is the analog of the unpolarized $D_1$ fragmentation function in SIDIS \cite{Bacchetta:2006tn} but integrated over the detected hadron momentum, and summed over all hadron flavors (indeed we are considering  inclusive DIS events, where the final state remains undetected).  The chiral-odd twist-3 coefficient $\zeta(k'^-)=\frac{\theta(k^-)}{2 (2\pi)^3} M_j/\Lambdatw$ is also independent of $\bm{k'_T}$, and $M_j=m_q+m^\text{corr}$ includes perturbative and non-perturbative ``jet mass'' contributions \cite{Accardi:2018gmh,Accardi:2020iqn}; it is the analog of the chiral-odd fragmentation function $E$ \cite{Bacchetta:2006tn}. The transverse momentum thus appears explicitly as a kinematic factor in the twist expansion \eqref{eq:J_TMD_II}.

We can now expand the trace appearing in Eq.~\eqref{eq:Wmunu_II}, which reads 
\begin{align}
    &2 (2\pi)^3 \int \dtwokT  \Tr \left[ \Phi(x,\bm{k_T}) \, \gamma^\mu\, J(k'^{-},\bm{k_T}) \gamma^\nu \right] 
    \nonumber \\
    & \quad\quad = \int \dtwokT  \bigg[\frac{1}{4} q(x,\kTsq) \Tr [\nbslash \, \gamma^\mu \,  \nslash \, \gamma^\nu] 
    + \frac{M \Lambdatw}{4 p^+ k'^-}\Big( \frac{M_j}{\Lambdatw} \, e(x,\kTsq) \, \Tr\big[\gamma^\mu \gamma^\nu\big] 
    +  q^\perp(x,\kTsq)  \frac{\Tr\big[\kTsl \, \gamma^\mu\, \kTsl \, \gamma^\nu\big]}{M\Lambdatw}\Big)  \bigg]
    \nonumber \\
    & \quad\quad = \underbrace{\frac{1}{4} \, q(x)\,  \Tr [\nbslash \, \gamma^\mu \,  \nslash \, \gamma^\nu]}_{\text{twist 2}} 
    + \frac{M \Lambdatw}{p^+ k'^-} \underbrace{\Big( \frac{M_j}{\Lambdatw} e(x) g^{\mu\nu} 
    + {\frac{2M}{\Lambdatw}} 
    q^{\perp(1)}(x) { \bar n^{\{\mu}n^{\nu\}} } 
    \Big)}_{\text{twist 4}}
    + O\big(\frac{M^2}{p^{+2}}\big)
    + O\big(\frac{\Lambdatw^2}{k'^{-2}}\big)
    + \text{HT} \ .
\label{eq:twist_expansion_trace}
\end{align}
up to higher-twist (HT) terms. 
At twist-2 level, and analogously to the asymptotic derivation outlined in Section~\ref{sec:asymptotic_CF}{}, the integration over $\dtwokT$ acts only on the quark distribution and produces the standard collinear PDF
\begin{align}
    q(x) \equiv  \int \dtwokT  q(x,\kTsq) \ .
\end{align}
Gauge invariance is guaranteed by $q_\mu \Tr[\,\nbslash\,\gamma^\mu\,\nslash\,\gamma^\nu] = 0$ despite having assumed a different light-cone virtuality $\vsqbar$ and $\vpsqbar$ for the scattering and recoiled partons. This result is only possible thanks to the twist expansion of the jet correlator, introduced in the handbag diagram instead of a single final-state quark line, and clearly goes beyond a naive parton-model treatment of the process.

At twist-4, we find a contribution from the collinear $e(x)\equiv\int \dtwokT e(x,\kTsq)$ distribution multiplied by the jet mass $M_j$, as well as from the first $\kTsq$ moment of the TMD $q^\perp(x,\kTsq)$ distribution, $q^{\perp(1)}(x)= \int \dkTtwo \frac{\kTsq}{2M^2}\, q^\perp(x,\kTsq)$ \cite{Bacchetta:2006tn,Bacchetta-lectures-2012}. One can thus explicitly see that the dynamics of the parton transverse momentum is not neglected in this approach, but rather included in twist-4 terms. The twist-4 terms appearing in Eq.~\eqref{eq:twist_expansion_trace} are, however, not gauge invariant by themselves. Gauge invariance can nonetheless be restored by properly summing these to contributions stemming from the inclusion of 4-parton matrix elements in the handbag diagram \cite{Ellis:1982cd,Jaffe:1983hp,Qiu:1988dn}. This is left for future work. 

We would also like to remark that the approach we have followed here is not entirely new. In fact it closely corresponds to the treatment of SIDIS cross sections in terms of transverse-momentum-dependent PDFs \cite{Bacchetta:2006tn,Bacchetta-lectures-2012}, and a detailed correspondence can be obtained through the use of the fragmentation function sum rules developed in \cite{Accardi:2017pmi,Accardi:2019luo,Accardi:2020iqn}. The SIDIS formalism is, however, at present fully developed only up to twist-3 level.

Finally, the factorized hadronic tensor at leading-twist (LT) can be written as a convolution of a hard scattering tensor $\mathcal{H}^{\mu \nu}$ and the collinear PDF $q(x)$,
\begin{align}
    2M W^{\mu\nu} \big|_\text{LT}
    & =
    \int \frac{dx}{x} \mathcal{H}^{\mu \nu} (x,\bar{x}) \, q(x) 
\label{eq:factorized_Wmunu_LT}
\end{align}
with 
\begin{align}
    \mathcal{H}^{\mu \nu} (x,\bar{x}) 
    & = \frac{1} {4} \, x \, \delta(x-\bar{x})\, \Tr \left[ \nbslash \, \gamma^\mu\, \nslash \,
\gamma^\nu \right] \frac{1}{|J_{x,k'^-}|}
\label{eq:Hardmunu}
\end{align}
In this equation, $\bar{x}$ arises from the manipulation of the delta function appearing in Eq.~\eqref{eq:Wmunu_II}, 
\begin{align}
    \bar{x} 
    &= \frac{\xi}{2} \Bigg(1 + \frac{\vpsqbar-\vsqbar}{Q^2} + \sqrt{\Big(1 + \frac{\vpsqbar-\vsqbar}{Q^2}\Big)^2+ 4\frac{\vsqbar}{Q^2}} \Bigg)\label{eq:x_sol_fact}\\
    &= \xi \Bigg(1 + \frac{\vpsqbar}{Q^2} - \frac{\vsqbar\vpsqbar}{Q^4} + O\Big(\frac{\mu^6}{Q^6}\Big) \Bigg) \ ,
\label{eq:x_sol_fact_I}
\end{align}
and depends on two mass scales, namely, the approximate incoming and outgoing light cone parton virtualities $\vsqbar$ and $\vpsqbar$, collectively denoted by $\mu^2$. Note that the incoming parton's virtuality $\vsqbar$ only contributes at $O(1/Q^4)$, and can be parametrically neglected. 
The Jacobian factor $J_{x,k'^-} = 1 - \frac{\vsqbar \vpsqbar}{4 (x p^+ k'^-)^2 }$, which also arises from manipulations of the mentioned delta functions, reads
\begin{equation}
	\frac{1}{J_{x,k'^-}} 
	= 1 + \frac{\vsqbar \vpsqbar}{Q^4} \Bigg(\frac{1}{(\frac{x}{\xi} + \frac{\vsqbar}{Q^2})^2-\frac{\vsqbar \vpsqbar}{Q^4}}\Bigg) \ , 
\label{eq:Jacobian}
\end{equation}
and deviates from 1 by a term scaling as the fourth inverse power of $Q$. 
The (average) light-cone virtualities $\vsqbar$ and $\vpsqbar$ are of non perturbative origin, so in general they follow $\vsqbar \sim \vpsqbar \sim \Lambda^2_\text{had}$, where $\Lambda_\text{had}$ is some hadronic scale. Using this one finds that $ J_{x,k'^-} = 1 + O \big( \frac{\Lambda_\text{had}^4}{Q^4} \big)$. Hence, the choice of virtualities will play a secondary role in $J_{x,k'^-}$ compared to the determination of $\bar x$. However, they play a very important role in the determination of $\bar{x}$.

\subsection{Kinematic approximations in QCD}
\label{sec:kin_approx_QCD}

As discussed in Section~\ref{sec:Derivation}, collinear factorization requires one to approximate the incoming and outgoing quark light-cone virtualities, namely to take $v^2 \approx \bar v^2$ and $\vpsq  \approx \vpsqbar$. 
However, $v^2$ and $\vpsq$ are not observable and cannot be experimentally controlled. One needs therefore  to resort to a physically or theoretically motivated \textit{Ansatz} to choose suitable $\vsqbar$ and $\vpsqbar$ values.
For this purpose, we will derive kinematic bounds on $v^2$ and $\vpsq$ valid at any order in perturbation theory, and use these to obtain good \textit{Ans\"atze} for $\bar v^2$ and $\vpsqbar$. We will then test these in Section~\ref{sec:testing_kin}.

\subsubsection{Choice of \texorpdfstring{$\vpsqbar$}{Lg}}

We start by considering the jet subdiagram on top of Figure~\ref{fig:DIS_jet}, in which the struck quark of momentum $k'$ is fragmenting in a number of particles. By fermion number conservation, the incoming quark line should also pass the cut and appear in the final state. Thus, by 4-momentum conservation, %
\begin{align}
    \vpsq 
    = k'^2 + \kpTsq 
    \geq m_q^2 + \kpTsq \ .
\label{eq:vp_inequality}
\end{align}
In fact, the quark needs to fragment into at least one hadron and therefore
one should take into account the mass of the lightest hadron the quark can quark hadronize into, that is, the pion. Hence, a tighter bound is
\begin{align}
    \vpsq
    \geq m_\pi^2 + \kpTsq \ .
\label{eq:vp_inequality_II}
\end{align}
We can then use the lower bound as a minimal approximation of the average $\langle \vpsq \rangle$. However, in a fully inclusive scattering, the transverse $\kpTsq$ momentum cannot be experimentally controlled. Therefore, we choose  
\begin{align}
    \vpsqbar = m_\pi^2 \ . 
    \label{eq:vpsqbar_choice_QCD}
\end{align}
Note that with this choice we depart from the standard derivations of collinear factorization, where $\vpsqbar$ is approximated to zero for light quarks. A similar argument for flavor tagged inclusive measurements such as of the charm structure function $F_2^c$ would lead to $\vpsqbar =m_D^2 \approx m_c^2$, with $m_D$ the $D$ meson mass and $m_c$ the $c$ quark mass, and eventually to scaling variables such as advocated in \cite{Aivazis:1993kh,Nadolsky:2009ge}.

As QCD transitions from perturbative to non-perturbative degrees of freedom at large $x_B$ it is also possible that the virtual photon couple not to a single quark but to a composite partonic substructure, and $\vpsqbar$ could also be considered to be the invariant mass of the latter \cite{Moffat:2019qll}. Without a specific model to guide one's choice, one could treat $\vpsqbar$ as a phenomenological parameter and determine this for example in a global QCD analysis of inelastic data. However, large values for this parameter, that Ref.~\cite{Moffat:2019qll} would interpret as evidence for nucleonic substructure, might instead emerge as the fit effectively subsumes the dynamics of the quark hadronization process or the kinematic shifts induced by the quark's transverse momentum into an effective $\vpsqbar \approx \langle m_Y^2 \rangle + \avekTsq$ parameter. We will come back to these considerations when we discuss the scaling variable $\bar x$ at the end of this subsection.

\subsubsection{Choice of \texorpdfstring{$\vsqbar$}{Lg}}

Considering now target fragmentation, \textit{i.e.}, the lower vertex in the DIS diagrams of Figure~\ref{fig:DIS_diagram} right we find that 
\begin{align}
	p_X^2 &= (p-k)^2 
	 = (1-x)\Big(M^2 - \frac{k^2 + \kTsq}{x} \Big) - \kTsq \ ,
\end{align}
hence the light cone virtuality $v^2 = k^2 + \kTsq$ reads
\begin{align}
    v^2
	& = -\frac{x}{1-x} \Big[ (p_X^2 - M^2) + x M^2 + \kTsq)\Big] \ .
  \label{eq:v2_QCD}
\end{align}
Note that the light-cone virtuality vanishes as $x\to0$, and becomes negative as $x\to1$. 
Imposing baryon number conservation in the right diagram of Figure~\ref{fig:DIS_diagram} requires the presence of at least one baryon in the final state. Assuming the baryon number flows into the target jet, imposes that the remnant $X$ minimally contains a nucleon, $p_X^2 \geq m_X^2 \geq M^2$. Hence,
\begin{align}
  v^2 & \leq -\frac{x}{1-x} \Big[x M^2 + \kTsq \Big]  \ ,
\end{align}
with the upper bound representing the case in which the target jet is made of just one nucleon. It is then reasonable to choose
\begin{align}
    \vsqbar = -\frac{x^2}{1-x} M^2 
\label{eq:vsqbar_QCD}
\end{align}
as a minimal approximation of the average $\langle v^2 \rangle$, where we also neglected the internal variable $\kTsq$ that cannot be experimentally controlled in inclusive lepton-proton scattering. 

A simplified choice can be obtained by noticing that in
Eq.~\eqref{eq:x_sol_fact_I}, the light-cone fraction $\bar{x} = \xi (1 + \vpsqbar/Q^2 - \vsqbar\vpsqbar/Q^4) + O(\mu^6/Q^6)$ depends on $\bar v^2$ only starting at order $O(\mu^4/ Q^4)$ and can be approximated in first instance as
$
    \bar x \approx \xi \Big(1 + \frac{\vpsqbar}{Q^2} \Big)
$. This is equivalent to effectively choosing 
\begin{align}
    \vsqbar = 0
\label{eq:vsqbar_choice_QCD}
\end{align}
in Eq.~\eqref{eq:x_sol_fact_I}. As one can expect from Eq.~\eqref{eq:v2_QCD} and we will numerically confirm in our model calculation, this is in fact a good approximation for $\langle v^2 \rangle$ at not too large values of $x_B$, and as long as one considers small enough light-cone virtualities $\vpsqbar$. It is also the approximation taken in the parton model, and in  standard derivations of collinear factorization, where $k^2 \approx m_q^2 \approx 0$ at the same time as $\bm{k_T} \approx 0$.

\subsubsection{Light cone fraction \texorpdfstring{$\bar{x}$}{Lg}}

Having discussed possible choices for \textit{Ans\"atze} for $\bar v^2$ and $\vpsqbar$, we can focus our attention on the light-cone fraction $\bar{x}$ derived in Eq.~\eqref{eq:x_sol_fact_I}.

Far from kinematic thresholds, namely for not too large values of $x_B$, one can choose $\vsqbar=0$ for the scattering quark, see Eq.~\eqref{eq:vsqbar_choice_QCD}. Using furthermore $\vpsqbar=m_\pi^2$ from \eqref{eq:vpsqbar_choice_QCD} for the recoiling quark, one obtains\footnote{The $\xi_\pi$ variable is in fact analogous to the $\chi$ scaling variable used in Ref.~\cite{Aivazis:1993kh,Nadolsky:2009ge} to study charm production in charged current $W+s \to c$ events. The derivation we offer here translates naturally to charm production, by replacing $\pi \rightsquigarrow D$ and approximating the mass of the $D$ meson with the mass of the charm quark, $M_D \approx m_c$.} 
\begin{align}
    \bar x \approx \xi \Big(1 + \frac{m_\pi^2}{Q^2} \Big) \equiv \xi_\pi\ ,
\label{eq:xbar_thr_QCD}
\end{align}
Note that with a non-zero $\vpsqbar$ like in Eq.~\eqref{eq:vpsqbar_choice_QCD} we are more closely respecting the internal kinematics of the handbag diagram than with $\vpsqbar=0$. Therefore, we can expect that $\bar{x} = \xi_\pi$ will provide a better approximation to the non-factorized diagram's than in standard collinear factorization.

At larger values of $x_B$, \textit{i.e.}, closer to the kinematic threshold, the virtuality \eqref{eq:v2_QCD} diverges to minus infinity, and a different approximation may be needed.  
In this regime, a suitable approximation to $v^2$ that is valid in both the small-$x_B$ and large-$x_B$ regimes is
\begin{align}
    \vsqbar(\bar x)
    = - \frac{\bar x^2}{1-\bar x} M^2  ,
\label{eq:v2_new_approx_QCD}
\end{align}
where we replaced $x=\bar{x}$ in Eq.~\eqref{eq:v2_QCD}.
Substituting Eq.~\eqref{eq:v2_new_approx_QCD} in Eq.~\eqref{eq:x_sol_fact_I}, solving the resulting equation for $\bar x$ perturbatively in powers of $\mu^2/Q^2$, and finally setting $\vpsqbar=m_\pi^2$ as in Eq.~\eqref{eq:vpsqbar_choice_QCD}, we find
\begin{align}
    \bar x 
    \approx \xi \Bigg[ 1 + \frac{m_\pi^2}{Q^2} 
    + \frac{\xi^2}{1-\xi} 
    \frac{m_\pi^2 M^2}{Q^4}\Bigg]
    \equiv \xipinew 
\label{eq:xbar_new_QCD}
\end{align}
up to corrections of $O\Big( \frac{\mu^6}{Q^6} \Big)$.
Note that at small $x_B$ the fourth order term quickly vanishes, and one recovers Eq.~\eqref{eq:xbar_thr_QCD}. Closer to the kinematic threshold, this new scaling variables accounts for the non-vanishing of the scattering quark's light-cone virtuality. The latter can be approximated by substituting Eq.~\eqref{eq:xbar_new_QCD} back in Eq.~\eqref{eq:v2_new_approx_QCD} we can also obtain an approximation for quark's virtuality, and written purely in terms of the external variables:
\begin{align}
    \vsqbarnew \equiv \vsqbar(\xipinew) \ . 
\label{eq:vbarsq_new}
\end{align}
This is the best approximation to the unobserved scattering parton's virtuality we can obtain without measuring the hadronic final state. However, as we will verify in Section~\ref{sec:testing_kin}, using Eq.~\eqref{eq:xbar_new_QCD} instead of Eq.~\eqref{eq:xbar_thr_QCD} has little effect on calculations of the factorized cross section, and Eq.~\eqref{eq:xbar_thr_QCD} is a sufficient approximation.

As already noted earlier, the approximated $\vpsqbar$ virtuality could also be considered as a free parameter and determined in a PDF fit utilizing $\bar x = \xi (1+\vpsqbar/Q^2)$ instead of the prescription \eqref{eq:xbar_thr_QCD}. The interpretation of the obtained $\vpsqbar$ value, however, may not be straightforward even if found to be substantially larger than $m_\pi^2$, which is, according to our analysis, the minimum expected value. On the one hand, Ref.~\cite{Moffat:2019qll} argues that any improvement in the fit would signal the emergence of composite partonic substructure in the nucleon target. On the other hand, the quark can in general hadronize to more than one particle and the fitted $\vpsqbar$ can naturally be expected to be larger than $m_\pi^2$. Furthermore, even apart from hadronization dynamics considerations, we will show in Section~\ref{sec:fact_limits} that an improvement in the CF description of a DIS structure function can also emerge if the unobserved quark's transverse momentum is kinematically accounted for by choosing $\vpsqbar = m_\pi^2 + \avekTsq$, with $\avekTsq$ a free parameter of $O(\Lambda_\text{QCD}^2)$. Thus, large values of $\vpsqbar$ may not necessarily indicate the presence of nucleon substructures other than asymptotically free partons.

\subsection{Discussion}

Despite its simplicity, formula \eqref{eq:Hardmunu} is non-trivial and it is worthwhile summarizing under what conditions it has been obtained. 

First of all, we emphasize once more that the transverse momentum is not approximated but rather included in higher-order terms in the twist expansion, which provides a \textit{controlled dynamical approximation}.
In fact, using the quark field's equation of motion relations, one can show that the twist-3 PDF $e$ can be decomposed as $xe = x\, \widetilde{e} + \frac{m_q}{M}q$, where $\widetilde{e}$ correspond to a ``pure'' twist-3  dynamical contribution~\cite{Bacchetta:2006tn}. Similarly, the jet mass can be decomposed as $M_j=m_q+m^\text{corr}$, where $m^\text{corr}$ is the dynamical mass of the jet~\cite{Accardi:2019luo,Accardi:2020iqn}. Thus the twist expansion \eqref{eq:twist_expansion_trace} does not only provide an expansion in the transverse momentum effects, but also in the quark mass contributions in a way that is reminiscent of Ref.~\cite{Boer:1996ay}. The quark mass expansion is appropriate as long as the quarks are light enough, otherwise one might want to include the term proportional to $m_q^2$ in the LT partonic tensor.

The only uncontrolled approximation we have performed is a purely kinematic one. Namely, we have fixed the value of the quark light-cone virtualities $v$ and $v'$ of the initial and final state quarks, such that their sub-dominant momentum components are approximated by $k^- \approx \vsqbar/2k^+$ and $k'^+ \approx \vpsqbar/2k'^-$. But, crucially, this approximation is \textit{only} taken inside the parton-level 4-momentum conservation delta function, which is part of the ``hard scattering'' graphically identified by red circles in Figure~\ref{fig:DIS_jet}. Thus, following the philosophy of \cite{Collins:2011zzd}, we have confined the only needed \textit{non-controlled} approximation to the hard interaction and kept the parton momenta otherwise unapproximated. The price to be paid for this approximation -- which is the \textit{minimal} kinematic approximation compatible with collinear factorization! -- is that transverse momentum conservation in the approximated hadronic tensor is effectively broken, and this sets an inescapable limit to the validity of the CF formula at large $x_B$ \cite{Moffat:2017sha}. We will numerically study this limit for the benchmark spectator model in Section~\ref{subsect:breakingCF}. 

Clearly, Eq.~\eqref{eq:Hardmunu} reduces to the parton model result for light quarks in the $\vsqbar \to 0$ and $\vpsqbar \to 0$ limit, in which the partons are taken to approximately travel on the light cone. But even then, the quarks need not be approximately on their $k^2 \approx k'^2 \approx m_q^2 \approx 0$ mass shell, as often stated in literature, unless one further assumes - with no need - that the quarks are real particles.
For generic values of the approximated $\vsqbar$ and $\vpsqbar$ quark virtualities, a gauge invariant factorized hadronic tensor can only be obtained if one considers the jet diagram in Figure~\ref{fig:DIS_jet} and its twist expansion. Had we worked in the parton model from the outset, or even in QCD but with a perturbative quark line instead of a jet correlator in the top part of the handbag diagram of Figure~\ref{fig:DIS_jet}, this would not have been possible.

We have thus found in Eq.~\eqref{eq:factorized_Wmunu_LT}-\eqref{eq:x_sol_fact_I} a gauge invariant generalization of the standard collinear factorization procedure for the hadronic tensor \eqref{eq:Wmunufact}, which is now also valid at sub-asymptotic values of $Q^2$ 
and does not require one to approximate to zero the virtuality of the scattering and recoiling partons. With this added flexibility, in Section~\ref{sec:testing_kin} we will study a range of choices for $\vsqbar$ and $\vpsqbar$ in order to maximally extend the range of validity of the LT factorization approximation towards large $x_B$ and low $Q^2$ values.

\section{Validation framework: the diquark spectator model}
\label{sec:model}

As we discussed in Section~\ref{sec:CF}, deeply inelastic lepton-nucleon scattering involves the fragmentation of the proton or neutron target, as illustrated in the left panel of Figure~\ref{fig:DIS_diagram}.
Target fragmentation in QCD is a complex, non-perturbative process that cannot be computed exactly, as yet. Instead, we wish to mimic this with a suitable proton-quark-meson vertex in a model theory, where full analytical calculations of the structure functions can be compared to their collinear approximation and thus validate the sub-asymptotic CF procedure derived in the previous Section.     

\begin{figure}[bt]
	\centering
	\includegraphics[width=9cm]{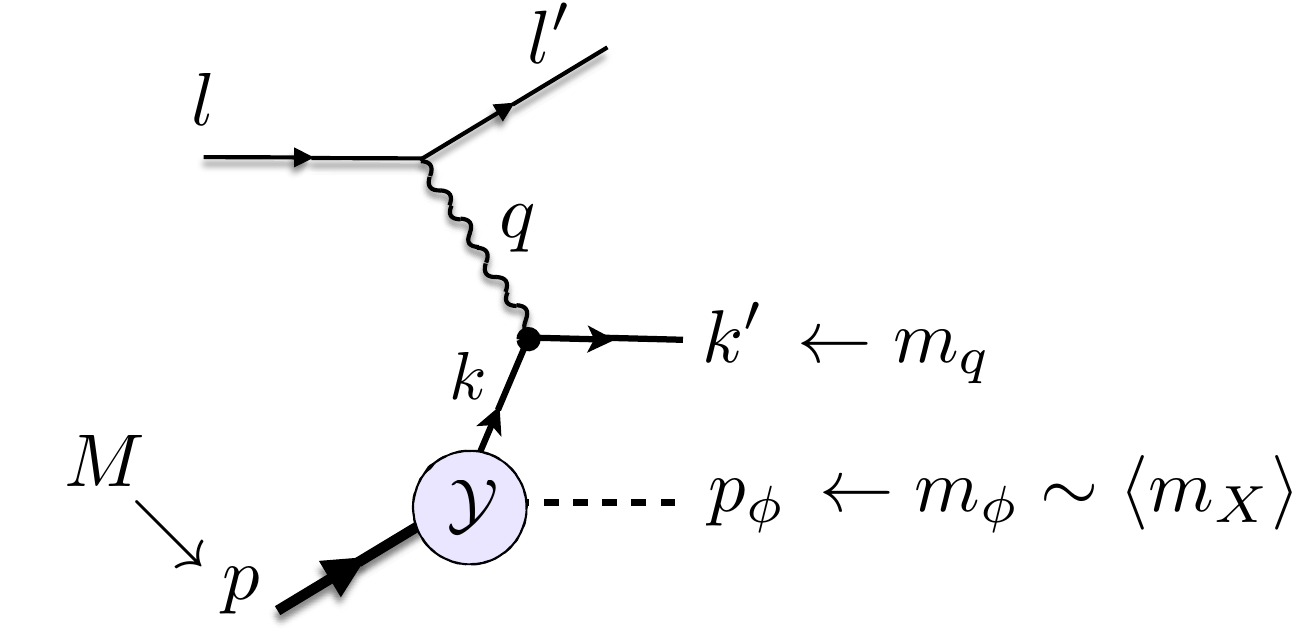}	
	\caption{Electron-proton DIS at leading order in the strong coupling constant in the spectator model, where we replaced the remnant by a scalar diquark $\phi$ of mass $m_\phi$, assumed to be of the order of the average remnant mass $m_X$, and $\mathcal{Y}$ is the model's proton-quark-diquark vertex, with ultraviolet modes cut off by a suitable form factor. The labels indicate the momenta of the particles involved in the collision: $q$ is the momentum of the photon; $p$ is the momentum target nucleon; $k$ and $k'$ are the momenta of the incoming and scattered quark participating in the hard scattering; $p_\phi$ is the momentum of the spectator diquark. The particle masses are also explicitly indicated. 
	}   
	\label{fig:DIS_model}
\end{figure}

We then consider an idealized field-theoretical model describing  an electrically charged spin 1/2 particle of mass $M$, that plays the role of a proton and contains a charged active quark of mass $m_q$ and a neutral scalar diquark spectator, $\phi$, of mass $m_\phi$ \cite{Bacchetta:2008af,Moffat:2017sha}. In this model the proton's remnant $X$ is mimicked by the spectator $\phi$, with $m_\phi$ of the order of the average remnant's invariant mass $\langle m_X \rangle$, and we can simulate $e+p$ collisions by studying the diagram in Figure~\ref{fig:DIS_model}. 

For the proton-quark-diquark vertex we choose the following structure:
\begin{align}
  \mathcal{Y}= i g(k^2) 1\!\!1 \ ,
  \label{eq:model-vertex}
\end{align}
where $g(k^2)$ generically denotes a form factor which takes into account that a diquark is, in fact, a composite field. In this paper, we choose for simplicity the dipolar form factor
\begin{equation}
  g(k^2) = g \frac{k^2-m_q^2}{|k^2-\Lambda^2|^2} \ ,
    \label{eq:dipolar-form-factor}
\end{equation}
where $g$ is a dimensionful coupling constant and $\Lambda$ a parameter.  This vertex is infrared safe, and smoothly cuts off ultraviolet modes in the quark leg when $k^2$ is much larger than $\Lambda^2$. This is an effective way of simulating confinement in the proton target, since the cutoff imposes a length scale of order $1/\Lambda$. The strong coupling constant $g$ does not play a significant role in our discussion, and we set this to $g=1$ GeV$^2$ for simplicity. The confinement scale $\Lambda$ and the spectator mass $m_\phi$, are considered free parameters of the model, and are meant to capture the salient non-perturbative features of the DIS process. Other possible choices of form factor, including an exponential form and a combination of scalar and axial diquarks to simulate up and down quarks have been discussed in Ref.~\cite{Bacchetta:2008af}. 

The model parameters can be determined by fitting the analytic calculations of parton distribution functions, which are possible in the model due to the relative simplicity of the vertex, to phenomenological extractions from experimental data \cite{Jimenez-Delgado:2013sma,Ethier:2020way}. Here we adopt the values fitted in Ref.~\cite{Bacchetta:2008af} to the PDFs determined by the ZEUS collaboration \cite{Chekanov:2002pv}, namely, 
\begin{alignat}{6}
  m_\phi&=0.822 \text{\ GeV},
  & \quad\quad
  \Lambda & = 0.609 \text{\ GeV},
  \label{eq:mphi_Lambda_default}
\end{alignat}
with quark and proton masses kept fixed at
\begin{alignat}{6}
  m_q & = 0.3 \text{\ GeV},
  & \quad\quad
  M & = 0.939 \text{\ GeV}.
  \label{eq:mq_M_default}
\end{alignat}
We will use these values as default, but we will also consider variations around these numbers in order to study the systematics of the sub-asymptotic collinear factorization scheme to be discussed later. We also note that with these mass parameters, $m_q+m_\phi > M$ and the proton is a stable particle as it happens in QCD. 

In this work, we consider $m_q \neq 0$ in order to study the kinematic dependence of the process on the mass generated in the final state, and how to retain this in the collinear factorization of the DIS cross section. 
Similar studies for the inclusive DIS process have been performed in Refs.~\cite{Aivazis:1993kh,Tung:2001mv,Nadolsky:2009ge} with a focus on heavy-quark production; for semi-inclusive DIS in Refs~\cite{Guerrero:2015wha,Guerrero:2017yvf}, that focused on kinematic corrections induced by the mass of the detected hadron; and in Ref.~\cite{Moffat:2019qll} with the aim of identifying “clustered” substructures within the target.
In all these works, masses are taken into account by a suitable rescaling of the struck quark's momentum light-cone fraction $x$. In this paper, we revisit the basis for these scaling approximations utilized and test these against the full analytic model calculation of the process. As in Ref.~\cite{Moffat:2017sha}, we will restrict our analysis to light quarks with mass much smaller than the charm's, $m_q \ll m_c$, and in particular consider values of the order of the strange quark mass, $m_q \sim m_s$. 

Lastly, we would like to stress that $m_\phi$ and $\Lambda$ are ``internal'', unobservable parameters of the model, in the same way that the QCD confinement scale or the remnant mass cannot be directly measured in electron-proton scattering. Conversely, even in an inclusive measurement, we treat the quark mass as a known ``external'' parameter. In QCD, this would be akin to what happens in measurements of the charmed structure function $F_2^c$ in DIS~\cite{Adloff:2001zj,Aubert:1982tt}, where the active charm quark can be tagged by identifying a heavy flavor hadron in the hadronic final state, without however measuring its momentum, or the momentum of any other hadron.

\subsection{Calculation of the hadronic tensor}
\label{sec:full_hadronic_tensor}

\begin{figure}
	\centering
	\includegraphics[width=15cm]{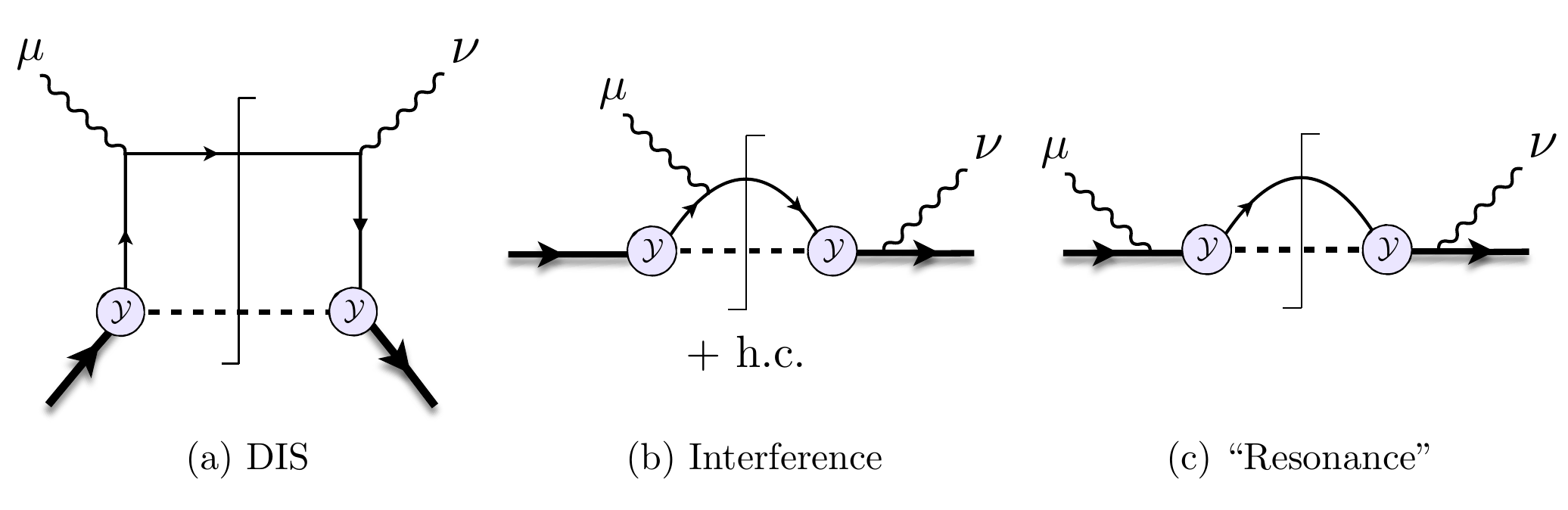}	
\caption{Diagrams contributing to $e+p$ scattering up to order $O(g^2)$ in the strong coupling constant. {\it Left:} DIS contribution.  {\it Center:} Interference term. {\it Right:} Proton resonance.}
\label{fig:ep_scattering_model}
\end{figure}

As opposed to QCD, where the matrix elements in Eq.~\eqref{eq:Wmunu_DIS} need to be parametrized because their non-perturbative nature, in the spectator model the nucleon-quark-diquark vertex is explicitly known, see Eqs.~\eqref{eq:model-vertex} and \eqref{eq:dipolar-form-factor}, and one can analytically compute the hadronic tensor.
At leading order in the strong coupling constant, the involved diagrams are collected in Figure~\ref{fig:ep_scattering_model}. 
Note that by electric charge conservation, as stressed in Ref.~\cite{Moffat:2017sha}, we need not only to consider the coupling of the photon to the quark, which is expected to dominate at large values of the invariant mass squared $W^2=(p+q)^2$, but also the photon-proton coupling. The LO cross section is therefore composed of 3 physical process that are observationally indistinguishable but theoretically separable, as we will discuss in the next subsection. The first one is photon-quark scattering, and mimics deeply inelastic scattering on the proton, see Figure~\ref{fig:ep_scattering_model}(a). The second one is the photo-induced excitation of the proton, that subsequently decays into into a quark and a diquark, see Figure~\ref{fig:ep_scattering_model}(c). This is akin to resonance excitation and subsequent decay in QCD, except the model as it stands does not include hadrons of higher mass than the nucleon. Finally, the hadronic tensor also receives a contribution from the interference of these two, see Figure~\ref{fig:ep_scattering_model}(b).

The hadronic tensor can be written as a sum of these three contributions:
\begin{equation}
W^{\mu\nu} =\sum_{(j)} W^{(j),\mu\nu}  = \sum_{(j)}  \int_{\Omega_k} \dfourk \, \mathcal{W}^{(j),\mu\nu} (k) \quad \quad (j) = \text{DIS, INT, RES} 
\label{eq:Wmunu}
\end{equation}
where $\Omega_k$ indicates the 4-D integration region in $k$ determined by four-momentum conservation and the external kinematics.  The individual, fully unintegrated contribution of each diagram in Figure~\ref{fig:ep_scattering_model}, $\mathcal{W}^{(j),\mu\nu} (k)$, can be calculated using the Feynman rules of the model, and at order $O(g^2)$ read:
\begin{align}
    2 M \mathcal{W}^{\DIS, \mu\nu}(k) &=
    \frac{1}{2} \frac{g^2(k^2)}{(2\pi)^3}  \frac{\Tr\big[(\psl+M)(\ksl+m_q)\gamma^\mu(\ksl+\qsl+m_q)\gamma^\nu(\ksl+m_q)\big]}{(k^2-m_q^2)^2}
    \nonumber\\
    & 
    \times\, \delta((k+q)^2-m_q^2) \delta((p-k)^2-m_\phi^2)\, ,
    \label{eq:WmunuDIS}
    \\
    2 M \mathcal{W}^{\text{INT}, \mu\nu}(k) &=  \, \frac{1}{2} \frac{2 \, g^2(k^2)}{(2\pi)^3} \frac{\Tr\big[(\psl+M)(\ksl+m_q)\gamma^\mu(\ksl+\qsl+m_q)(\psl+\qsl+M)\gamma^\nu\big]}{((p+q)^2-M^2)(k^2-m_q^2)}\nonumber\\
    &
    \times \delta((k+q)^2-m_q^2) \delta((p-k)^2-m_\phi^2)\, ,
    \label{eq:WmunuINT}
    \\
    2 M\mathcal{W}^{\text{RES}, \mu\nu}(k) 
    & =\frac{1}{2} \frac{g^2(k^2)}{(2\pi)^3}  \frac{\Tr\big[(\psl+M)\gamma^\mu(\psl+\qsl+M)(\ksl+\qsl+m_q)(\psl+\qsl+M)\gamma^\nu\big]}{((p+q)^2-M^2)^2} \nonumber \\
    &
    \times \delta((k+q)^2-m_q^2) \delta((p-k)^2-m_\phi^2)\, .
    \label{eq:WmunuRES}
\end{align}
Note that, following Ref.~\cite{Moffat:2017sha}, in the second and third equations we took the liberty of shifting the integration variable in order for the $\delta$-functions to match those appearing in the DIS diagram.

We remark from Eq.~\eqref{eq:WmunuINT} that the interference term contribution is nominally suppressed by a $1/(W^2-M^2)$ factor compared to the DIS term \eqref{eq:WmunuDIS}, due to the presence of one proton propagator. Similarly, the resonance term \eqref{eq:WmunuRES} contains two proton propagators, and is nominally suppressed by one more power of $W^2-M^2$. As we will demonstrate explicitly in Section~\ref{sec:FTcalcs}, this scaling holds to a very good degree for transverse structure functions -- hence the DIS term dominates the process at asymptotically large $Q^2$ or small Bjorken scaling variable $x_B= Q^2/(2p\cdot q)$, where $W^2 \sim Q^2/x_B$. However, this is not the case for the longitudinal structure function where, due to the off-shellness of the quark and proton propagators, all pieces are of similar magnitude and scale with $1/Q^2$.

Finally, it is important to stress that the individual contributions of the 3 diagrams considered in Figure~\ref{fig:ep_scattering_model} to the hadronic tensor are not gauge invariant by themselves, but their sum \eqref{eq:Wmunu} satisfies the electromagnetic Ward identities \cite{Moffat:2017sha}. Since, however, these diagrams contain heuristically distinguishable processes, in the next subsection we propose a method to isolate their gauge invariant parts, and to obtain a unique and physically meaningful (although only theoretical) decomposition of the hadronic tensor into a DIS, resonance, and interference contributions. This decomposition will also allow us to test in Sections \ref{sec:testing_kin} and \ref{sec:fact_limits} the validity of the collinear factorization procedure, that only purports to approximate the DIS component of the $e+p$ scattering cross section. In this respect, we differ from Ref.~\cite{Moffat:2017sha}, where factorized calculations were compared to the full DIS+INT+RES model calculation.

\subsection{Gauge invariant decomposition into DIS, resonance, and interference processes}
\label{sec:gauge_decomposition}

The strategy we follow to uniquely decompose a rank-2 tensor into a gauge invariant and a gauge breaking part is to define a complete set of orthogonal rank-2 projectors, $\left\{ P_\lambda^{\mu\nu} \right\}$, maximizing the number that satisfy the electromagnetic Ward identity, $q\cdot P_\lambda \equiv q_\mu P_\lambda^{\mu\nu} = 0$. Here we limit our treatment to the parity invariant tensors, such as those involved in unpolarized DIS, but will complete the discussion in Appendix~\ref{app:projectors_detailed}.

Following~\cite{Aivazis:1993kh,Accardi:2008ne}, we define longitudinal, transverse and scalar polarization vectors with respect to a longitudinal momentum $p$ (in our case the proton's momentum) and a reference vector $q$ (in our case the virtual photon's momentum) defining the longitudinal direction and the transverse plane:
\begin{align}
\begin{split}
  \varepsilon_0^\mu(p,q) 
    & = \frac{\hat{p}^\mu}{\sqrt{\hat{p}^2}}
    \\
  \varepsilon_\pm^\mu(p,q) 
    & = \frac{1}{\sqrt{2}} (0,\mp 1,-i,0) 
    \\
  \varepsilon_q^\mu(p,q) 
    & = \frac{q^\mu}{\sqrt{-q^2}} \ ,
\end{split}
\label{eq:polvect} 
\end{align}
where $\hat{p}^\mu \equiv p^\mu - \frac{p \cdot q}{q^2} q^\mu$.
We note that the transverse $\varepsilon_\pm$ polarization vectors indeed lie in the plane transverse to both $p$ and $q$ (since $\hat{p} \cdot \varepsilon_\pm = q\cdot \varepsilon_\pm = 0$) and that the ``longitudinal'' $\varepsilon_0$ polarization vector is transverse to the photon momentum ($q\cdot \varepsilon_0=0$).

The polarization vectors form an orthogonal basis in Minkowski space,
\begin{alignat}{2}
  & \varepsilon_\lambda \cdot \varepsilon_{\lambda'} = 0 
  & \qquad & \text{for\ } \lambda\neq\lambda' \nonumber \\
  & \varepsilon_\lambda \cdot \varepsilon_\lambda = 1 
  & \qquad & \text{for\ } \lambda = 0,+,- \\
  &\varepsilon_q \cdot \varepsilon_q = -1 \nonumber \ ,
\end{alignat}
and can be used to define parity invariant ``helicity projectors'' for rank 2 tensors. In particular, we define longitudinal, transverse, scalar, and mixed projectors $P_\lambda^{\mu\nu}$ , with $\lambda=L,T,S,\{LS\}$, as
\begin{align}
\begin{split}
  P_L^{\mu\nu}(p,q) & = \varepsilon_0^\mu(p,q) \varepsilon_0^{\nu*}(p,q) \\
  P_T^{\mu\nu}(p,q) & = \varepsilon_+^\mu(p,q) \varepsilon_+^{\nu*}(p,q) 
    + \varepsilon_-^\mu(p,q) \varepsilon_-^{\nu*}(p,q) \\
  P_S^{\mu\nu}(p,q) & = \varepsilon_q^\mu(p,q) \varepsilon_q^{\nu*}(p,q) \\
  P_{\{LS\}}^{\mu\nu}(p,q) 
    & = \varepsilon_0^\mu(p,q) \varepsilon_q^{\nu*}(p,q)
    + \varepsilon_q^\mu(p,q) \varepsilon_0^{\nu*}(p,q) \ .
\label{eq:helicityprojectors}
\end{split}
\end{align}
Taking advantage of
\begin{align}
\begin{split}
\varepsilon_+^\mu(p,q) \varepsilon_+^{\nu*}(p,q) 
+ \varepsilon_-^\mu(p,q) \varepsilon_-^{\nu*}(p,q) 
= -g^{\mu\nu} + \varepsilon_0^\mu(p,q) \varepsilon_0^{\nu*}(p,q)
- \varepsilon_q^\mu(p,q) \varepsilon_q^{\nu*}(p,q) 
\end{split}
\label{eq:transversepolsum}
\end{align}
and of the polarization vectors definition \eqref{eq:polvect}, the helicity projectors can be written in a more compact and suggestive way as
\begin{align}
\begin{split}
P_L^{\mu\nu}(p,q) & = \frac{\hat{p}^\mu \hat{p}^\nu}{\hat{p}^2} \\
P_T^{\mu\nu}(p,q) & = -\hat{g}^{\mu \nu} +\frac{\hat{p}^\mu \hat{p}^\nu}{\hat{p}^2}  \\
P_S^{\mu\nu}(p,q) & = -\frac{q^\mu q^\nu}{q^2} \\
P_{\{LS\}}^{\mu\nu}(p,q) 
& = \frac{\hat{p}^\mu q^\nu + q^\mu \hat{p}^\nu }{\sqrt{-q^2 \hat{p}^2}}\ ,
\label{eq:helicityprojectorsII}
\end{split}
\end{align}
where $\hat{g}^{\mu \nu} = g^{\mu\nu} - \frac{q^\mu q^\nu}{q^2} $ is also transverse to the photon's momentum. In either representation, it is straightforward to verify that these projectors are orthogonal. Indeed,
\begin{alignat}{2}
& P_\lambda \cdot P_{\lambda'} = 0 
& \qquad & \text{for\ } \lambda\neq\lambda' \nonumber \\
& P_\lambda \cdot P_{\lambda} =  1 
& \qquad & \text{for\ } \lambda = L, S \\
&P_T\cdot P_T = 2\nonumber\\
&P_{\{LS\}} \cdot P_ {\{LS\}} = - 2 \nonumber\, ,
\end{alignat}
where we have extended the use of the dot-product symbol to rank-2 tensors:  $P_\lambda \cdot P_{\lambda'} \equiv P_\lambda^{\mu \nu} P_{\lambda^\prime, {\mu \nu}}$.
From Eq.~\eqref{eq:helicityprojectorsII}, it is clear that the 4 defined projectors are also a complete orthogonal basis for symmetric tensors $T^{\mu\nu}=T^{\mu\nu}(p,q)$ that depend on the proton and photon momenta $p$ and $q$, such as the hadronic tensor for inelastic $e+p$ scattering.

Given a generic symmetric  tensor, $T^{\mu\nu}=T^{\mu\nu}(p,q)$, we can now define its helicity structure functions $F_\lambda$ as the projections of the tensor along the helicity basis defined in Eq.~\eqref{eq:helicityprojectors} or \eqref{eq:helicityprojectorsII}:
\begin{align}
F_\lambda(x_B,Q^2) \equiv c_\lambda P_\lambda (p,q) \cdot T(p,q) \, .
\label{eq:F_lambda}
\end{align}
with $c_\lambda = 1/(P_\lambda\cdot P_\lambda)$.
Thanks to the orthogonality and completeness of the helicity projectors, the tensor $T$ can then be decomposed as
$
 T^{\mu\nu}(p,q) = \sum_\lambda F_\lambda(x_B,Q^2) P^{\mu\nu}_\lambda (p,q)
$.
One can also go a step further, and separate this into a gauge invariant and gauge breaking components. Indeed, the longitudinal and transverse projectors satisfy the Ward identity,
\begin{align}
  q \cdot P_{L,T} \equiv q_\mu P_{L,T}^{\mu\nu} = 0 \ ,
\end{align}
and it is immediate to verify that no linear combination of the scalar and mixed projectors can satisfy that condition, either. Hence the longitudinal and transverse projectors form a complete orthogonal basis for the space of gauge invariant hadronic tensors. Similarly, the scalar and mixed projectors form a complete orthogonal basis for maximally gauge breaking tensors. As a result, 
\begin{equation}
  T^{\mu\nu}(p,q) = T_\text{inv.}^{\mu\nu}(p,q) + T_\text{g.b.}^{\mu\nu}(p,q) \ ,
\label{eq:Wmunu_gi_gb}
\end{equation}
where
\begin{align}
T_\text{inv.}^{\mu\nu}(p,q) &= P_T^{\mu\nu} F_{T}(x_B,Q^2) 
+P_L^{\mu\nu}F_{L}(x_B,Q^2)\\
T_\text{g.b.}^{\mu\nu}(p,q) &= P_S^{\mu\nu}F_{S}(x_B,Q^2)
+P_{\{LS\}}^{\mu\nu} F_{\{LS\}}  \ ,
\end{align}
are, respectively, the gauge invariant (inv.) and gauge breaking (g.b.) components of the tensor $T^{\mu\nu}$. Accordingly, we also call $F_T$ and $F_L$ ``gauge invariant structure functions'', and $F_S$ and $F_{\{LS\}}$ ``gauge breaking structure functions''. 

Coming back to our model electron-proton scattering, we can apply this decomposition to each of the 3 processes represented in Figure~\ref{fig:FT_pieces}, and define their gauge invariant and gauge breaking parts:
\begin{align}
2MW_{\text{inv.}}^{(j),\mu\nu}(p,q)
&= P_T^{\mu\nu} F_{T}^{(j)}(x_B,Q^2) 
+ P_L^{\mu\nu}F_{L}^{(j)}(x_B,Q^2)\, ,
\label{eq:Winv} \\
2MW_\text{g.b}^{(j),\mu\nu}(p,q)
&= P_S^{\mu\nu}F_{S}^{(j)}(x_B,Q^2)
+P^{\mu\nu}_{\{\!LS\}}\!F^{(j)}_{\{\!LS\}} \, ,
\label{eq:Wgb} 
\end{align}
for $j=\text{DIS, INT, RES}$. The gauge invariance of the hadronic tensor \eqref{eq:Wmunu} ensures that the sum of the gauge breaking parts of each diagram in Figure~\ref{fig:FT_pieces} vanishes,
\begin{equation}
  W_\text{g.b}^{\mu\nu}=
    \sum_{(j)}W_\text{g.b}^{(j),\mu\nu}
    =0\quad \quad (j) = \text{DIS, INT, RES} 
\label{eq:gbsum}
\end{equation}
as one can also explicitly verify utilizing the algebraic manipulations discussed in \cite{Moffat:2017sha}, even though the individual terms in the sum are different from zero.

In summary, the gauge-invariant $F_{T,L}$ structure functions of each individual diagram in Figure~\ref{fig:ep_scattering_model} are physically meaningful, and allow one to theoretically decompose each $e+p$ scattering structure functions into DIS, resonance and interference contributions. A more detailed discussion of this decomposition in the context of the scalar diquark model will be offered in Sections~\ref{sec:FTcalcs} and \ref{sec:FLcalcs}.

\subsection{Kinematics}
\label{sec:kinematics}

Like in $e+p$ scattering in QCD, the Bjorken invariant in the model is bounded, as we discuss in more detail in Appendix~\ref{app:kinematics}:
\begin{equation}
  0 < x_B \leq \frac{1}{1+\frac{(m_\phi + m_q)^2-M^2}{Q^2}} \equiv \xbmax \ .
\label{eq:xB_max}
\end{equation}
The lower bound is due to the fact that in an electron-proton scattering the photon momentum is spacelike. The upper bound, $\xbmax$, is determined by the on-shell condition for particles belonging to a minimal mass 2-particle final state, and is analogous to the ``pion threshold'' $x_\pi = 1/\Big[ 1+\frac{(M + m_\pi)^2-M^2}{Q^2}\Big]$ in inelastic $e+p$ collisions in QCD. Likewise, the maximum transverse momentum squared for the scattered quark, $\ktmaxsq$, is determined by the available invariant mass $W^2$ and the masses of the particles in the minimal mass final state, see Appendix~\ref{app:kinematics} for details:
\begin{equation}
  0 \leq \kTsq \leq \frac{\Big(W^2 - (m_\phi + m_q)^2\Big)
    \Big(W^2 - (m_\phi - m_q)^2\Big)}{4 W^2} \equiv \ktmaxsq \ .
\label{eq:maxkt2}
\end{equation}
It is interesting to note that the kinematic threshold in $x_B$ can only be reached at zero quark transverse momentum. Indeed, solving $\ktmaxsq = 0$ for $x_B$ one recovers the upper limit in \eqref{eq:xB_max}.

We can now explicitly write each of the contributions to the hadronic tensor \eqref{eq:Wmunu} as an integral over the parton's transverse and longitudinal momentum and virtuality:  
\begin{align}
    W^{(j),\mu\nu}(x_B,Q^2)  = 
    \int_0^{\ktmaxsq} \dkTtwo   \iint  dx \, dk^2 \, 
    \mathcal{W}^{(j),\mu\nu} (x, k^2, \kTsq)
    \label{eq:Wmunu_j}
\end{align}
where 
\begin{align}
    \mathcal{W}^{(j),\mu\nu} (x, k^2, \kTsq) = \frac{\pi}{2x} \, \mathcal{W}^{(j),\mu\nu}(k)
\end{align}
is the fully unintegrated hadronic tensor in the $x$, $k^2$ and $\kTsq$ variables, and the $\mathcal{W}^{j}(k)$ tensors on the right are defined in Eqs.~\eqref{eq:WmunuDIS}-\eqref{eq:WmunuRES}. 
It is important to note that the integral in $\dkTtwo$ appearing in Eq.~\eqref{eq:Wmunu_j} is limited from above by the maximum transverse momentum squared for the scattered quark defined in Eq.~\eqref{eq:maxkt2}.


At LO, due to the $\delta-$functions in the unintegrated hadronic tensors that originate, as mentioned, from the upper and lower cuts in the diagrams of Figure~\ref{fig:ep_scattering_model}, the integrals over $dx$ and $dk^2$ can be explicitly calculated, and we can explicitly calculate the  $\kTsq$-unintegrated (but $x$- and $k^-$-integrated) hadronic tensor $\mathcal W(\kTsq)=\int dx dk^2 \mathcal W(x,k^2,\kTsq)$:
\begin{align}
   \mathcal{W}^{(j),\mu\nu} (\kTsq) \equiv \frac{\pi}{2 x} \,  \frac{1}{|\Jfull|}\, \, 
   \widetilde{\mathcal{W}}
   ^{(j),\mu\nu} (k) \bigg\rvert_{x=x_\text{ex}(\kTsq), \,\, k^2 = k_\text{ex}^2(\kTsq)} \ ,
   \label{eq:Wmunu_j_kTsq}
\end{align}
where we notationally distinguished the integrated tensor from the unintegrated one only by their arguments. The Jacobian 
\begin{equation}
\Jfull = 
(\xi-1)\frac{(k^2+\kTsq)}{x^2} + \Big(1-\frac{1}{x}\Big)\frac{Q^2}{\xi} + \Big(1-\frac{\xi}{x}\Big) M^2 
\label{eq:Jacobian_x_k2}
\end{equation}
arises from the manipulation of the delta functions, and the tilde sign over the hadronic tensor symbol indicates the removal of these from Eqs.~\eqref{eq:WmunuDIS}-\eqref{eq:WmunuRES}. The whole expression~\eqref{eq:Wmunu_j_kTsq} is then evaluated at $x = x_\text{ex}(\kTsq)$ and $k = k_\text{ex}^2(\kTsq)$, which are the solutions of the said delta functions and are explicitly derived in Appendix~\ref{app:Delta_sol}, see Eqs.~\eqref{eq:x_sol_ex}-\eqref{eq:k2_sol_ex}.

In Eq.~\eqref{eq:Jacobian_x_k2} we have highlighted the role of the mass scales: beside the external proton mass $M^2$ and photon virtuality $Q^2$, the Jacobian also depends on the ``light-cone virtuality''
\begin{align}
    v^2 = k^2+\kTsq
\end{align}
of the struck quark, already defined in Eq.~\eqref{eq:v2_def}.

It is now instructive to look in more detail at the internal kinematics of the process. In fact, in an inclusive scattering, the 4-momentum of the scattered quark is not measured, and therefore neither $x$, nor $k^2$, nor $\kTsq$ can be experimentally determined. Nonetheless, 
in the spectator model we do have explicit control over these variables -- a major theme of this article -- in particular through the analysis of the momentum flow through the cuts in Figure~\ref{fig:ep_scattering_model}.
For example, the light-cone fraction $x$ is determined by the the cut of the quark line in the top part of the diagrams, that gives rise to the $\delta((k+q)^2-m_q^2)$ function in Eqs.~\eqref{eq:WmunuDIS}-\eqref{eq:WmunuRES}. This delta function imposes
\begin{align}
  x &= \frac{\xi}{2} \Bigg(1 + \frac{m_q^2-k^2}{Q^2}
        + \sqrt{\Big(1 + \frac{m_q^2-k^2}{Q^2}\Big)^2
        + 4\frac{k^2+\kTsq}{Q^2}} \Bigg)
        \label{eq:x_sol}\\
    &= \xi \Bigg(1 + \frac{m_q^2+ \kTsq}{Q^2}
        - \frac{v^2 \, (m_q^2+ \kTsq)}{Q^4}
        + O\Big(\frac{\mu^6}{Q^6}\Big) \Bigg) \ .
        \label{eq:x_sol_I}
\end{align}
In the second line, the constraint is expanded in inverse powers of $Q^2$ and acquires a suggestive form, that in fact holds at any order in the expansion. In detail, in Eq.~\eqref{eq:x_sol_I}, $x$ depends only on two mass scales: the transverse mass $m_{qT}^2 \equiv m_q^2 + \kTsq$ of the scattered quark and the quark's light-cone virtuality $v^2$, with no direct dependence on the diquark mass. The light-cone virtuality is, instead, fixed by the bottom cuts in Figure~\ref{fig:ep_scattering_model}:
\begin{align}
  v^2 =  -\frac{x}{1-x}
    \Big[ (m_\phi^2 - M^2) + x M^2 + \kTsq \Big] \ ,
\label{eq:v2_model} 
\end{align}
which can be obtained by replacing $p_X^2 \rightsquigarrow m_\phi^2$ in Eq.~\eqref{eq:v2_QCD} and depends on the target's fragmentation dynamics, encapsulated in $m_\phi$.  Notably, the light-cone virtuality vanishes when $x\to 0$, diverges to negative infinity as $x \to 1$, and is negative over the whole range in $x$ for physical choices of the $m_\phi$ parameter as can be expected of a bound particle\footnote{The only way the light-cone virtuality can become positive and large is if $m_\phi \ll M$, such that $v^2 \approx x M^2 - \kTsq/(1-x)$. Evaluating this at $\kTsq \sim \Lambda^2 \sim 0.4 M^2$, the maximum light-cone virtuality remains nonetheless small, only reaching $v^2_{max} \sim  0.1 M^2$ at $x \sim 0.4$ before dropping below 0 as $x \rightarrow 1$. This is, however, a quite unphysical choice of diquark mass because $\phi$ represents the proton's remnant, and thus one would expect $m_\phi \sim M$.}. 
This also means that the quark is off its $k^2=m_q^2$ mass shell over the whole range of $x_B$.  
This analysis will be substantiated in Section~\ref{sec:testing_kin}, with an explicit calculation of the average values of the internal kinematic variables as a function of Bjorken's $x_B$. 

It is finally important to note that the light-cone virtuality enters Eq.~\eqref{eq:x_sol_I} only starting at order $O(1/Q^4)$. Therefore, $x$ and $v^2$ are coupled essentially only through the quark transverse momentum squared $\kTsq$, which is the only variable left free in the loop integrations \eqref{eq:WmunuDIS}-\eqref{eq:WmunuRES}. Since, as discussed, $v^2$ remains small except close to the kinematic threshold,  where it can become substantially large and negative,
truncating Eq.~\eqref{eq:x_sol_I} to first order in the $1/Q^2$ expansion appears to be a meaningful approximation over much of the inclusive scattering's $(x_B,Q^2)$ phase space.
These considerations form the basis for the kinematic approximations needed to perform collinear factorization in DIS at sub-asymptotic energy, as discussed in Section~\ref{sec:CF} and numerically tested in Section~\ref{sec:testing_kin}.

\subsection{Transverse structure function}
\label{sec:FTcalcs}

Having discussed the hadronic and partonic kinematics, and isolated the gauge invariant part of the DIS, resonance, and interference diagrams in Figure~\ref{fig:ep_scattering_model} with the aid of Eqs.~\eqref{eq:Winv} and the projections \eqref{eq:F_lambda}, we can study their individual role in the $e+p$ scattering process. We focus first on the transverse structure function $F_T$, which, in our model, gives the dominant contribution to the inelastic cross section, and discuss $F_L$ in the next subsection.

\begin{figure}[tbh]
	\centering
	\includegraphics[width=0.48\linewidth]{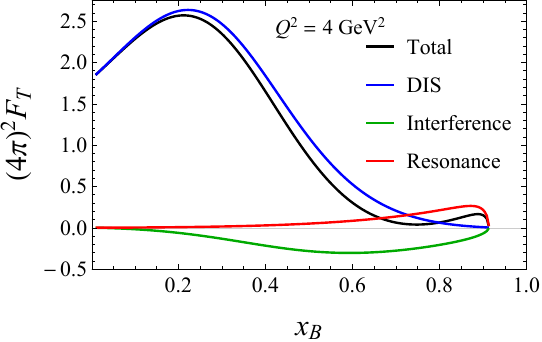}	
	\hspace*{0.01\linewidth}
	\includegraphics[width=0.48\linewidth]{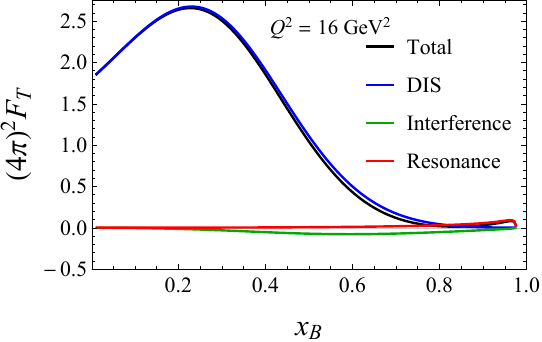}
\caption{
Gauge invariant decomposition of the transverse structure function $F_T$ at $Q^2 = 4$ GeV$^2$ (left plot) and  $Q^2 = 16$ GeV$^2$ (right plot) for the model parameters specified in Section~\ref{sec:model} [see Eqs.~\eqref{eq:mphi_Lambda_default}-\eqref{eq:mq_M_default}]. The kinematic thresholds are $\xbmax=0.913$ and 0.976, respectively. 
} 
\label{fig:FT_pieces}
\end{figure}

As discussed in Section~\ref{sec:gauge_decomposition}, the individual DIS, interference and resonance structure functions can be obtained by projecting the respective hadronic tensors:
\begin{align}
    F_{T}^{(j)} = M\, P_T^{\mu\nu} W_{\mu\nu}^{(j)} 
    \quad \quad (j) = \text{DIS, INT, RES} \ .
\label{eq:FT_full}
\end{align}	
In the case of the DIS contribution, $F_T^\DIS$ describes the scattering of a transversely polarized photon with a quark emitted by the proton (see Figure~\ref{fig:ep_scattering_model}(a)). Analogously, the resonance 
$F_T^\RES$ structure function describes a transverse photon that scatters on the proton as a whole (see Figure~\ref{fig:ep_scattering_model}(c)). Therefore we expect these structure functions to be non negative:
\begin{align}
    F_{T}^\DIS &\propto \big| \mathcal{M}_T^{\,q} \big|^2 \geq 0 \\ 
    F_{T}^\RES &\propto \big| \mathcal{M}_T^{\,p} \big|^2 \geq 0
\label{eq:FT_full_II}
\end{align}	
where $\mathcal{M}_T^{\,q,p}$ are the scattering amplitude of the transverse photon-quark and transverse photon-proton processes.
This is confirmed in Figure~\ref{fig:FT_pieces}, where we numerically evaluate
all components of $F_T$ in the spectator model at LO, and show that
$F_{T}^\DIS$ and $F_{T}^\RES$ are indeed positive at all values of $x_B$ and $Q^2$. In contrast, the interference structure function $F_{T}^\INT$ can have any sign, and in our model it turns out to be negative definite -- and not too small, either. 

In the left panel of Figure~\ref{fig:FT_pieces}, we use a value of $Q^2 = 4$ GeV$^2$  and notice that at small $x_B$ the DIS contribution is the dominant one. Nonetheless, the interference piece is large and non negligible even at intermediate and large $x_B$, and is largely responsible for the visible difference between the DIS curve (blue) and the total contribution (black). 

The resonance piece has an extended but small tail at lower $x_B$, and increases in magnitude as $x_B \to 1$, where in principle it would diverge. However, this divergence is cut off by the phase space, that limits $x_B<x_B^{\text{max}}<1$ when $m_q + m_\phi > M$, as in our model. The interplay of the resonance at $x_B=1$ and phase space limitations produces an asymmetric bump at $x_B \lesssim x_B^{\text{max}}$. This bump becomes visually more prominent and narrow in the total contribution, that seems separated into an inelastic contribution at $x_B \lesssim 0.75$ and a resonance peak at $x_B\gtrsim 0.75$. However, the trough at $x_B \approx 0.75$ is actually a combined effect of the resonance and interference pieces, whose influence extends beyond the edge of phase space, well into what one may consider to be the deep inelastic region. 
In Appendix~\ref{app:resonance}, we discuss for completeness the calculation for the $m_q+m_\phi < M$ case, where the resonant behavior of the proton excitation is in full display (see Figure~\ref{fig:FT_unstable_proton}). 

In $e+p$ scattering, it is important to separate the DIS contribution from the rest, because it is this one that can be factorized into a perturbatively calculable photon-parton hard scattering coefficient and a non-perturbative parton distribution function, thus giving one access to the partonic structure of the target:
\begin{align}
    F_{T}^\DIS \approx F_{T}^\CF = M\, P_T^{\mu\nu} W^{\mu\nu} \big|_\text{LT}
    & = q(\bar{x})
\end{align}
where $W^{\mu\nu} \big|_\text{LT}$ is given in Eq.~\eqref{eq:factorized_Wmunu_LT}.

What the model shows, however, is that the needed phenomenological separation of the DIS piece must be done carefully, and cannot rely only on phenomenological cuts (for example on $W^2$) to eliminate the apparent resonance ``peak''. Instead, one needs to also exploit the $Q^2$ dependence of the structure function, since, as already observed, the interference contribution is parametrically suppressed by a factor $\frac{1}{W^2-M^2}\sim 1/Q^2$ compared to DIS, and the resonance contribution is suppressed by $\frac{1}{(W^2-M^2)^2}\sim 1/Q^4$. (In the latter case the parametric scaling does not fully hold numerically, although the resonance piece is still suppressed compared to the resonance contribution, see Appendix~\ref{app:low_xB_scaling}.)

To illustrate the $Q^2$ suppression of the interference and resonance contributions, we show our calculations for $Q^2 = 16$ GeV$^2$ in the right panel of Figure~\ref{fig:FT_pieces}. In this case the interference term contribution is suppressed compared to the plot in the left panel, but still still appreciable. The resonance contribution is negligible except at very high $x_B$ values closer to the kinematic threshold, that moved to the right compared to the left panel. There it becomes the dominant contribution, and, combined with the interference, again gives rise to a narrow, but now smaller, ``peak''. Overall, the DIS piece dominates although a not yet necessarily negligible interference contribution is visible at intermediate $x_B$. At yet higher $Q^2$ values the DIS piece would become the dominant contribution over most of the available $x_B$ range, except right before the kinematic threshold.

While an experimental measurement is only sensitive to the full structure function, 
the scaling just described allows to phenomenologically control, for example, just 
the DIS component in a fit that includes power-suppressed $H(x)/Q^2$ terms, with $H(x)$ a suitable polynomial in $x$, and utilizes data in as large a $(x_B,Q^2)$ range as possible. This was first tried in Ref.~\cite{Virchaux:1991jc}, and more recently implemented in global fits of parton distributions by the CTEQ-JLab collaboration~\cite{Accardi:2009br,Owens:2012bv,Accardi:2016qay}, the JAM collaboration \cite{Cocuzza:2021rfn}, and by Alekhin and collaborators~\cite{Alekhin:2003qq,Alekhin:2017kpj}. A similar fit utilizing data generated from the spectator model goes beyond the scope of this paper, but will be presented elsewhere \cite{Krause-etal-inprep}. Instead, in Sections~\ref{sec:testing_kin} and~\ref{sec:fact_limits} we will compare the analytically isolated DIS component of the model's structure function to its factorized approximation to validate the latter.

\subsection{Longitudinal structure function}
\label{sec:FLcalcs}

\begin{figure}[tb]
	\centering
	\includegraphics[width=0.48\linewidth]{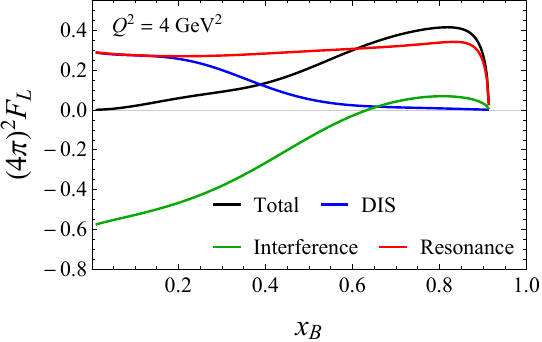}
	\hspace*{0.01\linewidth}
	\includegraphics[width=0.48\linewidth]{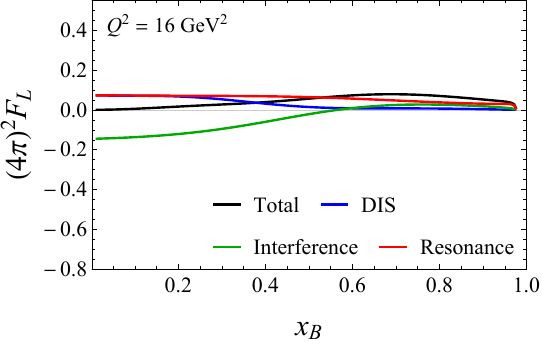}
\caption{Gauge invariant decomposition of the longitudinal $F_L$ structure function at $Q^2 = 4$ GeV$^2$ (left) and  $Q^2 = 16$ GeV$^2$ (right). The kinematic thresholds are $\xbmax=0.913$ and 0.976, respectively.
} 
\label{fig:FL_pieces}
\end{figure}

For completeness, in Figure~\ref{fig:FL_pieces} we present the model calculation of the longitudinal structure function $F_L$, even though we will not discuss this further in the rest of the paper, since in collinear factorization $F_{L}^\CF=0$ at LO.
The behavior of the longitudinal $F_L$ structure function, illustrated in Figure~\ref{fig:FL_pieces}, changes drastically from what we have discussed for its transverse $F_T$ counterpart in at least two respects.
Firstly, all 3 components scale approximately as $1/Q^2$, instead of displaying the hierarchy discussed for the transverse case. 
Secondly, $F_L \to 0$ also as $x_B \to 0$; however, each one of the three components remains different from zero, and, in fact, $F_L^{\text{DIS}} \to F_L^{\text{RES}}$. We will analytically study the highlighted features of $F_L$ in Appendix~\ref{app:low_xB_scaling}, and offer here a heuristic explanation for these. 

For the DIS component, the $1/Q^2$ scaling behavior can be understood by noticing that, if the scattered quark was on its mass shell, the Callan-Gross relation would be satisfied and one would find $F_L^{\text{DIS}}=0$. However, the quark is virtual, and we can expect the Callan-Gross relation to be broken by an amount proportional to the quark's average virtuality normalized by the scale of the process, which is provided by the invariant mass $W^2$: namely, $F_L^{\text{DIS}} \propto \langle k^2 \rangle/W^2 \propto \Lambda^2/Q^2$. Note that we have used $\langle k^2 \rangle = O(\Lambda^2)$, because this is the scale that determines the behavior of proton vertex's form factor $\cal Y$, and therefore determines the quark's nonperturbative dynamics in the model. This argument also justifies the much smaller size of the DIS component of $F_L$ compared to that of $F_T$.
For the resonance piece, the same argument can be applied to the scattered proton, whose virtuality is equal to $W^2$ by four-momentum conservation. The only scale left to neutralize this is $\Lambda^2$, hence we can expect
$F_L^\RES \propto \frac{1}{W^4} \times \frac{W^2}{\Lambda^2} \propto 1/(\Lambda^2 Q^2)$, where the fourth inverse power of the invariant mass is due to the proton propagator, as evident from Eq.~\eqref{eq:WmunuRES}.  The confinement scale now appears at the denominator, enhancing the resonance piece relative to the DIS contribution (in the transverse case it was much suppressed, instead). The interference piece is a mixture of these two, and we can expect $F_L \propto \frac{1}{W^2} \times \sqrt{\frac{\Lambda^2}{Q^2}} \sqrt{\frac{W^2}{\Lambda^2}} \propto \frac{1}{Q^2}$. In all cases, the three components of the longitudinal structure function scale as $1/Q^2$, as we will analytically corroborate in Appendix~\ref{app:low_xB_scaling}.

The limiting behavior of the full $F_L$, which vanishes as $x_B \to 0$, is a general consequence of gauge invariance. Indeed, one can easily see that the longitudinal projector satisfies
\begin{align}
P_L^{\mu\nu}(p,q) \xrightarrow[x_B \to 0]{} P_S^{\mu\nu}(p,q) \ .
\end{align}
Noticing that $P_S$ is one of the two gauge-breaking operators discussed in Section~\ref{sec:gauge_decomposition}, we obtain 
\begin{align}
F_L \xrightarrow[x_B \to 0]{} 2M\, P_S \cdot W
= 2M\, P_S \cdot W_{g.b.} = 0 \ .
\label{eq:FL_limit}
\end{align}
The last equality is in fact valid for any gauge invariant tensor, see Eq.~\eqref{eq:gbsum}, and therefore also for inelastic scattering at LO in our model, as plotted in Figure~\ref{fig:FL_pieces}. Unitarity then imposes specific small-$x_B$ constraints for the resonance and interference components of $F_L$:
\begin{align}
  F_L^\RES & \xrightarrow[x_B \to 0]{} F_L^\DIS \ , 
\label{eq:DISRES_limit}  \\   
  F_L^\INT & \xrightarrow[x_B \to 0]{} -2 F_L^\DIS \ .  
\label{eq:INT_limit}
\end{align}
Indeed, since the two individual $e+p$ scattering amplitudes involving the longitudinal photon-quark and the longitudinal photon-proton interactions are imaginary, ${\cal M}_L^{q,p}= i T_{q,p}$, the vanishing of $F_L$ in that limit imposes $T_q^2 + T_p^2 - 2 T_q T_p = (T_q - T_p)^2 \rightarrow 0$. 

Let us stress that Eqs.~\eqref{eq:FL_limit}-\eqref{eq:INT_limit}, are consequence of electromagnetic gauge invariance rather than special features of the model we have considered. What is not constrained by general principles is the limiting behavior of the DIS contribution, that at small $x_B$ could as well tend to zero or diverge (in which case, we note that the interference term should also diverge but with opposite sign). The fact that it is, instead finite, is a feature of the chosen spectator model, as we will analytically demonstrate in Appendix~\ref{app:low_xB_scaling}, where we will also prove the common $1/Q^2$ scaling behavior.

In closing this section, we recall that the spectator model we are considering  is only designed to account for the quark dynamics inside the proton, and therefore can only mimic electron-proton scattering in the ``valence quark region'' at large $x_B \gtrsim 0.2$.
In QCD, the DIS longitudinal structure function at small $x_B$ is instead dominated by photon-gluon fusion interactions, that the model as it stands cannot describe\footnote{A generalization of the model to include gluon dynamics has been discussed in Ref.~\cite{Bacchetta:2020vty}.}. Nonetheless, the conclusion that $F_L \to 0$ as $x_B \to 0$ is a consequence of electromagnetic gauge invariance, and as such is model independent. 
Therefore we can also expect this to happen in nature. In fact, data on inelastic $e+p$ scattering gathered at HERA~\cite{Andreev:2013vha,Abramowicz:2014jak,Abramowicz:2015mha} show that $F_L$ first grows as $x_B$ decreases towards $x_B\sim 10^{-4}$, then falls off as $x_B$ becomes smaller than that value, contrary to expectations from perturbation theory.
Many explanations have been advanced for this observation, including higher twist effects~\cite{Bartels:2000hv,Abt:2016vjh,Harland-Lang:2016yfn,Motyka:2017xgk,Alekhin:2017kpj} and deviations from perturbative QCD evolution~\cite{Abdolmaleki:2018jln,Ball:2017otu,Bartels:2002cj,Gelis:2010nm}.
Here, we are suggesting that a further source of deviation from perturbative calculations of $F_L$ is due to the limiting behavior of this structure function, that is forced by gauge symmetry to vanish at small $x_B$.


\section{Testing the kinematic approximations}
\label{sec:testing_kin} 

In deriving the factorized hadronic tensor \eqref{eq:factorized_Wmunu_LT} in Section~\ref{sec:CF}, we have distinguished \textit{dynamical} and \textit{kinematical} approximations, namely, the inclusion of the parton transverse momentum loop integral contribution into the twist expansion, and the choice of approximate light cone virtualities $\vsqbar$ and $\vpsqbar$ entering the calculation of the collinear hard scattering tensor $\mathcal{H}^{\mu\nu}$.

The twist expansion is controlled by powers of $\mu^2/Q^2$ and can be calculated up to any desired order (although each order eventually needs the introduction of one or more new non-perturbative functions). 
The light cone virtualities, instead, are not observable nor can they be experimentally controlled. Thus, as discussed in Section~\ref{sec:kin_approx_QCD}, we can only resort to a physically or theoretically motivated \textit{Ansatz} to choose suitable $\vsqbar$ and $\vpsqbar$ values. Contrary to the twist expansion, this approximation cannot be systematically improved. 
In the spectator model, however, one also has full control of the unobserved (or ``internal'') variables, and our choices can be in fact verified by comparing the approximated internal variables with their average value. 

Let us then define the average of a generic function ${\cal O} = {\cal O}(x,k^2,\kTsq)$ of the internal variables as:
\begin{equation}
    \langle \mathcal{O} \rangle (x_B,Q^2) 
    = 
    \frac{\int^{\ktmaxsq}_0  d\kTsq dk^2 dx \, \mathcal{O}(x,k^2,\kTsq) \mathcal{F}^\DIS_T(x,k^2,\kTsq)_{x_B,Q^2}}
    {\int^{\ktmaxsq}_0 d\kTsq dk^2 dx \, \mathcal{F}^\DIS_T(x,k^2,\kTsq)_{x_B,Q^2}}\, ,
\label{eq:avg_obs}
\end{equation}
where $\ktmaxsq$ was defined in Eq.~\eqref{eq:maxkt2}, and the fully unintegrated structure function $\mathcal{F}^\DIS_T$ is defined as the transverse projection of the fully unintegrated DIS hadronic tensor \eqref{eq:WmunuDIS}:
\begin{align}
\mathcal{F}^\DIS_T(x,k^2,\kTsq)_{x_B,Q^2}  &= M\,P_T^{\mu\nu} \mathcal{W}^\DIS_{\mu\nu}(x,k^2,\kTsq)_{x_B,Q^2}\, .
\label{eq:unintegrated_FT}
\end{align}
The dependence of $\langle \mathcal{O} \rangle$ on $x_B$ and $Q^2$ is due to the hard scattering delta function inside $\mathcal{F}_T$, and is symbolically denoted in the subscript following the functional dependence on the internal variables.
The denominator at the right hand side of Eq.~\eqref{eq:avg_obs} is easily recognized as the inclusive DIS transverse function $F_T^\DIS$. 

Furthermore, in the model one can calculate not only the hard-scattering term, but also the PDFs themselves. Hence one can perform a detailed comparison of the factorized vs. full cross section test the range of validity of the collinear factorization approximation. This will be discussed in Section~\ref{sec:fact_limits}. In the remaining of this Section we will focus on the validity of the kinematic approximations, to which we now turn our attention.

\subsection{Kinematic approximations in the model}
\label{sec:kinematic_approximations}

In Section~\ref{sec:kin_approx_QCD}, we discussed a theoretical~\textit{Ansatz} to justify the choice of the approximated $\bar v^2 \approx \langle v^2 \rangle$ and $\vpsq \approx \langle v'^2 \rangle $ parton light-cone virtualities, and of the approximate $\bar x$ scaling variable that enter the collinear factorization formula \eqref{eq:factorized_Wmunu_LT}-\eqref{eq:x_sol_fact_I}.
That discussion, that was tailored to QCD, can be repeated for the spectator model with simple modifications.

To start with, we consider the approximate light-cone virtuality of the scattered quark, $\vpsqbar \approx \langle k'^2 + \kpTsq \rangle$. In the model, isolated quarks exist as asymptotic states. Hence the virtuality $k'^2$ must be at least as large as $m_q^2$, and since we do not measure the quark's transverse momentum in an inclusive process we can only choose
\begin{align}
    \vpsqbar =  m_q^2 \ .    
\label{eq:vpsqbar_choice}
\end{align}

Next, we consider the scattering quark's virtuality $\vsqbar$, and notice that the target's remnant is represented in the model by the spectator $\phi$. Thus, we can substitute $p_X \rightsquigarrow p_\phi$ and $m_X^2 \rightsquigarrow m_\phi^2$ in Eq.~\eqref{eq:v2_QCD}, repeat the ensuing derivation, and eventually find
\begin{align}
    \vsqbar = 0 \ 
\label{eq:vsqbar_choice}
\end{align}
as our standard choice. Analogously to the QCD case, we expect to be a good approximation for small enough $x_B$ and not too large quark masses.

Finally, utilizing in Eq.~\eqref{eq:x_sol_fact_I} the light-cone virtuality choices \eqref{eq:vpsqbar_choice} and \eqref{eq:vsqbar_choice}, one obtains 
\begin{align}
    \bar x \approx \xi \Big(1 + \frac{m_q^2}{Q^2} \Big) \equiv \xi_q\ .
\label{eq:xbar_AOT}
\end{align}
With a non-zero $\vpsqbar$ like in Eq.~\eqref{eq:vpsqbar_choice} we are more closely respecting the internal kinematics of the handbag diagram than with $\vpsqbar=0$. Therefore, we can expect $\bar{x} = \xi_q$ to provide a better approximation to the non-factorized diagram's than in standard collinear factorization. In fact, our $\xi_q$ is  analogous to the $\chi$ scaling variable used for example in Ref.~\cite{Aivazis:1993kh} to study charm production in charged current $W+s \to c$ events, which was indeed found to capture much of the heavy quark production kinematics even when setting the quark masses equal to zero in the calculation of the hard scattering coefficient~\cite{Nadolsky:2009ge}.

As in QCD, the model's virtuality diverges to minus infinity at large values of $x_B$, see Eq.~\eqref{eq:v2_model}, and a different approximation may be needed. 
Analogously to the detailed discussion in Section~\ref{sec:kin_approx_QCD}, a suitable approximation to the model's $v^2$ that is valid in both the small and large-$x_B$ regimes is
\begin{align}
    \vsqbar(\bar x)
    = - \frac{\bar x}{1-\bar x} \big( m_\phi^2 + (\bar x -1) M^2 \big) ,
\label{eq:v2_new_approx}
\end{align}
Substituting Eq.~\eqref{eq:v2_new_approx} in Eq.~\eqref{eq:x_sol_fact_I}, solving the resulting equation for $\bar x$ perturbatively in powers of $\mu^2/Q^2$, and finally setting $\vpsqbar=m_q^2$ we find
\begin{align}
    \bar x 
    \approx \xi \Bigg[ 1 + \frac{m_q^2}{Q^2} 
    + \frac{\xi}{1-\xi} 
    \frac{m_q^2 \big(m_\phi^2+(\xi-1)M^2\big)}{Q^4}\Bigg]
    \equiv \xiqnew 
\label{eq:xbar_new}
\end{align}
up to corrections of $O\Big( \frac{\mu^6}{Q^6} \Big)$.
Note that at small $x_B$ the fourth order term quickly vanishes, and one recovers Eq.~\eqref{eq:xbar_AOT}. Closer to the kinematic threshold, this new scaling variables accounts for the non-vanishing of the scattered quark's light-cone virtuality. Substituting this back in Eq.~\eqref{eq:v2_new_approx} we can also write the corresponding virtuality purely in terms of external variables:
\begin{align}
    \vsqbarnew \equiv \vsqbar(\xiqnew) \ . 
\label{eq:vbarsq_new_model}
\end{align}

The scaling variable we just obtained are suitable for collinear factorization of inclusive scattering processes. By necessity, however, they neglect the kinematic effect of the internal transverse momentum, because no particles other than the scattered lepton is measured in the final state, so that no measured scale is available to control the size of $\avekTsq$. Nonetheless, Eq.~\eqref{eq:x_sol_I}, shows that $\kTsq$ contributes to the scaling variables at the same order as $m_q^2$, and it would be desirable to estimate the size of its contribution. 

Contrary to the QCD case, an estimate of the average transverse momentum squared is actually possible in the model, where $\avekTsq$ can be explicitly calculated using Eq.~\eqref{eq:avg_obs}. The result is presented in Figure~\ref{fig:avg_kT2} as a function of $x_B$ at 
several values of $Q^2$. The upper left panel utilizes the default model parameter choices discussed in Section~\ref{sec:model}, and the other two panels increase, respectively, the values of the target mass and the quark mass\footnote{This layout will be used also when studying other internal variables in this Section, and for the study of factorized vs. full structure function in Section~\ref{sec:fact_limits}.}. A comprehensive study of the systematic dependence on the model parameters is presented in Appendix~\ref{app:systematics}.
As one can expect, $\avekTsq = O(\Lambda^2)$ is of non perturbative origin and determined by the ``confinement'' scale $\Lambda$, with  a mild dependence on the model parameters (see also Figure~\ref{fig:avg_kT2_II}). The average transverse momentum squared is independent of the scattering kinematics at small $x_B$, then decreases at large $x_B$, where four momentum conservation limits the amount of invariant energy in the final state and forces this to vanish: $\avekTsq \to 0$ as $x_B \to x_B^{\max}$ (see Appendix~\ref{app:kinematics} for more detail).
Leveraging the detailed calculations in Appendix~\ref{app:kinematics} we can then define theoretical, transverse-momentum-improved scaling variables simply by substituting $m_q^2 \rightsquigarrow m_q^2 + \avekTsq$ and $m_\phi^2 + (x-1) M^2 \rightsquigarrow m_\phi^2 + (x-1) M^2 + \avekTsq$ in Eq.~\eqref{eq:xbar_new} and define:
\begin{align}
    \xiqTnew 
    & \equiv \xi \Bigg( 1 + \frac{m_q^2 +  \avekTsq}{Q^2} 
    + \frac{\xi}{1-\xi} \frac{\big( m_q^2 + \avekTsq \big) \, \big(m_\phi^2 + (\xi-1) M^2 + \avekTsq \big)}{Q^4}\Bigg) 
\label{eq:xiqTstar_model}
    \\
    \vsqbarnewT 
    & \equiv v^2(\xiqTnew) \ ,
\label{eq:vbarsq_newT}
\end{align}
With these we will theoretically estimate the importance of including in the calculation the transverse momentum dynamics, that is not included in collinear factorization if one limits the analysis to leading twist level.

\subsection{Numerical validation}
\label{sec:numerical_validation}

In order to numerically study the validity of the kinematic approximations discussed in Section~\ref{sec:kinematic_approximations}, we will consider the sequence of $\bar x$ scaling variables summarized in Table~\ref{tab:xapprox}. Starting with Eq.~\eqref{eq:xbar_AOT}, we neglect at first the scattered quark virtuality as well as all external mass scales (i.e., $M=0, m_q=0$), then consecutively switch on the target mass and the quark mass. Subsequently, we consider the non-zero parton virtuality effects captured in Eq.~\eqref{eq:xbar_new}. Finally we analyze the kinematic impact of the parton transverse momentum by utilizing Eq.~\eqref{eq:xiqTstar_model}.

\begin{table}[bt]
\renewcommand{\arraystretch}{1.4}
\newcommand{\cm}{\checkmark}
\begin{tabular}{ c | c c | c c }
    $\bar{x}$ & $M$ & $m_q$  & $\vsqbar$ & $\kTsq$  \\
    \hline\hline
    \ $x_B$ \  & $0$   & $0$      & $0$ & $0$ \\
    $\xi$      & $\cm$ & $0$      & $0$ & $0$ \\
    $\xi_q$    & $\cm$ & $\cm$    & $0$ & $0$ \\
    $\xiqnew$  & $\cm$ & $\cm$    & $\vsqbarnew$ & $0$ \\
    \hline 
    $\xiqTnew$ & $\cm$ & $\cm$ & $\vsqbarnewT$ & $\avekTsq$ \\
\end{tabular}
\caption{Sequence of kinematic approximations considered for the validation of the $\bar x$ scaling variables choices discussed in Section~\ref{sec:kinematic_approximations}. 
A check mark indicates the inclusion of the particle masses  $M$ or $m_q$ in \eqref{eq:xbar_AOT}, and the approximation used for the internal $v^2$ and $\kTsq$ variables is indicated in the last two columns. Below the thin horizontal line, we included an approximation that can be considered for benchmarking purposes in the spectator model, where the dynamics of internal degrees of freedom is explicitly known, but not in QCD.
(The $\vsqbarnew$ and $\vsqbarnewT$ light-cone virtualities are defined in Eqs.~\eqref{eq:vbarsq_new_model} and \eqref{eq:vbarsq_newT}, respectively, and the corresponding scaling variables in Eqs.~\eqref{eq:xbar_new} and \eqref{eq:xiqTstar_model}. The average transverse momentum $\avekTsq$ is calculated in the model according to Eq.~\eqref{eq:avg_obs}.)
}
\label{tab:xapprox}
\end{table}

\begin{figure}[tb]
	\centering
	
	\includegraphics[trim=0.7cm 1cm 0.725cm 1cm,clip,width=0.85\textwidth]{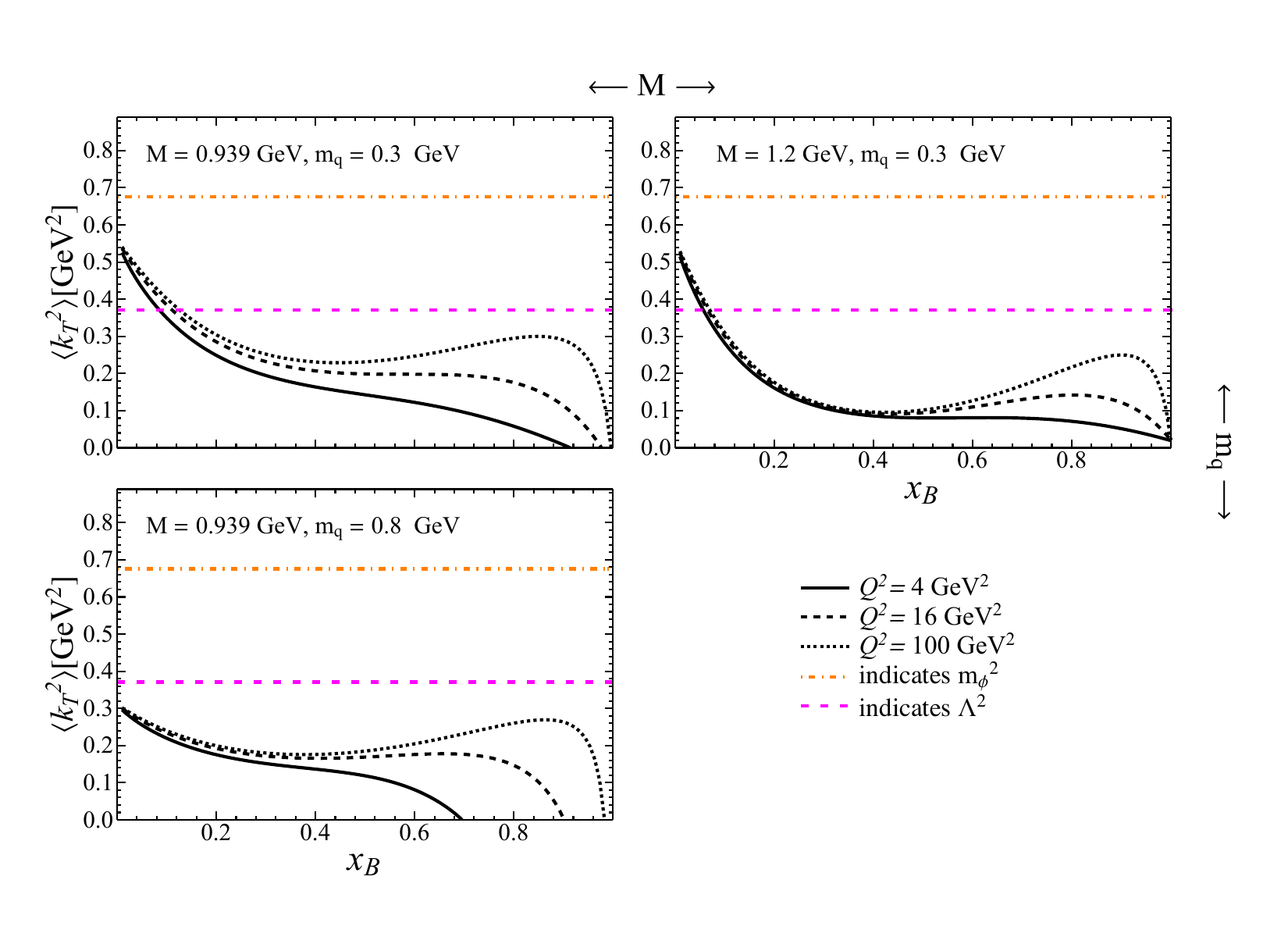}	
	\caption{
	Average unobserved $\kTsq$ of the incoming quark calculated in the full model as a function of $x_B$ at various $Q^2$ values and for different values of the model parameters $M$ and $m_q$. For scale comparison, the orange dot-dashed line indicates the model's value of $m_\phi^2$, while the magenta dashed line indicates the model's value of $\Lambda^2$, see Eq.~\eqref{eq:mphi_Lambda_default}.
	} 
	\label{fig:avg_kT2}
	
	\includegraphics[trim=0.45cm 1cm 0.5cm 1cm,clip,width=0.85\textwidth]{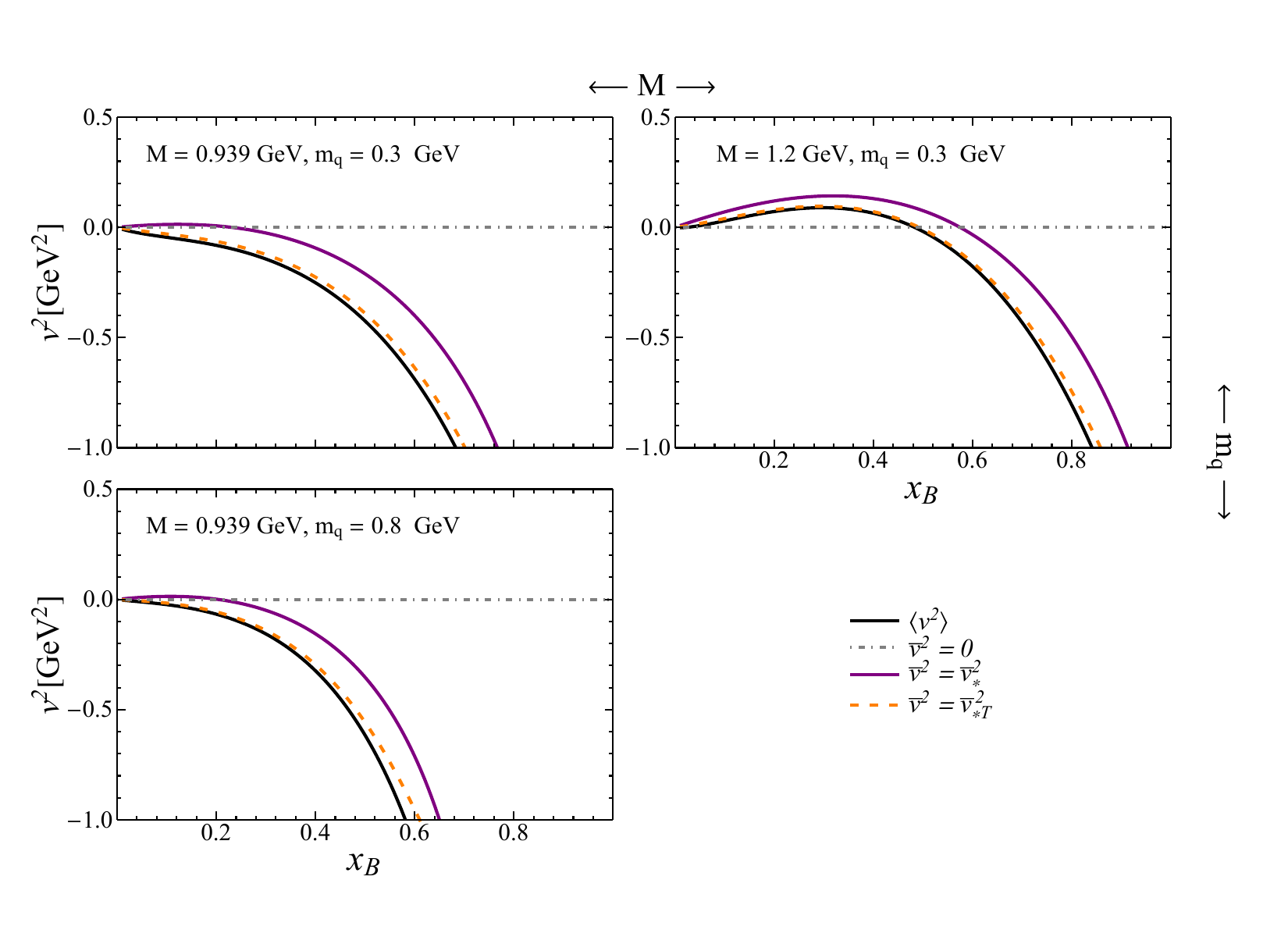}
	\caption{
	Average light cone virtuality $v^2=k^2+\kTsq$ calculated in the full model for $Q^2 = 4 \text{\ GeV}^2$, compared to the $\vsqbar$ approximations discussed in the text. In the top left panel, the default model parameters are used, and the other panels show the effect of increasing, respectively, $M$ and $m_q$. } 
	\label{fig:avg_k2_kT2}

\end{figure}

We start from the average light-cone virtuality $\langle v^2 \rangle = \langle k^2 + \kTsq \rangle$, shown as a black line in Figure~\ref{fig:avg_k2_kT2} and compared to the 3 choices of approximated $\vsqbar$ defined in Eqs.~\eqref{eq:vsqbar_choice}, \eqref{eq:vbarsq_new_model} and \eqref{eq:vbarsq_newT}. At small $x_B$, all vanish and the parton can be effectively considered collinear to the nucleon in the plus light-cone direction. As $x_B$ increases, however, the light cone virtuality becomes more and more negative until the kinematic threshold $\xbmax$ is reached. The only exception occurs at larger target mass values, for which a modest rise with $x_B$ is followed again by a fast dive as $x_B$ approaches its threshold. This behavior was already expected from the analysis of Eq.~\eqref{eq:v2_model} offered in Section~\ref{sec:kinematics}. The deviation of $\langle v^2 \rangle$ from $\vsqbar = 0$ is qualitatively captured by $\vsqbarnew$, and the remaining small gap is closed if the transverse momentum is taken into account by including the calculated, but unobserved, $\avekTsq$ into $\vsqbarnewT$.

It is worthwhile remarking that the plots in Figs.~\ref{fig:avg_kT2}-\ref{fig:avg_k2_kT2}, combined, also show that the incoming quark virtuality $\langle k^2 \rangle = \langle v^2 \rangle - \avekTsq$ is negative and of order $O(\Lambda^2)$ at all $x_B$ values: the quark is indeed a bound state of the proton. Clearly, the $k^2 \approx 0$ approximation utilized in many parton model calculations and derivations of factorization formulas is not accurate, and should rather be replaced, by $v^2 \approx 0$. Remarkably, as we will see, this approximation is quite sufficient in describing the average value of the quark light-cone momentum fraction $x$ and there will be no need to use a non-zero virtuality such as $\vsqbarnew$ in the calculation of inclusive observables.

\begin{figure}[tbh]
	\centering
	
	\includegraphics[trim=0.5cm 1cm 0.5cm 1cm,clip,width=0.85\textwidth]{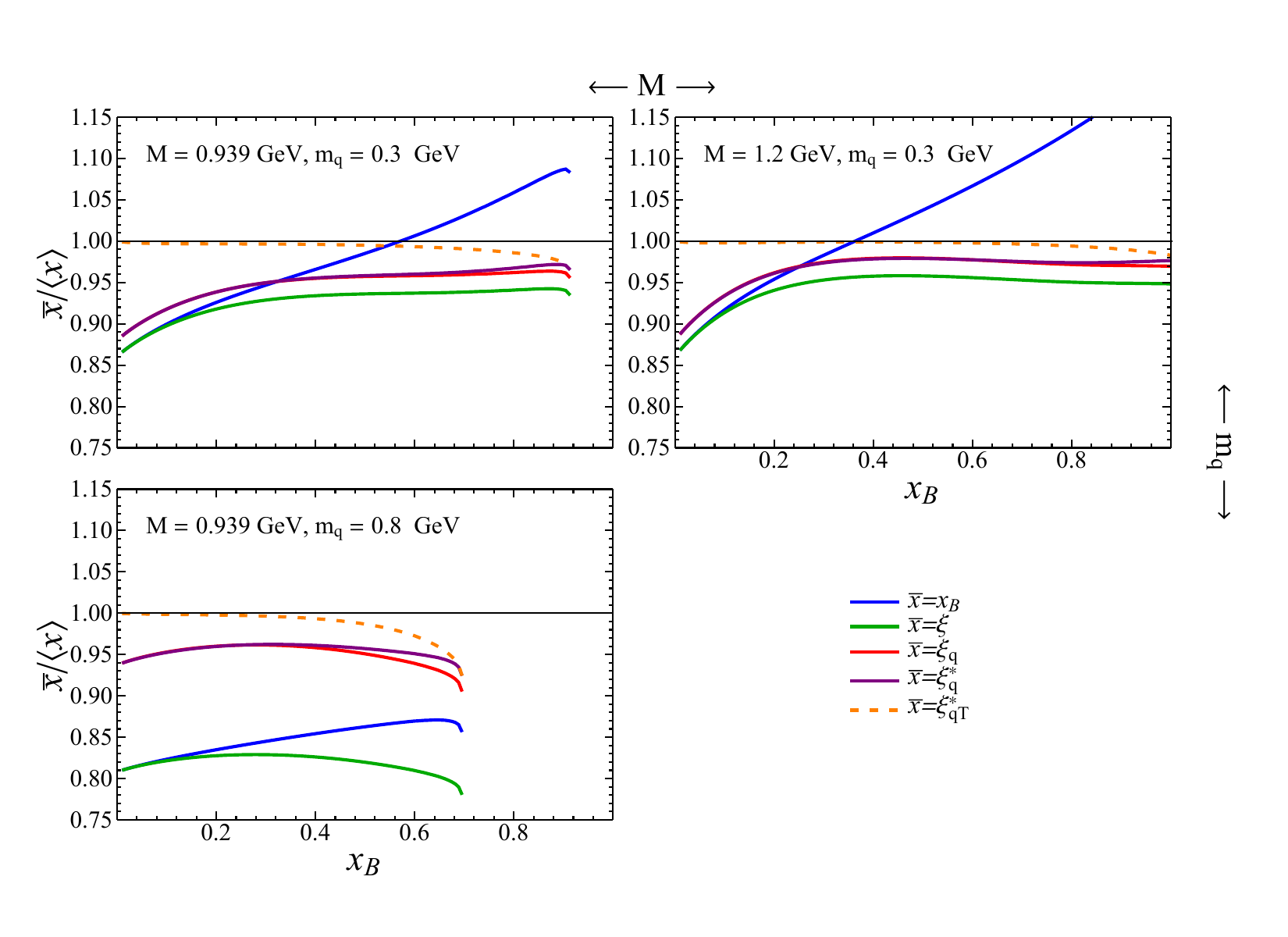}	
	\caption{
    Ratio of approximated $\bar x$ parton light-cone momentum fraction to the full $\langle x \rangle$ calculated in the model at $Q^2 = 4 \text{\ GeV}^2$. In the top left panel, the default model parameters are used, and the other panels show the effect of increasing, respectively, $M$ and $m_q$.} 
	\label{fig:avg_x}
\end{figure}

The average light-cone fraction $\langle x \rangle$ is another important internal variable that can be calculated in the model but can only be indirectly controlled experimentally by measuring the Bjorken invariant $x_B$. The comparison to the 5 approximations listed in Table~\ref{tab:xapprox} is performed for clarity by plotting the corresponding $\bar x/\langle x \rangle$ ratios as a function of $x_B$ in Figure~\ref{fig:avg_x}, where we adopted the same choice of kinematics and model parameters as in the average light-cone virtuality just discussed.
One can immediately see that the standard massless collinear approximation $\bar x=x_B$ (blue line) very poorly approximates the parton's average fractional momentum, and results in a $\bar x/\langle x \rangle$ ratio that is very sensitive to the model parameters. While this is not a problem at small $x_B$, where a small shift in the $x$ value at which the PDFs are calculated does not significantly alter the cross section, this can lead to significant under- or over-estimation of the structure functions at larger $x_B$.
On the contrary, keeping into account the target's mass inside the $\bar x = \xi$ scaling variable stabilizes the large $x_B$ approximation, even though the new scaling variable can still significantly underestimate $\langle x \rangle$, especially if the quark is heavy, see the green line. This is remedied by including the quark mass inside $\bar x = \xi_q$ (red line), which produces a good approximation for the model parameter values considered in Figure~\ref{fig:avg_x}. In fact, the approximation degrades for higher values of the ``confinement'' scale $\Lambda$ or the spectator mass $m_\phi$, see Appendix~\ref{app:systematics}. The latter case is particularly meaningful because, in reality, the proton's remnant after the hard scattering is a multi-particle state with a distribution in invariant mass not fully captured by a fixed $m_\phi \sim \langle m_X \rangle$. Finally, one can consider the effects of a non-vanishing virtuality $\vsqbarnew$, which however are suppressed by $1/Q^4$ and do not improve much the approximation, see the purple line compared to the red one. 

In summary, there seems indeed to be a limit to what any experimentally controllable inclusive DIS kinematic approximation to $\langle x \rangle$ can achieve. However, as it was the case for $\langle v^2 \rangle$, if one was able to control the partonic $\kTsq$ accessed in the hard scattering reaction, the approximation would become nearly perfect, as the orange dashed lines representing $\xiqTnew/\langle x \rangle$ shows. In Section~\ref{sec:beyond_LT} we will briefly discuss how this could in fact be achieved, either by utilizing the TMD factorization formalism, or including HT terms in the collinear calculation.

\section{Testing the limits of factorization}
\label{sec:fact_limits}

The model's hadronic tensor can be factorized according to Eqs.~\eqref{eq:factorized_Wmunu_LT}-\eqref{eq:Jacobian}. It is then possible to compute the factorized transverse structure function $F^\CF_T$ by contracting the tensor with the transverse projector $P_T$ of Eq.~\eqref{eq:helicityprojectorsII}. One finds:
\begin{equation}
F^{\CF}_T(x_B,Q^2) =   q(\bar x) \, ,
\label{eq:FT_approx}
\end{equation}
which at LO coincides with the quark PDF evaluated at a suitable $\bar x$ variable, that replaces $x_B$ as scaling variable away from the Bjorken limit. In the numerical analysis to follow, the $\bar x$ scaling variable chosen as summarized in Table~\ref{tab:xapprox}. 
Working analogously with the longitudinal projector $P_L$ of Eq.~\eqref{eq:helicityprojectorsII}, the longitudinal structure function vanishes at LO, $F_L^\CF = 0$ and will not be further considered.

At variance with QCD, where $q(x)$ is a non-perturbative quantity and needs to be extracted from the data or from lattice QCD measurements, the model's quark PDF can be analytically calculated, see for example Refs~\cite{Bacchetta:2008af,Gamberg:2014zwa}. From the definition \eqref{eq:phi_int_minus_II}, we find
\begin{align}
    q(x)
    & = \int \dtwokT dk^-\Tr \left[\frac{\gamma^+}{2} \Phi(p,k) \right]_{k^+=xp^+}
    = \int \frac{\dtwokT dk^-}{(2\pi)^4} \Tr \Bigg[ \vcenter{\hbox{\includegraphics[scale=0.25]{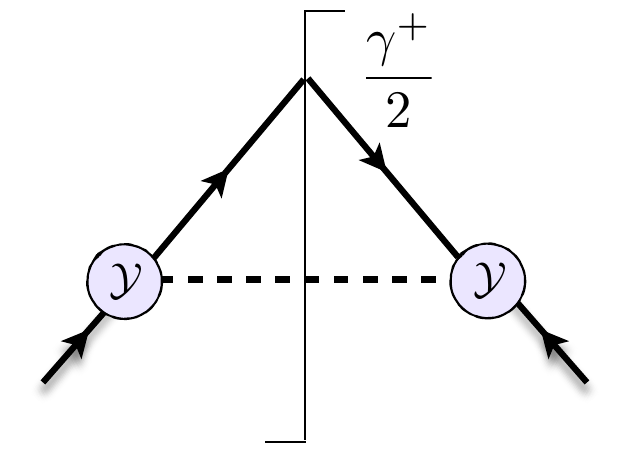}}} \, \Bigg]_{k^+=xp^+} \ ,
\end{align}
Using the model's Feynman rules to calculate the contribution of the diagram inside the trace, we obtain
\begin{align}
    q(x) 
    &= \int \frac{\dtwokT dk^-}{(2\pi)^4} \, g^2(k^2) \frac{1}{2} \frac{\Tr\big[(\psl+M)(\ksl+m_q)\frac{\gamma^+}{2}(\ksl+m_q)\big]}{(k^2-m_q^2)^2} (2 \pi) \, \delta((p-k)^2-m_\phi^2)\\
    &= \frac{g^2}{(2 \pi)^2} \frac{\big[2(m_q+xM)^2 + L^2(\Lambda^2)\big](1-x)^3}{24 L^6(\Lambda^2)},
\label{eq:model_PDF}
\end{align}
where 
\begin{align}
    L^2(\Lambda^2) 
    = x m_\phi^2 + (1-x)\Lambda^2 - x(1-x)M^2 \ .
\end{align}
In the model, we can therefore test the validity of the proposed sub-asymptotic collinear factorization by comparing the factorized and full calculations of the DIS transverse structure functions. By also comparing $\bm{k_T}$-unintegrated structure functions, we will furthermore be able to numerically explore the conditions for the breakdown of this approximation.

\subsection{Validity of the factorization approximations}

We first study to what degree collinear factorization provides a good approximation of the DIS process in the model by comparing the full $F_T^\DIS$ calculated as in Eq.~\eqref{eq:FT_full} with the factorized $F_T^\CF$ calculated with the model PDF \eqref{eq:model_PDF} according to Eq.~\eqref{eq:FT_approx}.
(The longitudinal structure function identically vanishes at LO, and an analogous test would require one to extend the model calculation to NLO, which goes beyond the scope of this article.)

\begin{figure}[p]

	\centering
	\includegraphics[trim=0.7cm 1cm 0.725cm 1cm,clip,width=0.85\textwidth]{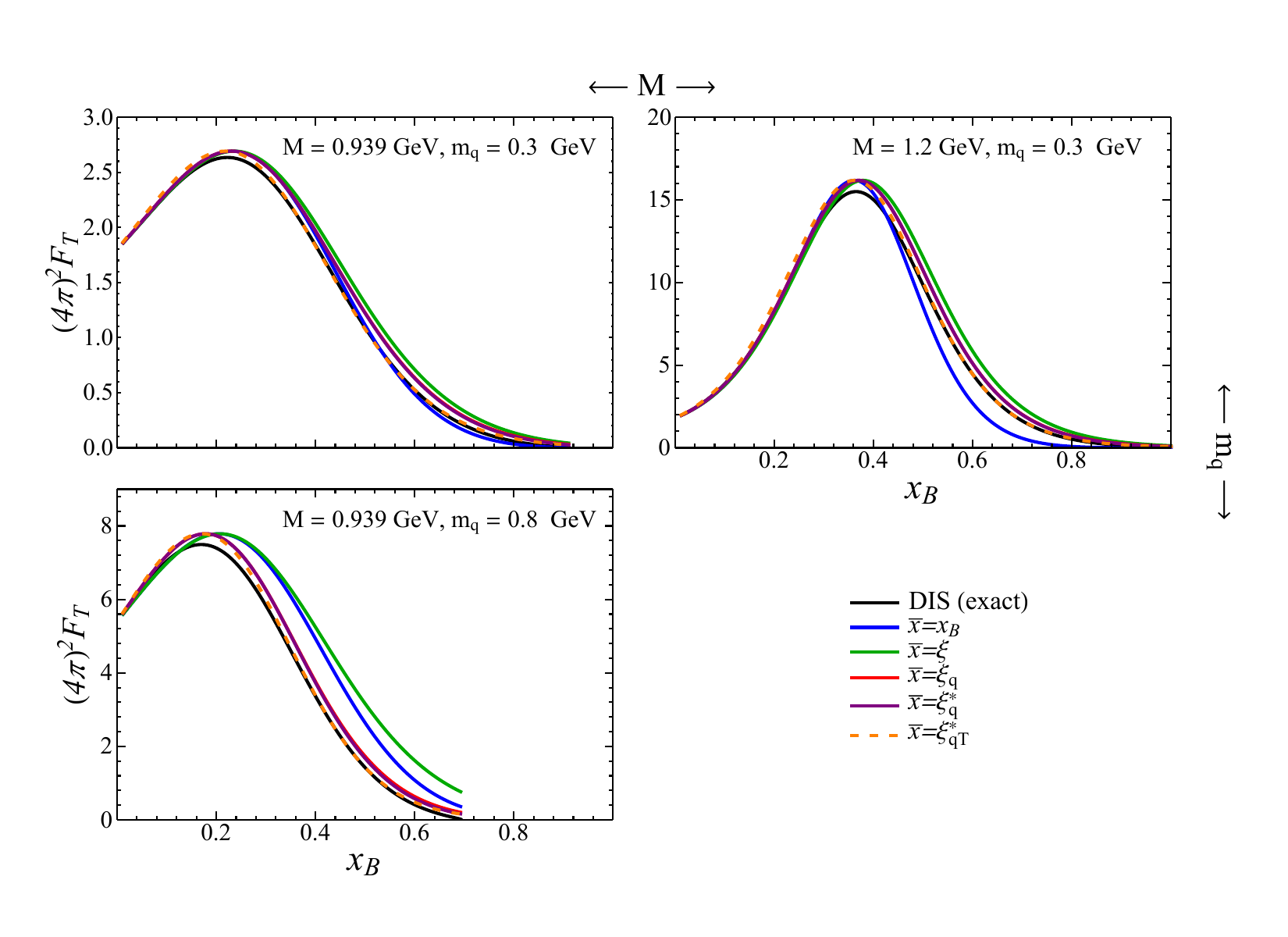}	
	\caption{
	Transverse $F_T^{\DIS}$ structure function calculated in the model at $Q^2=4$ GeV$^2$ compared to sub-asymptotic collinear factorization $F_T^{\CF}$ calculations that utilize the kinematic $\bar x$ approximations listed in Table~\ref{tab:xapprox}. In the top left panel, the default model parameters are used, and the other panels show the effect of increasing, respectively, $M$ and $m_q$.
	}
	\label{fig:FT}

	\centering
	\includegraphics[trim=0.7cm 1cm 0.725cm 1cm,clip,width=0.85\textwidth]{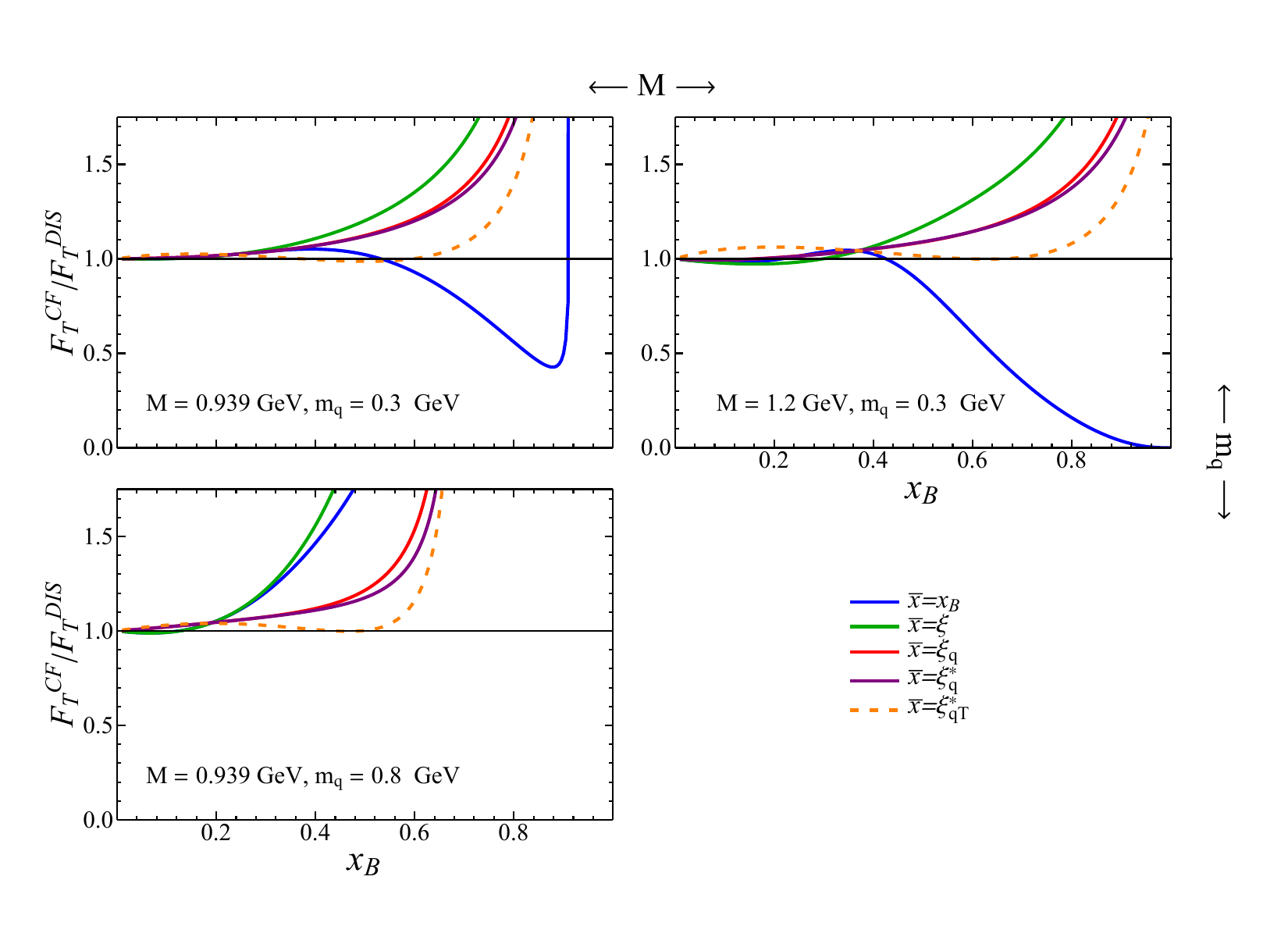}	
	\vskip-0.1cm
	\caption{
	Ratio of the factorized to exact DIS transverse structure functions presented in Figure~\ref{fig:FT}.
	In the top left panel, the default model parameters are used, and the other panels show the effect of increasing, respectively, $M$ and $m_q$.}
	\label{fig:ratio_FT}

\end{figure}

We do this first in  Figure~\ref{fig:FT}, where $F_T^\DIS$ is plotted as a black line, and various factorized $F_T^\CF$  structure functions computed at $Q^2=4$ GeV$^2$ with the scaling variables collected in Table~\ref{tab:xapprox} are represented by colored lines. We adopt the same choice of model parameters as in the study of the internal kinematics performed in Section~\ref{sec:numerical_validation}, and present a more detailed study of model parameter variations in Appendix~\ref{app:systematics}. 

One can immediately notice  the same pattern observed in Section~\ref{sec:numerical_validation}. Firstly, the asymptotic kinematic choice $\bar x=x_B$ (blue line) is very unstable with respect to the model parameters, as already observed in Ref.~\cite{Moffat:2017sha} for $k_T$-dependent structure functions. Secondly, taking into account mass corrections by using the $\bar x = \xi_q$ or $\bar x = \xiqnew$ scaling variables provides a much more stable and numerically closer approximation of $F_T^\DIS$. Finally, the $\kTsq$ corrections theoretically estimated with $\bar x = \xiqTnew$ (but not controllable experimentally) explain most of the remaining differences. One can also see that none of the factorized cross sections respect the kinematic $x_B \leq \xbmax$ bound (unless one worked strictly in the Bjorken $Q^2 \to \infty$ limit), which is a first illustration of the fact that the adopted factorization approximation breaks 4-momentum conservation, as already remarked in Section~\ref{sec:CF}.

The hierarchy of approximations just discussed can be better appreciated in Figure~\ref{fig:ratio_FT}, where we show the ratio of the $F_T^\CF$ collinear structure functions to their exact $F_T^\DIS$ counterpart. The fact that $F_T^\text{CF}|_{\bar x = x_B}$ does not approximate $F_T^\text{DIS}$ in any accurate way, can be traced to the failure of $x_B$ to approximate the average partonic momentum fraction $\langle x \rangle$. For example looking at the blue line in Figure~\ref{fig:ratio_FT}(a), where the default model parameters are utilized, we see that the CF structure function overestimates the exact DIS calculation at small $x_B$, then suddenly drops and  underestimates it even down to the $\sim 50$\% level before surging again closer to the kinematic threshold. This behavior can be understood from Figure~\ref{fig:avg_x}(a), where the choice $\bar x = x_B$ first underestimates the average $\langle x \rangle$ at $x_B \lesssim 0.6$ then overestimates it at $x_B \gtrsim 0.6$, right around the value where $F_T^\text{CF}|_{\bar x=x_B}$ becomes smaller than $F_T^\text{DIS}$. This clearly highlights the need of a more accurate kinematic approximation, which is indeed reached by fully taking into account the target and quark masses in $\bar x = \xi_q$. The order $O(1/Q^4)$ corrections included in $\bar x = \xiqnew$ do not qualitatively improve the agreement of $F_T^\CF$ with $F_T^\DIS$.

\begin{figure}[tbh]
	\centering
	\includegraphics[trim=0.7cm 1cm 1cm 1.75cm,clip,width=\textwidth]{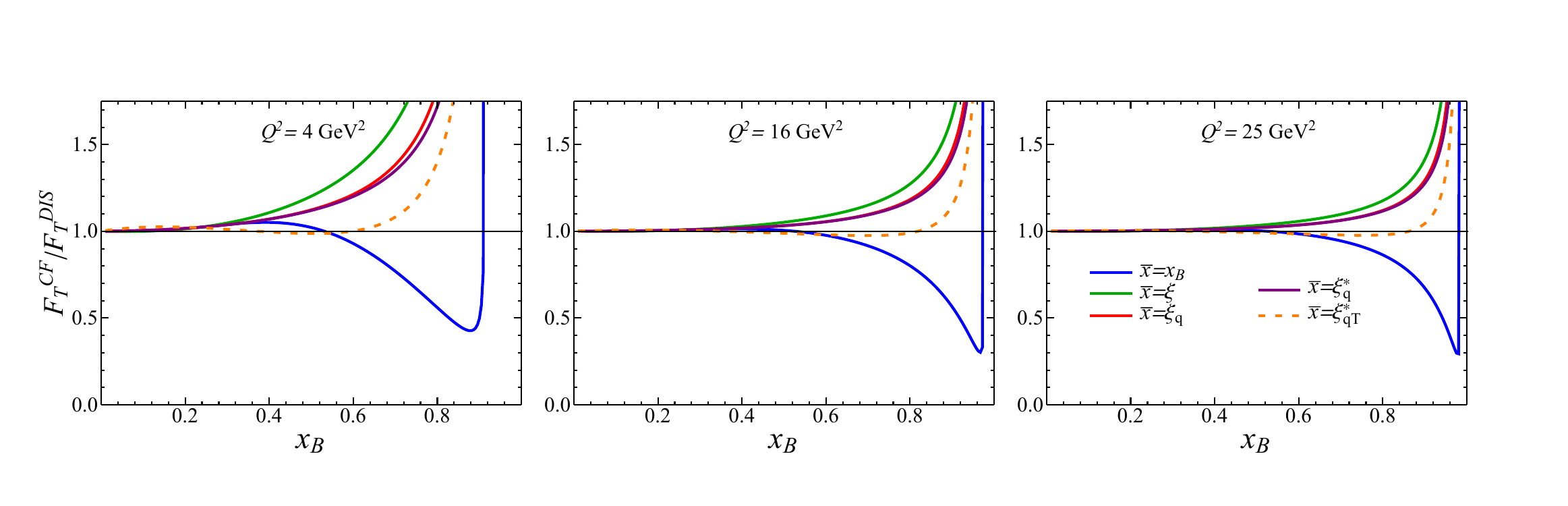}	
	\vskip-0.1cm
	\caption{
    Ratio of the factorized to exact DIS structure functions calculated in the model for several $Q^2$ values, and the default model parameters. The sub-asymptotic collinear factorization calculation was performed utilizing the kinematic approximations listed in Table~\ref{tab:xapprox}.
	}
	\label{fig:ratio_FT_Q2}
\end{figure}

In summary, it is clear that the use of the $\bar x=\xi_q$ scaling variable effectively provides the closest $F_T^{\CF}$ approximation to the real one, and extends the kinematic range of validity of CF to the largest $x_B$ value that can be achieved considering only external variables. The validity of CF is naturally extended at larger $Q^2$, where mass effects are suppressed by an inverse of power of the photon virtuality and become sizable only in an increasingly narrow region close to the kinematic threshold, see Figure~\ref{fig:ratio_FT_Q2}.

As observed many times, however, the CF cross section at leading twist misses transverse momentum effects, which are of the same order of magnitude, parametrically, as the mass effects successfully incorporated in $\xi_q$. The ``missing'' $\kTsq$ corrections can nonetheless be theoretically estimated by choosing $\bar x = \xiqnew$, and are shown as a dashed orange line in Figures~\ref{fig:ratio_FT} and \ref{fig:ratio_FT_Q2}. The model DIS calculation is largely reproduced in this way, showing that maximizing the range of validity of a factorized calculation requires one to control transverse-motion-induced power corrections. Even though these are beyond the scope 
of our leading twist calculation, one can handle their contribution to the cross section either by extending collinear factorization to higher-twist~\cite{Qiu:1988dn,Ellis:1982cd} or by utilizing a TMD formalism, where these are taken into account at parton distribution level \cite{Bacchetta:2006tn,Collins:2011zzd}. We will discuss this point more extensively in Section~\ref{sec:beyond_LT}.

\subsection{Breaking of Collinear Factorization}
\label{subsect:breakingCF}

Even after considering transverse momentum contributions, one can notice in Figures~\ref{fig:ratio_FT} and \ref{fig:ratio_FT_Q2} that the ratio $F_T^\CF/F_T^\text{DIS}$ strongly deviates from 1 at 
a Bjorken momentum fraction $x_B$ larger than a scale-dependent, ``factorization breaking'' threshold $\xBbreak = \xBbreak(Q^2)$. Defining breaking of factorization as a $\sim 10\%$ deviation of the CF structure function from the exactly calculated one,  
we obtain the values of $\xBbreak$ collected in Table~\ref{tab:xbreak}, where we also indicate the corresponding value of $W^2$ in parentheses, denoted by $\Wsqbreak$. A more complete study of the parameter systematics can be found in Appendix~\ref{app:systematics}.

Phenomenologically, one can notice that the dependence of $\xBbreak$ on $Q^2$ and on the model parameters, both internal and external, is largely determined by that of $\xbmax$, derived in Appendix~\ref{app:kinematics}:
\begin{align}
    \xbmax = \frac{1}{1+\frac{(m_\phi + m_q)^2-M^2}{Q^2}} \ .
\end{align}
Indeed, the larger the target mass $M$ the higher the $x_B$ value where factorization breaks down, and the larger the quark mass $m_q$ (or indeed the remnant's $m_\phi$) the smaller the breaking threshold. We can thus see that breaking of factorization is mostly a kinematic effect, and occurs when the process is too close to the edge of phase space. As a rule of thumb, we can see that this happens when
\begin{align}
    W^2 \lesssim 4 \text{\ GeV}^2 \ ,
\end{align}
with the quoted numerical value being a slight overestimate of the worst case combination of parameters shown here or in Appendix~\ref{app:systematics}. Even if this value was onained by a numerical analysis of calculations performed in the adopted spectator model, the parameters of latter have been chosen to phenomenologically reproduce the PDFs extracted in QCD by means of global fits and thus we can take the model to be a fair proxy for the latter. This makes us hopeful that the region of applicability of collinear factorization may be controlled at least in first approximation by a simple kinematic cuts even in QCD (and in that case, it looks like one would simply need to neglect the hadron resonance region, which is anyway removed from global QCD fits). 

\begin{table}[bt]
\renewcommand{\arraystretch}{1.4}
\begin{tabular}{ c c || c c c | c c c | c c c }
              &             & \multicolumn{3}{c|}{$Q^2=4$ GeV$^2$} 
                            & \multicolumn{3}{c|}{$Q^2=16$ GeV$^2$} 
                            & \multicolumn{3}{c}{$Q^2=25$ GeV$^2$} \\
    $M$ [GeV] & $m_q$ [GeV] & $\xBbreak$ & $\Wsqbreak$ [GeV$^2$] & $R_\text{break}$
                            & $\xBbreak$ & $\Wsqbreak$ [GeV$^2$] & $R_\text{break}$
                            & $\xBbreak$ & $\Wsqbreak$ [GeV$^2$] & $R_\text{break}$ \\
    \hline\hline
    \bf 0.94  & \bf 0.3     & 0.72 & (2.4) & 0.29 & 0.89 & (2.9) & 0.29 & 0.92 & (3.1) & 0.27 \\
    \bf 1.2   & 0.3         & 0.82 & (2.3) & 0.21 &      &        & &      &        &\\
        0.94  & \bf 0.8     & 0.59 & (3.7) & 0.24 &      &        & &     &         & \\ 
\end{tabular}
\caption{Factorization breaking thresholds $\xBbreak$, the corresponding values of invariant mass $\Wsqbreak$ (in parentheses) and of $\Rbreak$ for the choices of target mass $M$ and quark mass $m_q$ discussed in the main text and a selection of $Q^2$ values.}
\label{tab:xbreak}
\end{table}

Theoretically, the breaking of factorization can be traced back to the violation of momentum conservation in the kinematic approximations \eqref{eq:ktilde}--\eqref{eq:kptilde}, and in particular to the lack of transverse momentum conservation in the factorized diagram~\cite{Moffat:2017sha}. To see this explicitly, let us define the $\bm{k_T}$-unintegrated DIS structure function
\begin{align}
    \mathcal{F}_T^{\text{DIS}} (\kTsq)
    & \equiv
    \iint dx \, dk^2 \, \mathcal{F}_T^{\text{DIS}} (x,k^2,\kTsq)_{x_B,Q^2} \ ,
\end{align}
where the fully unintegrated DIS structure function appearing in the integrand was defined in Eqs.~\eqref{eq:unintegrated_FT} as the $P_T^{\mu\nu} \mathcal{W}_{\mu\nu}^\DIS(k)$ projection of the fully unintegrated DIS hadronic tensor. Similarly, we can define the $\bm{k_T}$-unintegrated CF structure function as
\begin{align}
    \mathcal{F}^\text{CF}_T(x,k^2,\kTsq)_{x_B,Q^2} 
    & \equiv M\, P_T^{\mu\nu} 
    \mathcal{W}^{\text{CF}}_{\mu\nu}(x,k^2,\kTsq)_{x_B,Q^2} \, ,
\label{eq:unintegrated_FT_CF}
\end{align}
with the fully unintegrated collinearly factorized  hadronic tensor defined as
\begin{align}
    2M \mathcal{W}^{\text{CF}}_{\mu\nu}(x,k^2,\kTsq)_{x_B,Q^2} 
    \equiv 
    \frac{\pi}{2x} \frac{1}{2 k^+} 
    \Tr\big[\nslash \Phi(k)\big] 
    \mathcal{H}_{\mu \nu} (x,\bar{x})\, . 
\end{align}
In either the DIS or the CF case, the structure function can then be obtained as an integral over $d\kTsq$:
\begin{align}
    F_T^\text{DIS}(x_B,Q^2) 
    & =
    \int_{0}^{{\ktmaxsq}} \dkTtwo \, \mathcal{F}^\text{DIS}_T(\kTsq)_{x_B,Q^2} 
    \\
    F_T^\text{\CF}(x_B,Q^2) 
    & =
    \int_{0}^{\infty} \dkTtwo \, \mathcal{F}^\CF_T(\kTsq)_{x_B,Q^2}
\end{align}
The crucial difference is the limit of integration: in the DIS case, $\kTsq$ is limited by 
four momentum conservation at any fixed value of the invariant mass $W^2$, and the integrand does not have support beyond $\ktmaxsq$; in the CF calculation, $\kTsq$ can run up to infinity after being effectively decoupled from the light-cone plus and minus directions by the approximations  \eqref{eq:ktilde}--\eqref{eq:kptilde}.

\begin{figure}[b]
	\centering
	\includegraphics[height=5.25cm]{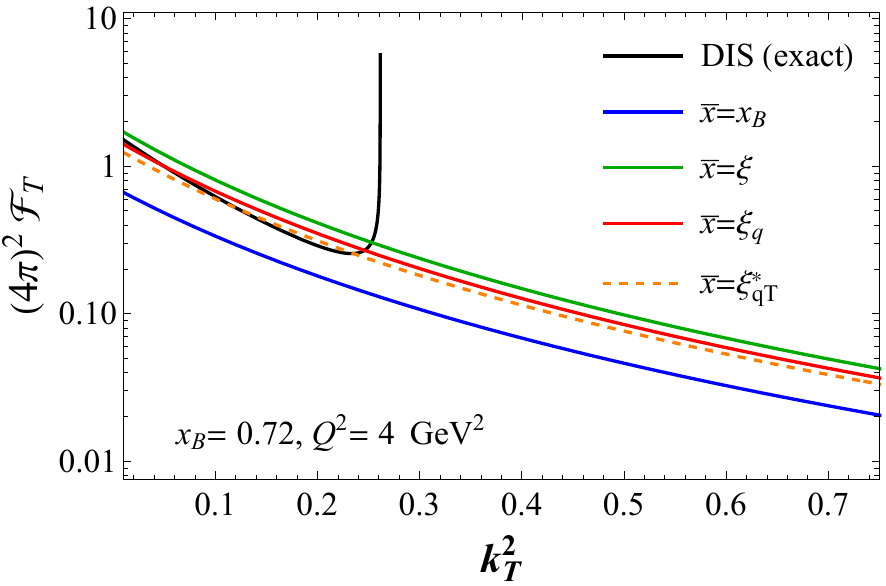}
	\includegraphics[height=5.25cm]{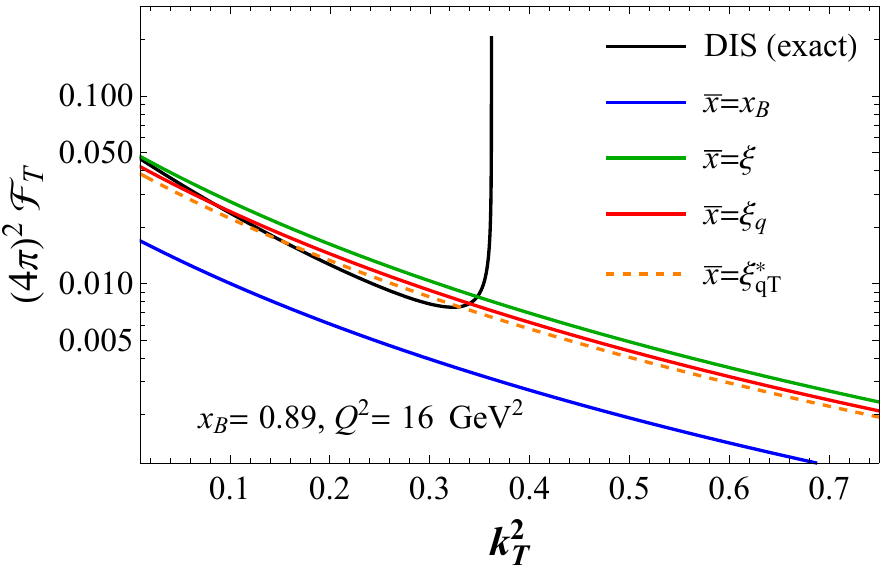}
	\caption{
	$\bm{k_T}$-dependent transverse structure function for the default parameters of the model. The two panels correspond to a low $Q^2=4$ GeV scale and a higher $Q^2=16$ GeV$^2$ scale. Their respective factorization breaking thresholds, $\xBbreak = 0.72$ and $\xBbreak = 0.89$ are chosen to maximize the difference between the factorized and full model calculations.}
	\label{fig:kT_FT}
\end{figure}

The difference between the handling of $\bm{k_T}$ in the DIS and the CF calculations is evident from Figure~\ref{fig:kT_FT}, where the $\bm{k_T}$-dependent $\mathcal{F}_T$ structure functions are plotted as a function of $\kTsq$ for $x_B= 0.6$, $Q^2 =4$ GeV$^2$, and nominal model parameter values.

At small $\kTsq$, we have a generally good enough approximation, except when using the $\bar x = x_B$ scaling variable (blue line) that consistently underestimates the transverse structure function. On the contrary, taking into account both the target and the quark mass in the $\bar x = \xi_q$ scaling variable provides one with the best approximation achievable considering only external variables. Nonetheless, this calculation slightly overestimates the full one and displays a less steep slope. This cannot be remedied without handling the transverse motion of the quark in detail, which one cannot do at LT in collinear factorization; however, including the average $\avekTsq$ in the theoretical $\bar x = \xiqTnew$ scaling variable, one can obtain a good description \textit{on average} of the full $\mathcal{F}_T^\DIS$, slightly underestimating it at small $\kTsq$ values, and slightly overestimating it at intermediate $\kTsq$ values. This is the reason why the $\bar x = \xiqTnew$ choice gives the best approximation for the $k_T$-integrated $F_T$ previously discussed.  

At even larger values of $\kTsq$ the factorized calculation is radically different: the black DIS curve  displays an integrable divergence at $\ktmaxsq$, while the CF colored curves have noticeable tails beyond that. At high $Q^2$ values, this happens at large enough $\kTsq$ that he unintegrated $\mathcal F_T$, dropping approximately as an inverse power of $\kTsq$, does not contribute much to the integrated $F_T$. However, at smaller $Q^2$ values the CF tail exceeding the kinematic limit provides a non negligible contribution to $F_T$. 
In order to quantify this contribution, we define the ratio
\begin{align}
R(x_B,Q^2) = \frac{\int_{\ktmaxsq}^\infty \dkTtwo \, \mathcal{F}^\CF_T(\kTsq)_{x_B,Q^2} }{ \int_{0}^\infty \dkTtwo \, \mathcal{F}^\CF_T(\kTsq)_{x_B,Q^2} } \ ,
\label{eq:R_ratio}
\end{align}
that provides one with the relative amount of integrated $F_T^\CF$ coming from the 4-momentum non-conserving $\kTsq > \ktmaxsq$ region. In Table~\ref{tab:xbreak}, we collect the $R$ values corresponding to each $\xBbreak$ and denote these by $\Rbreak$. As a rule of thumb we found that the CF calculation strongly deviates from the full calculation and factorization breaks down for $(x_B,Q^2)$ values such that 20\% or more of the integrated $F_T^\CF$ resides in the tail at $\kTsq > \ktmaxsq$.

The breaking of factorization just described is unavoidable, as it is rooted in the very approximation that allows one to define the quark PDF as an integral of the quark correlator $\Phi$ over the sub-leading parton momentum components $\bm{k_T}$ and $k^+$, and to factorize these from the parton-level, hard scattering. Being rooted in kinematics, however, the values $x_B$ and $Q^2$ at which the breakdown occurs can be at least qualitatively, if not semi-quantitatively, controlled by calculating the kinematic limits and estimating how far from these one should safely stay. As argued in the previous subsection in the context of our spectator model, a simple cut in $W^2$ may indeed be sufficient for inclusive structure functions. However, similar issues appear and are amplified in the case of semi-inclusive DIS measurements, and in particular when the hadron's transverse momentum is also measured \cite{Boglione:2016bph,Gonzalez-Hernandez:2018ipj,Boglione:2019nwk}. More refined analyses are urgently called for, especially in the context of the Jefferson Lab SIDIS program at the 12 GeV upgraded CEBAF facility \cite{Dudek:2012vr}, where the kinematics is inherently sub-asymptotic due to the relatively low beam energy.

\section{Theory and phenomenology beyond the leading twist}
\label{sec:beyond_LT}

Summarizing the numerical validation performed in the last 2 sections, there is a limit to the degree of accuracy with which a CF calculation can reproduce the inclusive DIS structure functions. We have identified 2 main drivers for this:
\begin{itemize}
\item  
There is an intrinsic limitation in the approximation of high-$\bm{k_T}$ scattering contributions, due to the neglect of transverse momentum conservation in the factorization of the non-perturbative parton correlator; 
\item 
the kinematic $\bar x$ scaling variable in inclusive processes can only be determined using external variables, and cannot incorporate the unobserved transverse momentum scale.
\end{itemize}

The first limitation is unavoidable, and one should try to identify in what region of phase space this occurs, in order to ensure that factorization is properly applied to describe experimental data and, conversely, to accurately extract parton distribution from these. In this paper, we suggest that, at least for inclusive observables, a measure of control over the validity of the factorized description of the scattering can be reached by kinematic considerations, and in particular by estimating $\xbmax$. The latter can be calculated exactly in the model, where all masses are known and all particles are asymptotic states, but in QCD one would would need some phenomenological input to obtain useful estimates. Identifying the region of applicability of factorization in semi-inclusive measurements is, instead, significantly more intricate \cite{Boglione:2016bph,Gonzalez-Hernandez:2018ipj,Boglione:2019nwk}. In both cases, we believe this is an important endeavor and more detailed studies are needed to avoid interpreting as partonic features of the data that are not. This is of capital importance, for example, to fully exploit the vast mess of data expected from the Jefferson Lab 12 GeV program \cite{Dudek:2012vr}.

Even within the region of validity of factorization, an imperfect description of the DIS cross section at large $x_B$ arises due to the neglect of transverse momentum in the factorized calculation at leading twist, see Figures~\ref{fig:ratio_FT}, \ref{fig:ratio_FT_Q2} and \ref{fig:ratio_FT_syst}.
However, after incorporating all external mass scales in the $\bar x = \xi_q$ scaling variable, the difference between the LT and the full cross section is small, and quite independent of the $\Lambda$ and $m_\phi$ internal parameters that mimic the nonperturbative QCD dynamics of the target. As demonstrated by incorporating in the scaling variable the (unmeasured, but calculable in the model) $\avekTsq/Q^2$ corrections, the calculation seems systematically improvable. 

The needed control of the partonic $\kTsq$ accessed in the hard scattering reaction can be theoretically achieved either by utilizing the TMD factorization formalism, where parton distributions themselves depend on $\bm{k_T}$ \cite{Bacchetta:2006tn,Collins:2011zzd}, or by performing collinear factorization up to higher twist level, in which case the transverse momentum dynamics is effectively included in multi-parton matrix elements \cite{Qiu:1988dn}.

Remaining in the context of collinear factorization, on can also try to absorb the contributions from the partonic transverse momentum into a phenomenological HT term to be fitted to the data:
\begin{equation}
 F_T^\DIS(x_B,Q^2) \approx F_T^\CF(x_B,Q^2) + H(x_B)/Q^2 \, .  
\end{equation}
Given a large enough leverage in $Q^2$, the $H(x)$ coefficient can clearly absorb the difference between the solid read and dashed orange curves in, say, Figure~\ref{fig:ratio_FT_Q2} and take care of the missing $O(\avekTsq/Q^2)$ transverse momentum corrections. However, a fit to model pseudo-data is needed to establish the accuracy to which PDFs can be extracted given the still imperfect, and not improvable, estimate of the internal partonic kinematic provided by the $\xi_q$ scaling variable, see Figure~\ref{fig:avg_x}. We leave this phenomenological investigation, as well as a determination of the level to which the DIS contribution to the lepton-nucleon cross section can be separated from the interference and resonance contributions discussed in Section~\ref{sec:FTcalcs}, to future work \cite{Krause-etal-inprep}.

\section{Summary and conclusions}
\label{sec:conclusions}

In this work, we have revisited QCD factorization with the aim to maximally extending its validity to sub-asymptotic values of the hard scale $Q^2$, where particle masses and other nonperturbative scales cannot be altogether neglected as customarily done in the Bjorken limit \cite{Aivazis:1993kh,Moffat:2017sha,Nadolsky:2009ge}. 
We have furthermore validated our findings by comparing factorized and analytically calculable structure functions in a spectator model designed to simulate the essential aspects of lepton-proton scattering in QCD.
The focus of this paper has been on inclusive lepton-nucleon observables as a first step towards collinear, and then transverse momentum dependent semi-inclusive scattering processes. 
The specific findings of our inclusive scattering analysis are many faceted, and it is worthwhile reviewing them in some detail before closing this paper.

\subsection{Collinear Factorization in DIS at sub-asymptotic \texorpdfstring{$\bm{Q^2}$}{p}}
\label{sec:summary-sub_asymptotic_CF}

In the context of inclusive DIS, we have re-derived the QCD factorization formula with the aid of a final state quark jet function, explicitly separating the partonic internal kinematics, which needs to be approximated, from the hard scattering dynamics, that can be systematically expanded in powers of $|\vect{k}_T/Q|$ and, for light quarks, of $m_q/Q$. In particular, we have obtained a gauge invariant factorized formula for the leading power (or ``leading-twist'') contribution to the DIS cross section, that in fact coincides with the asymptotic formula evaluated at a rescaled variable $\bar x$ instead of $x_B$. Using a jet diagram proved essential for the gauge invariance of the result, and no change in the operator definition of parton distributions proved necessary.

While our proof has been obtained at LO, a generalization to NLO should not present essential difficulties. Indeed, any diagram at any order can be written as a trace term multiplied by an overall 4-momentum conservation delta function, as we have done in Eq.~\eqref{eq:Wmunufact}, thus isolating the part of the cross section in need of kinematic approximation from the terms in which partonic dynamics comes into play. The latter can be expanded in a power series around collinear partons, and transverse momentum and parton virtuality contributions are included in higher-twist terms. Let us stress that this is the central idea of our approach: isolating the kinematics of the external legs of the hard scattering diagram from the rest, and performing a \textit{minimal} set of (uncontrolled) approximations only there. 

\subsection{Validation at Leading Twist}

Working at leading twist level, the goodness of the kinematic approximations was checked by utilizing a detailed spectator model calculation, were we could separate the DIS lepton-quark scattering contribution from resonance and interference contributions. The factorized model transverse function, fully calculable analytically, was the compared to the exact DIS model calculation. 

As shown by the theoretical factorization analysis, the choice of the scaling variable $\bar x$ is arbitrary, in principle, but can be guided by considering the kinematics of the hard vertex. In particular, using the spectator model, we have verified that choosing $\bar x = \xi_q$ with $\xi_q \equiv \xi\big(1 + \frac{m_q^2}{Q^2}\big)$ one can reproduce the DIS cross section with only a relatively small overestimation at the large values of $x_B$ close to the inelastic scattering kinematic threshold.
The variable $\xi_q$, advocated for example in \cite{Aivazis:1993kh,Nadolsky:2009ge} for heavy quark production and adapted here to scattering on light quarks, captures the kinematic effect of the initial state target mass and of the final state quark mass. It is also analogous to the SIDIS scaling variable  $\xi_h = \xi\big(1 + \frac{m_h^2}{ \zeta_h Q^2}\big)$ proposed in \cite{Accardi:2009md,Guerrero:2015wha} and successfully utilized in \cite{Guerrero:2017yvf} to incorporate corrections due to the mass of the observed $h$ hadron in the final state. The freedom of rescaling variable choice guaranteed by our approach has also allowed us to explore the effects of $O(\mu^4/Q^4)$ mass corrections, which are not necessarily small due to the parton becoming more and more offshell as one approaches the boundary of phase space at large $x_B$. Fortunately, these corrections turn out to be small for inclusive processes, unless the mass of the quark is increased beyond the range suitable for our light-quark treatment.  

Finally, as already noticed by \cite{Moffat:2017sha}, we have explicitly illustrated an inherent limitation of collinear factorization, that breaks down at large $x_B$  due to the neglect of momentum conservation in the transverse direction. Here, we suggest that the region in $x_B$ and $Q^2$ where this happens might be semi-quantitatively controlled by kinematic considerations, and in fact a cut in invariant mass of the order of $W^2 \gtrsim 4$ GeV$^2$ might circumscribe the factorization breaking region. Given that in QCD the excluded region corresponds to resonance excitations, which are outside the scope of the DIS handbag diagram, such a requirement does not seem to be an unduly restrictive condition.

\subsection{Transverse momentum effects beyond leading twist}

The remaining difference between the CF and full calculation of the transverse structure function is mostly explained by incorporating the average $\avekTsq$ in the scaling variable, and considering corrections only up to $O(\avekTsq/Q^2)$. For inclusive observables, this study is only possible in the model but we found that the result is stable against variations of the internal confinement parameter $\Lambda$ and mass of the target remnant $m_\phi$. This strongly suggests that the range of validity of collinear factorization can be maximized in the large-$x_B$ and low-$Q^2$ region by extending the calculations to twist-4 level, either theoretically (including contributions from multi-parton diagrams) or phenomenologically (by adding a fitting a $H(x)/Q^2$ term to the LT factorized calculation). A fit of model pseudo-data is under way \cite{Krause-etal-inprep} to verify if a phenomenological extraction of PDFs can be pushed to as large values of $x_B$ as the factorized calculation, which is still underestimating by $\sim10\%$ the values of $\langle x \rangle$ approximated by $\xi_q$.

A finer treatment of the partonic transverse momentum can only be obtained in SIDIS processes, where this can be controlled by a measurement of a final state hadron but hadron mass corrections are more complex. Nevertheless, the results of our paper are very encouraging regarding the procedure advocated in Refs.~\cite{Guerrero:2015wha,Guerrero:2017yvf} for SIDIS and in Ref.~\cite{Accardi:2014qda} for semi-inclusive hadron annihilation in $e^+ + e^-$ scattering. In those works the scattered quark virtuality is bound by $\vpsq \geq m_h^2/\zeta$ (with $m_h$ the identified hadron mass and $\zeta$ its light-cone fractional momentum) instead of $\vpsq \geq m_q^2$ as in the DIS case. 
It remains an important exercise to perform a model validation for these two cases, and study the range of validity of collinear factorization for semi-inclusive processes, before extending our analysis to kinematically more complex case of transverse momentum dependent measurements \cite{Boglione:2016bph,Gonzalez-Hernandez:2018ipj,Boglione:2019nwk}.

\subsection{Beyond Deep Inelastic Scattering}

The model used in this study has richer structure than exploited in the validation of collinear factorization of the cross section's DIS component. Indeed the model also includes $e+p$ resonant quasi-elastic scattering and the interference between this and the DIS scattering. (Suitably generalized to higher mass resonances, the model may thus also prove to be an useful tool in the study of quark-hadron duality~\cite{Melnitchouk:2005zr}, potentially allowing one to access PDFs in lepton-proton scattering processes at even higher $x_B$ than currently possible.)

With a gauge-invariant decomposition of the physical process into DIS, resonance and interference processes, we have shown that the interference and resonance contribution to the transverse $F_T$ structure function exhibit an approximate $1/Q^2$ and $1/Q^4$ scaling. Thus one can also envision using model pseudo-data to test how effectively a phenomenological HT piece can decouple these from the DIS contribution in a PDF fit \cite{Krause-etal-inprep}.

The longitudinal $F_L$ structure function, instead, is further constrained by gauge invariance and does not exhibit the strongly ordered power-law scaling observed in the case of $F_T$. Rather, its DIS, resonance and interference components are all of the same order of magnitude -- an effect we traced back to the small-$x_B$ behavior of the longitudinal $P_L^{\mu\nu}$ projector. This is therefore a model-independent result and may have a bearing also for a CF analysis of real $F_L$ measurements, with the factorized $F_L^\text{CF}$ free to increase as an inverse power of $x_B$ at small values of the Bjorken invariant while the full $F_L \to 0$ as $x_B \to 0$. In fact, recent measurements at HERA \cite{Andreev:2013vha,Abramowicz:2014jak,Abramowicz:2015mha} have revealed small-$x_B$ tensions between data and CF calculations, part of which may be explained by our observation. A more detailed CF analysis of $F_L$ at small $x_B$ would require a NLO model calculation that is outside of the scope of the present article but remains an interesting exercise for the future.

\begin{acknowledgments}
We thank A. Bacchetta, L. Gamberg, M. Radici and A. Signori for helpful discussions and remarks.
This work was supported in part by the  U.S. Department of Energy contract DE-AC05-06OR23177, under which Jefferson Science Associates LLC manages and operates Jefferson Lab, and within the framework of the TMD Topical Collaboration.
AA also acknowledges support from DOE contract DE-SC0008791. JVG also acknowledges partial support from the Jefferson Science Associates (JSA) 2018-2019 Graduate Fellowship Program.  
\end{acknowledgments}

\begin{appendix}

\newpage
\section{Invariant and helicity structure functions, and symmetry constraints at small $x_B$}
\label{app:projectors_detailed}	

In Section~\ref{sec:full_hadronic_tensor}, we discussed how to uniquely decompose a rank-2 tensor, $T^{\mu \nu}$, into a gauge invariant and a gauge breaking part. 
The procedure we adopted consists in defining a complete set of orthogonal rank-2 projectors, $\left\{ P_\lambda^{\mu\nu} \right\}$, maximizing the number that satisfy the electromagnetic Ward identity, $q_\mu P_\lambda^{\mu\nu}=0$. For simplicity, in the main text we focused only on the parity invariant tensors, which are involved in neutral current exchanges such as in the model. 
In this appendix we complete that discussion by also consider the parity breaking tensors. We will also examine the $x_B \to 0$ limit of the structure functions, completing the analysis initiated in Section~\ref{sec:FLcalcs} for the longitudinal $F_L$ structure function. 

We can now use the polarization vectors~\eqref{eq:polvect} to define the parity breaking helicity projectors for rank two tensors as
\begin{align}
	\begin{split}
		P_A^{\mu\nu}(p,q) & = \varepsilon_+^\mu(p,q) \varepsilon_+^{\nu*}(p,q) 
		- \varepsilon_-^\mu(p,q) \varepsilon_-^{\nu*}(p,q) \\
		P_{[LS]}^{\mu\nu}(p,q) 
		& = \varepsilon_0^\mu(p,q) \varepsilon_q^{\nu*}(p,q)
		- \varepsilon_q^\mu(p,q) \varepsilon_0^{\nu*}(p,q).
		\label{eq:helicityprojectors_breaking}
	\end{split}
\end{align}
Using 
\begin{align}
	 \varepsilon_+^\mu(p,q) \varepsilon_+^{\nu*}(p,q) 
	- \varepsilon_-^\mu(p,q) \varepsilon_-^{\nu*}(p,q)
	 = \frac{-i\varepsilon^{\mu\nu\alpha\beta} p_\alpha q_\beta}
	{\sqrt{(p \cdot q)^2 - p^2 q^2}} 
\end{align}
and the explicit polarization vectors definition~\eqref{eq:polvect}, the parity breaking projectors can be more compactly written in ``Lorentz'' representation as
\begin{align}
	\begin{split}
		P_A^{\mu\nu}(p,q) &= \frac{ -i\varepsilon^{\mu\nu\alpha\beta} \hat{p}_\alpha q_\beta}
		{\sqrt{-q^2\hat{p}^2}}\\
		P_{[LS]}^{\mu\nu}(p,q) 
		& = \frac{\hat{p}^\mu q^\nu - q^\mu \hat{p}^\nu }{\sqrt{-q^2 \hat{p}^2}}
		\label{eq:helicityprojectors_breaking_II}
	\end{split}
\end{align}
where $\hat{p}^\mu \equiv p^\mu - \frac{p \cdot q}{q^2} q^\mu$. In this form it is easy to verify that these two projectors are orthogonal to each other. Moreover, the parity breaking projectors are antisymmetric in the indices, and therefore also orthogonal to the parity invariant projectors~\eqref{eq:helicityprojectorsII}, that are symmetric in the indices and we report here for reading convenience:
\begin{align}
\begin{split}
    P_L^{\mu\nu}(p,q) & = \frac{\hat{p}^\mu \hat{p}^\nu}{\hat{p}^2} 
    \\
    P_T^{\mu\nu}(p,q) & = -\hat{g}^{\mu \nu} +\frac{\hat{p}^\mu \hat{p}^\nu}{\hat{p}^2}  
    \\
    P_S^{\mu\nu}(p,q) & = -\frac{q^\mu q^\nu}{q^2} 
    \\
    P_{\{LS\}}^{\mu\nu}(p,q) 
    & = \frac{\hat{p}^\mu q^\nu + q^\mu \hat{p}^\nu }{\sqrt{-q^2 \hat{p}^2}} \ ,
\label{eq:helicityprojectorsII_bis}
\end{split}
\end{align}
with $\hat{g}^{\mu \nu} = g^{\mu\nu} - \frac{q^\mu q^\nu}{q^2}$. In summary, 
\begin{alignat}{2}
	& P_\lambda \cdot P_{\lambda'} = 0 
	& \qquad & \text{for\ } \lambda\neq\lambda' \nonumber \\
	& P_\lambda \cdot P_{\lambda} =  1 
	& \qquad & \text{for\ } \lambda = L, S \\
	& P_\lambda \cdot P_{\lambda} =  2
	& \qquad & \text{for\ } \lambda = T  \nonumber \\
	& P_\lambda \cdot P_{\lambda} =  -2
    & \qquad & \text{for\ } \lambda = A, \{LS\}, [LS] \nonumber
\end{alignat}
In the Lorentz representation \eqref{eq:helicityprojectors_breaking_II} and \eqref{eq:helicityprojectorsII_bis}, it is easy to convince oneself that these 6 defined projectors are also a complete orthogonal basis for the space of rank-2 tensors $T^{\mu\nu}=T^{\mu\nu}(p,q)$ built of the proton and photon momenta $p$ and $q$, such as the hadronic tensor for inelastic $e+p$ scattering.

Exploiting the the completeness of this basis, and noticing that the axial projector satisfy the Ward identity,
\begin{align}
q \cdot P_{A} \equiv q_\mu P_{A}^{\mu\nu} = 0 \ ,
\end{align}
we can extend to the anti-symmetric sector the decomposition~\eqref{eq:Wmunu_gi_gb} of a $T^{\mu\nu}$ tensor into a gauge invariant and maximally gauge breaking piece by defining 
\begin{align}
    T^{\mu\nu} = T^{\mu\nu}_\text{inv.} + T^{\mu\nu}_\text{g.b.} \ ,
\end{align} 
with 
\begin{align}
    T_\text{inv.}^{\mu\nu}(p,q) 
    & = P_T^{\mu\nu} F_{T}(x_B,Q^2) 
    + P_L^{\mu\nu}F_{L}(x_B,Q^2)
    + P_A^{\mu\nu} F_{A}(x_B,Q^2)
    \\
    T_\text{g.b.}^{\mu\nu}(p,q) 
    & = P_S^{\mu\nu}F_{S}(x_B,Q^2)
    + P_{\{LS\}}^{\mu\nu} F_{\{LS\}}(x_B,Q^2) 
    + P_{[LS]}^{\mu\nu} F_{[LS]}(x_B,Q^2) \ .
\end{align}
The $F_\lambda$ structure function are defined as in Eq.~\eqref{eq:F_lambda}, \textit{i.e.}, 
\begin{align}
F_\lambda(x_B,Q^2) \equiv c_\lambda P_\lambda (p,q) \cdot T(p,q) \, ,
\end{align}
but with the index now ranging over the full range $\lambda = L,T,A,S,\{LS\}, [LS]$ and $c_\lambda = 1/P_\lambda\cdot P_\lambda$.

As demonstrated in Section~\ref{sec:FLcalcs}, it is instructive to study the $x_B \to 0$ limit of the helicity projectors. Utilizing $\hat p^\mu \to - (p \cdot q / q^2 ) q^\mu = (1/2x_B) q^\mu$ and $\hat{p}^2 \to -(p\cdot q)^2/q^2 = -q^2/(2x_B)^2$, it is easy to see that 
\begin{alignat}{2}
    P_T^{\mu\nu} & \xrightarrow[x_B \to 0]{} -g^{\mu\nu} 
    \nonumber \\
    P_L^{\mu\nu} & \xrightarrow[x_B \to 0]{}  P_S^{\mu\nu} \hspace*{1cm} 
      & P_{\{LS\}}^{\mu\nu} &\xrightarrow[x_B \to 0]{} -2P_S^{\mu\nu} 
    \\
    P_A^{\mu\nu} &\xrightarrow[x_B \to 0]{} 0
      & P_{[LS]}^{\mu\nu} &\xrightarrow[x_B \to 0]{} 0 \ .
    \nonumber 
\end{alignat}
Therefore, all structure functions except $F_T$ are constrained at small $x_B$:
\begin{alignat}{2}
    F_L & \xrightarrow[x_B \to 0]{} F_S \hspace*{1cm}
      & F_{\{LS\}} & \xrightarrow[x_B \to 0]{} F_S \\
    F_A & \xrightarrow[x_B \to 0]{} 0 
      & F_{[LS]} & \xrightarrow[x_B \to 0]{} 0 \ .
\end{alignat}
For gauge conserving interactions, such as electromagnetism, we furthermore find that
\begin{align}
    F_L & \xrightarrow[x_B \to 0]{} 0 \ . 
\label{eq:FL_to_zero}
\end{align}
An interesting consequence of this constraint, 
already discussed in Section~\ref{sec:FLcalcs}, is that
\begin{align}
    F_L^\DIS \xrightarrow[x_B \to 0]{} -F_L^{\text{\,non-DIS}} \ ,
\end{align}
which alters the usual scaling expectations based on perturbative arguments, and introduces a novel source of ``higher-twist'', or rather inverse $Q^2$ power corrections. In particular, we have explicitly demonstrated for electromagnetic interactions in our model that $F_L^\DIS \propto 1/Q^2$ in the small $x_B$ limit, see also Appendix~\ref{app:low_xB_scaling}. Let us stress that Eq.~\eqref{eq:FL_to_zero} is purely a consequence of gauge invariance, and therefore we expect this remark to be important for the QCD interpretation of small-$x_B$ measurements of the proton's $F_L$. Similar considerations also apply to $F_A$ and $F_{[LS]}$.

\newpage
\section{Sub-asymptotic kinematics in detail}
\label{app:kinematics}

We discuss here in some detail the kinematics of the model, and derive a number of bounds on relevant internal and external variables. Note that the derivation of the bounds does not rely on model peculiarities, nor it is confined to LO diagrams. A generalization to QCD will also be presented. 

\subsection{Limits on \texorpdfstring{$\bm{x_B}$}{p}}

We start by considering the total invariant mass, $W^2$,
\begin{equation}
	W^2 = (p+q)^2 = M^2 + Q^2\Big(\frac{1}{x_B} - 1\Big) \ .
\end{equation}
In the model, by four momentum and lepton number conservation we can write 
\begin{equation}
W^2 \geq (p_\phi + k')^2 \ ,
\label{eq:W2_finalstate}
\end{equation}
where we have exploited the fact that a quark and a meson are the minimal mass final state that can be produced in an inelastic scattering. These particles are asymptotic states of the model, and we can use 
$p_\phi^2 = m_\phi^2$, $k^{\prime\,2} = m_q^2$ and  $p_\phi \cdot k' \geq m_\phi m_q$ to derive an upper limit for $x_B$:
\begin{equation}
x_B \leq \frac{1}{1+\frac{(m_\phi + m_q)^2-M^2}{Q^2}} \equiv \xbmax \ .
\label{eq:xbmax}
\end{equation}
The threshold $\xbmax$ corresponds to the production of the scattered quark and spectator meson at rest in the target rest frame. Note that the upper bound can be larger than 1 if the nucleon is unstable, that is, if $m_\phi + m_q < M$. A resonance will then appear in the scattering cross section at $x_B=1$ as shown, for example, in Figure~\ref{fig:FT_unstable_proton}.

In QCD, the quark needs to minimally hadronize into a pion, and by baryon number conservation a proton must also be minimally present in the final state. The QCD threshold value can then be obtained from Eq.~\eqref{eq:xbmax} substituting $m_\phi \rightsquigarrow M$ and $m_q \rightsquigarrow m_\pi$, and is known as the ``pion threshold'' $x_\pi$:
\begin{align}
    x_{B|\QCD} \leq \frac{1}{1+\frac{2m_\pi M + m_\pi^2 }{Q^2}} \equiv x_\pi \ .
\end{align}
The proton in QCD is a stable particle, and the phase space for inelastic scattering is indeed cut off before $x_{B|QCD}$ can reach 1.

\subsection{Limits on \texorpdfstring{$\bm{\kTsq}$}{p}}

Given a finite final state invariant mass $W^2$, the transverse momentum squared of the scattered quark is also limited. To derive its bounds, we consider the center of mass frame, where $\boldsymbol{k}' = -\boldsymbol{p_\phi}$ and $\boldsymbol{p_{\phi,T}} =  - \boldsymbol{k_T}' = - \boldsymbol{k_T}$. Then,
\begin{align}
W^2 \geq (p_\phi^0 + k'^0)^2 
&= \Big(\, \sqrt{m_\phi^2 + \kTsq + (p_\phi^z)^2} +  \sqrt{m_q^2 + \kTsq + (k'^z)^2}\, \Big)^2\\
&\geq 	 \Big(\, \sqrt{m_\phi^2 + \kTsq} +  \sqrt{m_q^2 + \kTsq}\, \Big)^2
\end{align}
where we have used $(p_\phi^z)^2 \geq 0$ and $(k'^z)^2 \geq 0$. Now, solving the last inequality imposes an upper limit in $\kTsq$,
\begin{equation}
	\kTsq \leq \frac{\Big(W^2 - (m_\phi + m_q)^2\Big)\Big(W^2 - (m_\phi - m_q)^2\Big)}{4 W^2} \equiv \ktmaxsq
\label{eq:ktmaxsq}
\end{equation}
Note that solving this for $x_B$ when $\ktmaxsq =0$ one recovers the upper limit Eq.~\eqref{eq:xB_max}. This is as it should, since $\xbmax$ corresponds to the production of the recoiled quark and spectator both at rest in the target rest frame.

As with the derivation of $\xbmax$, the transverse momentum bound in QCD can be obtained by substituting $m_\phi \rightsquigarrow M$ and $m_q \rightsquigarrow m_\pi$ in Eq.~\eqref{eq:ktmaxsq}.


%
\subsection{Internal variables: parton fractional momentum and virtuality}
\label{app:Delta_sol}
At LO, the unintegrated hadronic tensor contains two delta functions, see Eqs.~\eqref{eq:WmunuDIS}-\eqref{eq:WmunuRES}, which originate from the upper and lower cuts in the diagrams of Figure~\ref{fig:ep_scattering_model}. These delta functions impose
\begin{align}
    (k+q)^2 - m_q^2 &= 0 \label{eq:delta_hard}\\
    (p-k)^2 - m_\phi^2&= 0 \label{eq:delta_target}
\end{align}
and the exact solutions of this system of equations, denoted by $x_\text{ex}$ and $k^2_\text{ex}$, enter in the calculation of the $k_T$-unintegrated hadronic tensor~\eqref{eq:Wmunu_j_kTsq}. Note that Eqs.~\eqref{eq:delta_hard} and~\eqref{eq:delta_target} are coupled. However, as we already discussed in the main text, the upper cut \eqref{eq:delta_hard} provides the main constraint on the light cone fraction $x$, since $k^2$ only contributes to it at next to leading order in $1/Q^4$ (or rather $1/W^4$). The lower cut \eqref{eq:delta_target} then constraints $k^2$ as a function of $x$.

In order to highlight the role of the $W^2$ scale in determining the kinematics of the process, we can express the hard scattering invariant mass, $(k+q)^2$, as
\begin{align}
    (k+q)^2 &= W^2 + (k-p)^2 + 2(p+q)\cdot(k-p) \nonumber\\
            &= W^2 + (k-p)^2 + (1-\xi)\Big(\frac{k^2+\kTsq}{x}-M^2\Big)+\frac{x-1}{1-\xi}W^2
\end{align}
where in the last line we have used $W^2 = (1-\xi)\big(M^2+\frac{Q^2}{\xi}\big)$. 
Substituting this in Eq.~\eqref{eq:delta_hard}, and solving the resulting system, we obtain a relatively compact solution of Eqs.~\eqref{eq:delta_hard}-\eqref{eq:delta_target}:
\begin{align}
    x_\text{ex} &= \frac{1+\xi}{2} - \frac{1-\xi}{2}\sqrt{\frac{4(\ktmaxsq-\kTsq)}{W^2}}  + \frac{(1-\xi)(m_q^2-m_\phi^2)}{2W^2}
  \label{eq:x_sol_ex}\\
    k^2_\text{ex} &=  x_\text{ex}M^2 - \frac{W^2}{2(1-\xi)}\Bigg[1-\sqrt{\frac{4(\ktmaxsq-\kTsq)}{W^2}}-\frac{m_q^2+(1-2\xi)m_\phi^2}{W^2}\Bigg]
  \label{eq:k2_sol_ex}
\end{align}
where $\ktmaxsq$ is given by Eq.~\eqref{eq:ktmaxsq}. Note that
Eqs.~\eqref{eq:x_sol_ex} and \eqref{eq:k2_sol_ex} highlight the role of the $\ktmaxsq$ transverse momentum threshold, and  explicitly show how the kinematics of the process is determined by the $W^2$ scale. It is therefore interesting to study their expansion in powers of $1/W^2$. Using 
\begin{equation}
    \sqrt{\frac{4(\ktmaxsq-\kTsq)}{W^2}} = 1 - \frac{m_q^2+m_\phi^2+2\kTsq}{W^2} - \frac{2(m_q^2+\kTsq)(m_\phi^2+\kTsq)}{W^4} + O\Big(\frac{1}{W^6}\Big) \ ,
\end{equation}
we find 
\begin{align}
    x_\text{ex} &= \xi + (1-\xi)\frac{m_q^2+\kTsq}{W^2} + (1-\xi)\frac{(m_q^2+\kTsq)(m_\phi^2+\kTsq)}{W^4} +  O\Big(\frac{1}{W^6}\Big)
  \label{eq:x_sol_ex_II} \\
    k_\text{ex}^2 &= \xi M^2 - \frac{\xi m_\phi^2+\kTsq}{1-\xi} + O\Big(\frac{1}{W^2}\Big) \ .
  \label{eq:k2_sol_ex_II}
\end{align}
We do not need to expand $k^2_\text{ex}$ beyond the leading order because, as mentioned, the parton virtuality plays a role in the determination of $x_\text{ex}$ only starting at $O(1/W^4)$. In the high-energy $W^2 \to \infty$ limit (reached, for example, when $x_B \to 0$ at fixed $Q^2$, or when $Q^2 \to \infty$ at fixed $x_B$ or $\xi$) one then easily sees that
\begin{align}
    x_\text{ex} & \xrightarrow[W^2\to\infty]{} x_B \\
    k_\text{ex}^2 & \xrightarrow[W^2\to\infty]{} x_B M^2 - \frac{x_B m_\phi^2+\kTsq}{1-x_B} \ .
\end{align}
In other words, the quark's light-cone momentum fraction becomes independent of the internal partonic kinematics, while the quark remains off its mass shell at any value of $x_B$ (and by quite a large amount if $x_B \sim 1$).

A more compact expansion of $x_\text{ex}$ can be obtained in terms of a $1/Q^2$ instead of $1/W^2$ power series, and in terms of the light-cone virtuality $v^2_\text{ex} = k^2_\text{ex} + k_T^2$ instead of the virtuality $k^2_\text{ex}$. Indeed, 
\begin{align}
    x_\text{ex} 
    & =\xi \bigg( 1 + \frac{m_q^2+\kTsq}{Q^2} 
        - \frac{(m_q^2+\kTsq)v^2_\text{ex}}{Q^4}  
        + O\Big(\frac{1}{Q^6}\Big) \bigg) \ ,
    \label{eq:x_sol_ex_III}
\end{align}
where
\begin{align}
  v^2_\text{ex} 
    = -\frac{\xi}{1-\xi}
    \Big[ (m_\phi^2 - M^2) + \xi M^2 + \kTsq \Big] + O\Big(\frac{1}{Q^2}\Big) \ .
\end{align}
Eq.~\eqref{eq:x_sol_ex_III} coincides with the expansion \eqref{eq:x_sol_I} discussed in the main text, and clearly shows that $v^2_\text{ex}$ and therefore $k^2_\text{ex}$ only play a role at $O(1/Q^4)$ or higher, as mentioned at the beginning of this Appendix. 

Finally, note that the light-cone virtuality vanishes at small Bjorken $x_B$,
\begin{align}
    v^2_\text{ex} \xrightarrow[x_B \to 0]{} 0 \ ,
\end{align}
while the ordinary virtuality $k^2$, as becomes a bound particle, does not:
\begin{align}
    k^2_\text{ex} \xrightarrow[x_B \to 0]{} -\kTsq \ . 
\end{align}

\newpage
\section{Resonant scattering in the spectator model}
\label{app:resonance}	

In the main text, we focused on studying the case in which the target, as it also happens for a QCD proton, is a stable particle and cannot in particular decay into a quark-diquark pair.

\begin{figure}[bth]
	\centering
	\includegraphics[width=0.48\linewidth]{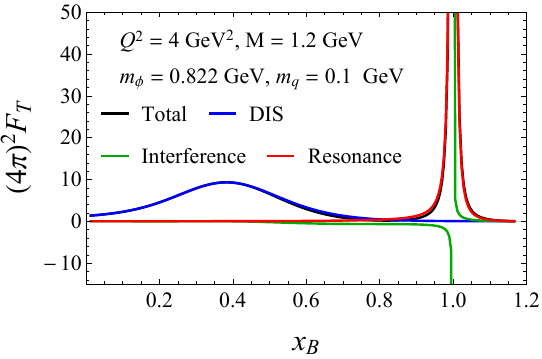}	
\caption{
Same as Figure~\ref{fig:FT_pieces} but for a choice of parameters such that $m_q+m_\phi<M$, which places the kinematic threshold at $x_B=\xbmax=1.173$. In this case the proton is an unstable particle, and the $F_T$ structure function displays a resonance peak at $x_B=1$. This corresponds to a $W^2=M^2$ four momentum squared exchanged in the $s$ channel, see the diagram (c) in Figure~\ref{fig:ep_scattering_model}. An analogous divergence can be seen for the interference contribution. 
} 
\label{fig:FT_unstable_proton}
\end{figure}

For completeness, in Figure~\ref{fig:FT_unstable_proton}, we show a calculation of the DIS, resonance and interference contributions to the structure function $F_T$ for a choice of parameters such that $m_q + m_\phi < M$ and the target is no longer stable. A peak then appears around $x_B \to 1$ in both the resonance and interference contributions, corresponding to the $W^2 \to M^2$ photon-proton resonant scattering. 

When simulating a stable proton in the main text, we chose $m_q + m_\phi > M$ and the peaks were kinematically cut at $\xbmax$, see Eq.~\eqref{eq:xbmax}. The result was a small positive bump in the total $F_T$, which should not be mistaken for a higher mass resonance such as the $\Delta(1232)$ in QCD. Nevertheless, it should not be difficult to conceive an extension of the model that also takes into account such a possibility.

\newpage
\section{Structure function scaling at small \texorpdfstring{$\bm{x_B}$}{p}}
\label{app:low_xB_scaling}

In this appendix, we justify from an analytic point of view the $Q^2$ scaling and limiting small-$x_B$ behavior observed numerically in Figures~\ref{fig:FT_pieces} and \ref{fig:FL_pieces} for the model's transverse and longitudinal structure functions.
Namely, the DIS, interference and resonance components of the transverse $F_T$
are strongly ordered in the photon's virtuality at all values of $x_B$, \textit{i.e.}, $F_T^\DIS \sim Q^2 F_T^\INT \sim Q^4 F_T^\RES$. On the contrary, all components of $F_L$ are of order $1/Q^2$. Moreover, $F_L \rightarrow 0$ as $x_B \to 0$, whereas $F_T$ shows no smal-$x_B$ constraint. As discussed in Appendix~\ref{app:projectors_detailed}, this behavior of the longitudinal structure function is a general consequence of gauge invariance, that ties together the components of $F_L$ but leaves $F_T$ unconstrained. Nonetheless, it is instructive to see how this explicitly arises from the analytical model calculations.

\subsection{Small-\texorpdfstring{$\bm{x_B}$}{p} scaling of \texorpdfstring{\bm{$F_L$}}{p}}

To compute $F_L^{(j)}$, we have to contract the longitudinal tensor $P_L$ with the individual contributions to the hadronic tensor, 
\begin{align}
    F_{L}^{(j)}
    &= 2M\, P_L^{\mu\nu} W_{\mu\nu}^{(j)} \nonumber \\
    &= \frac{\pi}{4(2\pi)^3} \int_0^{\ktmaxsq} \dkTtwo   \iint  \frac{dx}{x} \, dk^2 \, g^2(k^2) \frac{P_L \cdot \text{Tr}_{(j)}}{\text{Den}_{(j)}} \frac{1}{|\Jfull|} \delta{(x-x_\text{ex})} \delta(k^2-k^2_\text{ex})
\label{eq:FL_app}
\end{align}	
where $\text{Tr}_{(j)} = \Tr [... \gamma^\mu .... \gamma^\nu ...]$ are the traces appearing in Eqs.~\eqref{eq:WmunuDIS}-\eqref{eq:WmunuRES}) for each $j=$DIS, INT, RES, 
and $\text{Den}_{(j)}$ are the respective denominators. $\Jfull$ is the Jacobian appearing after rewriting the final state delta functions in terms of $x_\text{ex}$ and $k^2_\text{ex}$, see Eq.~\eqref{eq:Jacobian_x_k2} and Appendix~\ref{app:Delta_sol}.

We are now interested in studying Eq.~\eqref{eq:FL_app} as $x_B\to 0$ at fixed values of $Q^2$. We will also expand our results in powers of $1/Q^2$ as needed. In this limit, we find that 
\begin{align}
    k^2 &\,\longrightarrow\, - \kTsq \\
    x_\text{ex} &\,\longrightarrow\, x_B \Big(1 + \frac{\omega^2}{Q^2}\Big)\\
    \Jfull &\,\longrightarrow\, -\frac{Q^2}{x \, x_B}\\
    g(k^2) &\,\longrightarrow\, -g\, \frac{m_q^2 + \kTsq}{|\Lambda^2 + \kTsq|^2} \equiv g(\kTsq)\\
    \ktmaxsq &\,\longrightarrow\, \frac{Q^2}{4 x_B}
\end{align}
where we use $\omega^2 = m_q^2 - k^2 \approx m_q^2+\kTsq$ as a convenient shorthand. The longitudinal structure function contributions then read
\begin{align}
    F_{L}^{(j)} 
    &\,\longrightarrow\,\frac{\pi}{4(2\pi)^3} \, \frac{1}{Q^2} \, \int_0^{\ktmaxsq} \dkTtwo \,   
    g^2(\kTsq) \bigg[ x_B \frac{P_L \cdot \text{Tr}_{(j)}}{\text{Den}_{(j)}} \bigg]_{x= x_B(1+\omega^2/Q^2),\,k^2=-\kTsq} 
    \, .
\label{eq:FL_j_low_x}
\end{align}	
Since a factor $\frac{1}{Q^2}$ already appears explicitly in this equation, and the integrable $g=g(\kTsq)$ does not contribute additional factors, the $Q^2$ scaling behavior of $F_L^{(j)}$ is determined by the scaling of $x_B\,P_L \cdot \text{Tr}_{(j)}\,/\,\text{Den}_{(j)}$, which we will study case by case.

We start with DIS contribution:
\begin{align}
P_L \cdot \text{Tr}_\DIS &= \frac{1}{\hat{p}^2} \Tr\big[(\psl+M)(\ksl+m_q) \hatpsl (\ksl+\qsl+m_q) \hatpsl (\ksl+m_q)\big]\nonumber\\
&= 4 \omega^2 \overbrace{\Big( 2 k \cdot \hat{p} - k\cdot p - p\cdot q  
	\Big)}^{= \frac{\omega^2}{2 x_B}+O(x_B^0)}	
\nonumber\\ & + 4 m_q M 
\underbrace{\Big( 2 \frac{\big( k \cdot \hat{p} \big)^2}{\hat{p}^2}	- k\cdot q \Big)}_{O(x_B^0)} + 
4 \underbrace{(2 k \cdot p + M m_q) \Big( 2 \frac{\big( k \cdot \hat{p} \big)^2}{\hat{p}^2}	- k\cdot q -(k^2 - m_q^2)\Big)}_{O(x_B^0)}\nonumber\\
&= 2 \frac{\omega^4}{x_B} + O(x_B^0)\, .
\end{align}
Since the denominator in the DIS diagram is $\text{Den}_\DIS= (k^2-m_q^2)^2 = \omega^4$, we obtain
\begin{align}
    x_B\,\frac{P_L \cdot \text{Tr}_\DIS}{\text{Den}_\DIS}
    &=   2 + O(x_B)\, .
\label{eq:trace_DIS}
\end{align}
This term doesn't introduce any additional $Q^2$ dependence in Eq.~\eqref{eq:FL_j_low_x}, therefore $F_L^\DIS$ scales as $1/Q^2$ at small $x_B$ as observed numerically.

We can now repeat a similar calculation for the resonance contribution:
\begin{align}
    P_L \cdot \text{Tr}_\RES &= \frac{1}{\hat{p}^2} \Tr\big[(\psl+M)\hatpsl(\psl+\qsl+M)(\ksl+\qsl+m_q)(\psl+\qsl+M)\hatpsl\big]\nonumber\\
    &=-4(W^2 - M^2)
    \overbrace{\Big( 2 k \cdot \hat{p} - k\cdot p  - p\cdot q  
	\Big)}^{= \frac{\omega^2}{2 x_B}+O(x_B^0)}	
    + 4 m_q M 
    \underbrace{\Big( 2 \hat{p}^2 - p \cdot q + W^2 \Big)}_{O(1/x_B^2)}
    \nonumber\\
    & + 4 \underbrace{\big(2 (k+q) \cdot (p+q) + M m_q \big)}_{= W^2-M^2 + O(x_B^0)} \underbrace{ \big( 2 \hat{p}^2	- p\cdot q \big)}_{=\frac{1}{2 x_B}(W^2-M^2)}\nonumber\\
    &= 2 \frac{(W^2-M^2)^2}{x_B} + O\big(\frac{1}{x_B^2}\big)\, ,
\end{align}
where $(W^2-M^2)^2 = Q^4/x_B^2 +  O(1/x_B)$ in the leading term cancels exactly the denominator $\text{Den}_\RES= (W^2-M^2)^2$. We then obtain 
\begin{align}
    x_B\,\frac{P_L \cdot \text{Tr}_\RES }{\text{Den}_\RES}  
    &= 2 + O(x_B) \ ,
\label{eq:trace_RES}
\end{align}
which corroborates the fact that $F_L^\DIS \to F_L^{\text{RES}}$ when $x_B\to0$. 

Inserting the trace terms \eqref{eq:trace_DIS} and \eqref{eq:trace_RES} in Eq.~\eqref{eq:FL_j_low_x}, we find that 
\begin{equation}
    F_{L}^{\DIS,\RES} \,\longrightarrow\,
    \frac{\pi}{2(2\pi)^3} \frac{\xi}{x_B} \frac{1}{Q^2}  \int_{0}^{\infty} \dkTtwo \, g^2(\kTsq)
\end{equation}
which explains their common $1/Q^2$ scaling observed in Figure~\ref{fig:FT_pieces} at small $x_B$. 
An analogous calculation can be done for the interference contribution to show that
\begin{equation}
    F_{L}^\text{INT} \,\longrightarrow\,
    -\frac{\pi}{(2\pi)^3} \frac{\xi}{x_B} \frac{1}{Q^2}  \int_{0}^{\infty} \dkTtwo \, g^2(\kTsq)  \ ,
\end{equation}
that explains why $F_L \to 0$ in the small-$x_B$ limit.

\subsection{Small-\texorpdfstring{$\bm{x_B}$}{p} scaling of \texorpdfstring{$\bm{F_T}$}{p}}

The calculation for $F_T$ is analogous to the one just performed for $F_L$, 
but we need to project the hadronic tensor with $P_T$:
\begin{align}
F_{T}^{(j)} 
&= \frac{\pi}{8(2\pi)^3} \, \frac{\xi}{Q^2} \, \int_0^{\ktmaxsq} \dkTtwo   \iint  dx \, dk^2 \, g^2(k^2)  \frac{P_T \cdot \text{Tr}_{(j)}}{\text{Den}_{(j)}}  \delta{(x-x_\text{ex})} \delta(k^2-k^2_\text{ex})
\label{eq:FT_j_low_x}
\end{align}	
Again, we need to focus on the $Q^2$ scaling of 
$ \frac{P_L \cdot \text{Tr}_{(j)}}{\text{Den}_{(j)}}$. 
These DIS, INT and RES contributions can be computed and expanded in $x_B$, and  read
\begin{align}
\frac{P_T \cdot \text{Tr}_\DIS}{\text{Den}_\DIS} &= \frac{4 Q^2}{\omega^2 x_B} + O(x_B^0)\\
\frac{P_T \cdot \text{Tr}_\INT}{\text{Den}_\INT} &= -\frac{8 m_q^2}{\omega^2} + O(x_B)\\
\frac{P_T \cdot \text{Tr}_\RES}{\text{Den}_\RES} &= \frac{4(\omega^2-2M^2)}{Q^2} + O(x_B^2)
\end{align}

Now, we can obtain the leading term for each of the contributions to the transverse structure function in the low$-x_B$ limit. The DIS and interference contributions to the transverse structure function read
\begin{align}
    F_{T}^\DIS &\,\longrightarrow\, 
    \frac{\pi}{2(2\pi)^3} \int_{0}^{\infty} \dkTtwo \, \frac{g^2(\kTsq)}{\omega^2}
\label{eq:FT_DIS_low_x}
    \\
    F_{T}^\INT &\,\longrightarrow\,
    -\frac{\pi}{(2\pi)^3}\frac{x_B}{Q^2} m_q^2 \int_{0}^{\infty} \dkTtwo \, \frac{g^2(\kTsq)}{\omega^2} \  
\label{eq:FT_INT_low_x}
    \\
    F_{T}^\RES &\,\longrightarrow\,
    \frac{\pi}{2(2\pi)^3}\frac{x_B^2}{Q^4} \bigg[ \int_{0}^{\ktmaxsq } \dkTtwo g^2 \frac{(m_q^2+\kTsq)^3}{(\Lambda^2+ \kTsq)^4} - 2 M^2  \int_{0}^{\infty} \dkTtwo \,g^2(\kTsq)\bigg] 
\label{eq:FT_RES_low_x}
\end{align}
The interference $F_T^\text{INT}$ contribution is explicitly suppressed by a factor $1/Q^2$ compared to the DIS term, as we numerically observed in Figure~\ref{fig:FT_pieces} and expected from the structure of the hadronic tensor~\eqref{eq:WmunuINT}. Moreover, the factor $x_B$ in the interference contribution explains why $F_L^\INT \to 0$ when $x_B\to0$.

The resonance $F_T^\text{RES}$ has an overall $x_B^2/Q^4$ factor, which explains why this vanishes quicker than $F_T^\text{INT}$ at small $x_B$. However, the first term in square brackets is proportional to $\log \big(\frac{Q^2}{4 x_B \Lambda^2}\big)$ and alters the naive $\frac{1}{W^2-M^2}\sim 1/Q^4$ scaling, as first observed numerically in Figure~\ref{fig:FT_pieces}. Note that this logarithmic term term is a direct consequence of the choice of dipole form factor used to simulate confinement; had we used an exponential form factor the resonance contribution would have followed the naive expectation.

\newpage
\section{Model systematics}
\label{app:systematics}

\begin{figure}[p]
	\centering
	\includegraphics[trim=0.7cm 1cm 0.75cm 1cm,clip,width=0.85\textwidth]{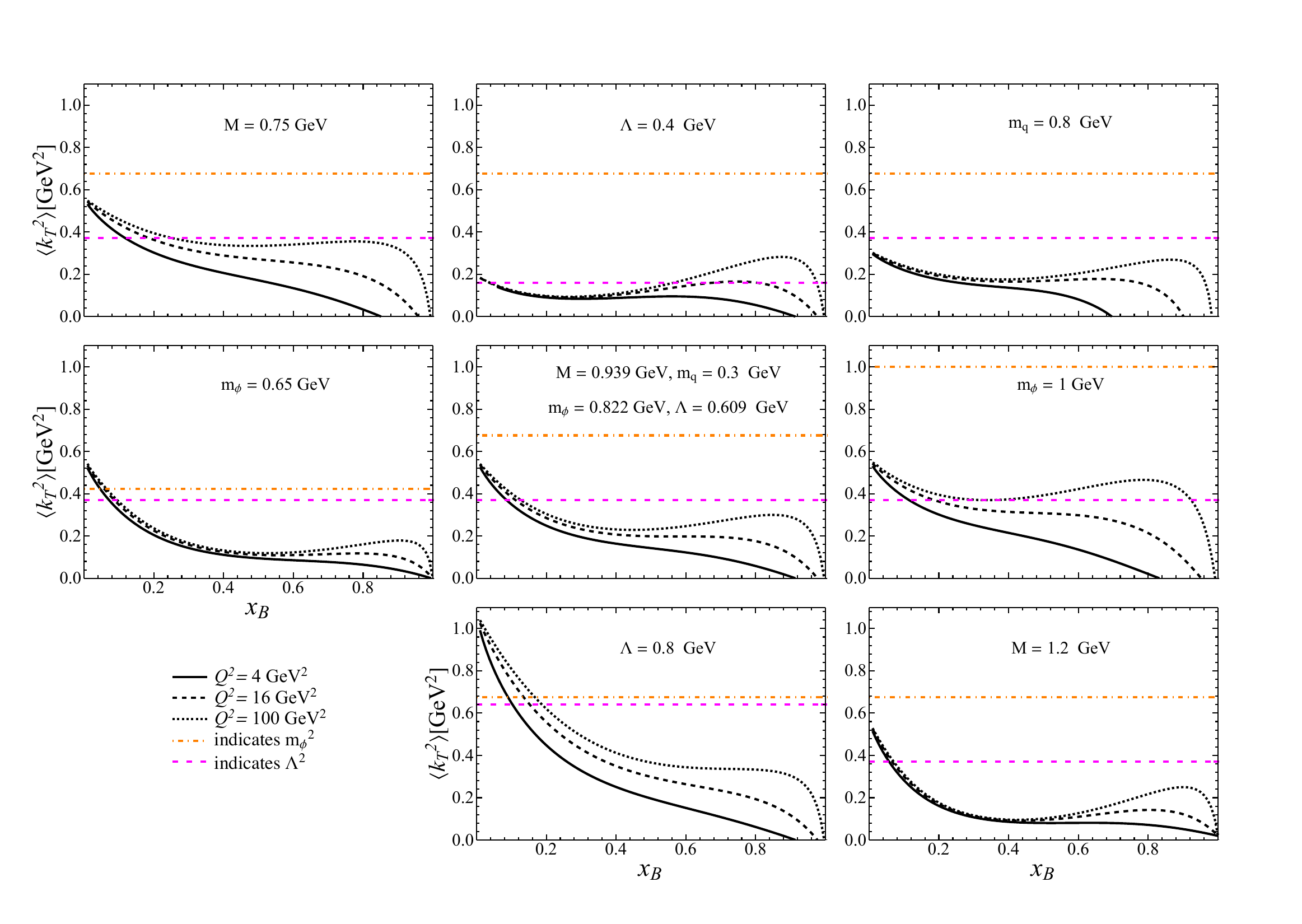}
		\vskip-0.1cm
	\caption{Average unobserved $\kTsq$ of the incoming quark calculated in the full model as a function of $x_B$ for various choices of $Q^2$ (black lines). For reference, the orange dot-dashed line marks the chosen model $m_q^2$ value, and the magenta dashed line $\Lambda^2$. The center plot corresponds to the default model parameters, while the others show variations of one parameter at a time.} 
	\label{fig:avg_kT2_II}
	\includegraphics[trim=0.45cm 1cm 0.75cm 1cm,clip,width=0.875\textwidth]{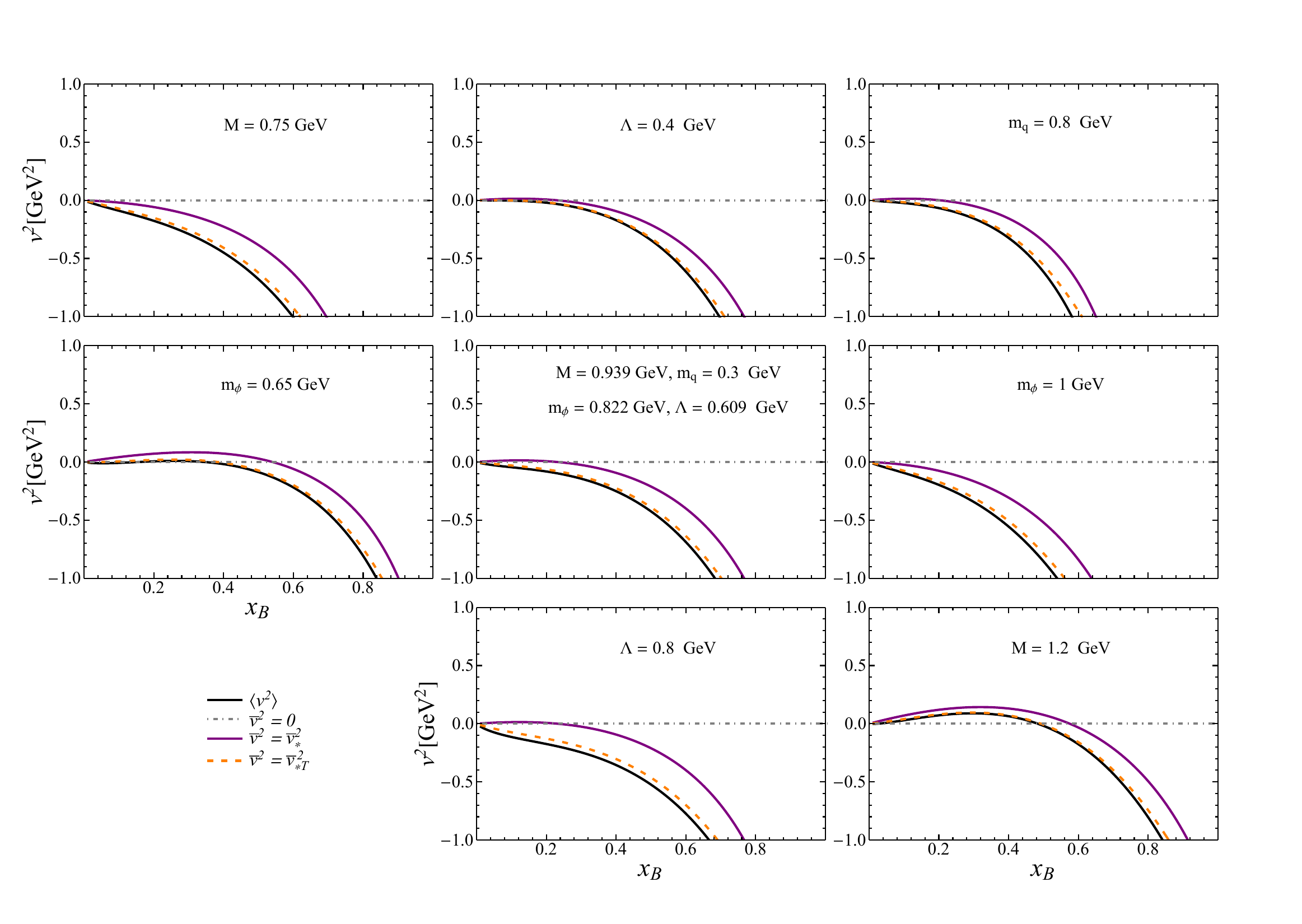}	
		\vskip-0.1cm
	\caption{Average light cone virtuality $v^2=k^2+\kTsq$ in the full model, compared to various collinear approximations. The center plot corresponds to the default model parameters, while the others show variations of one parameter at a time.} 
	\label{fig:avg_k2_kT2_II}
\end{figure}

\begin{figure}[p]
	\centering
	\includegraphics[trim=0.5cm 1.cm 0.75cm 1cm,clip,width=0.85\textwidth]{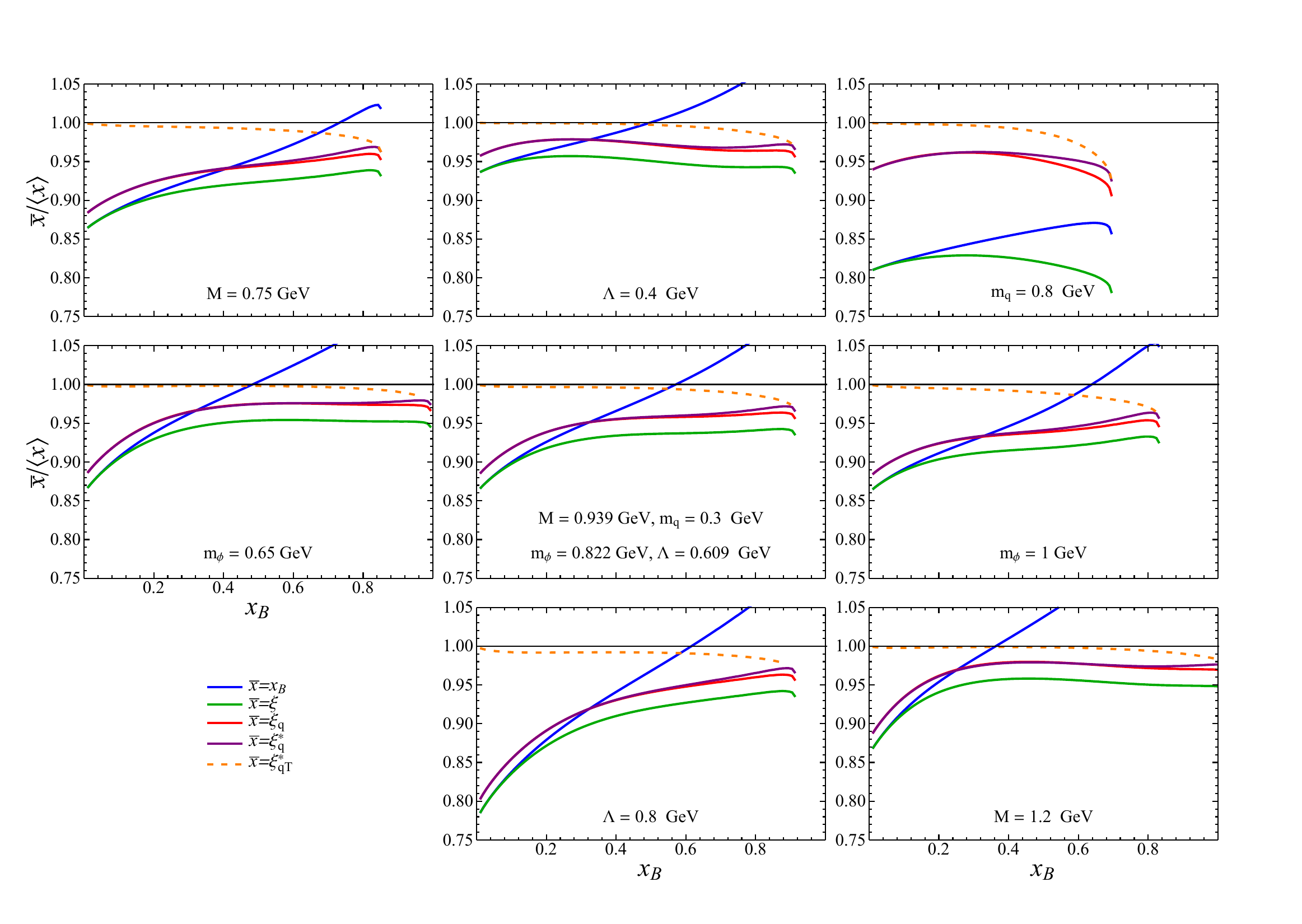}	
	\caption{
	Ratio of approximated $\bar x$ parton light-cone momentum fraction to the full $\langle x \rangle$ calculated in the model at $Q^2 = 4 \text{\ GeV}^2$. In the center panel plot, the default model parameters are used, and the other panels show the effect of varying one parameter at a time.
	} 
	\label{fig:avg_x_II}
	\includegraphics[trim=0.5cm 1cm 0.75cm 1cm,clip,width=0.85\textwidth]{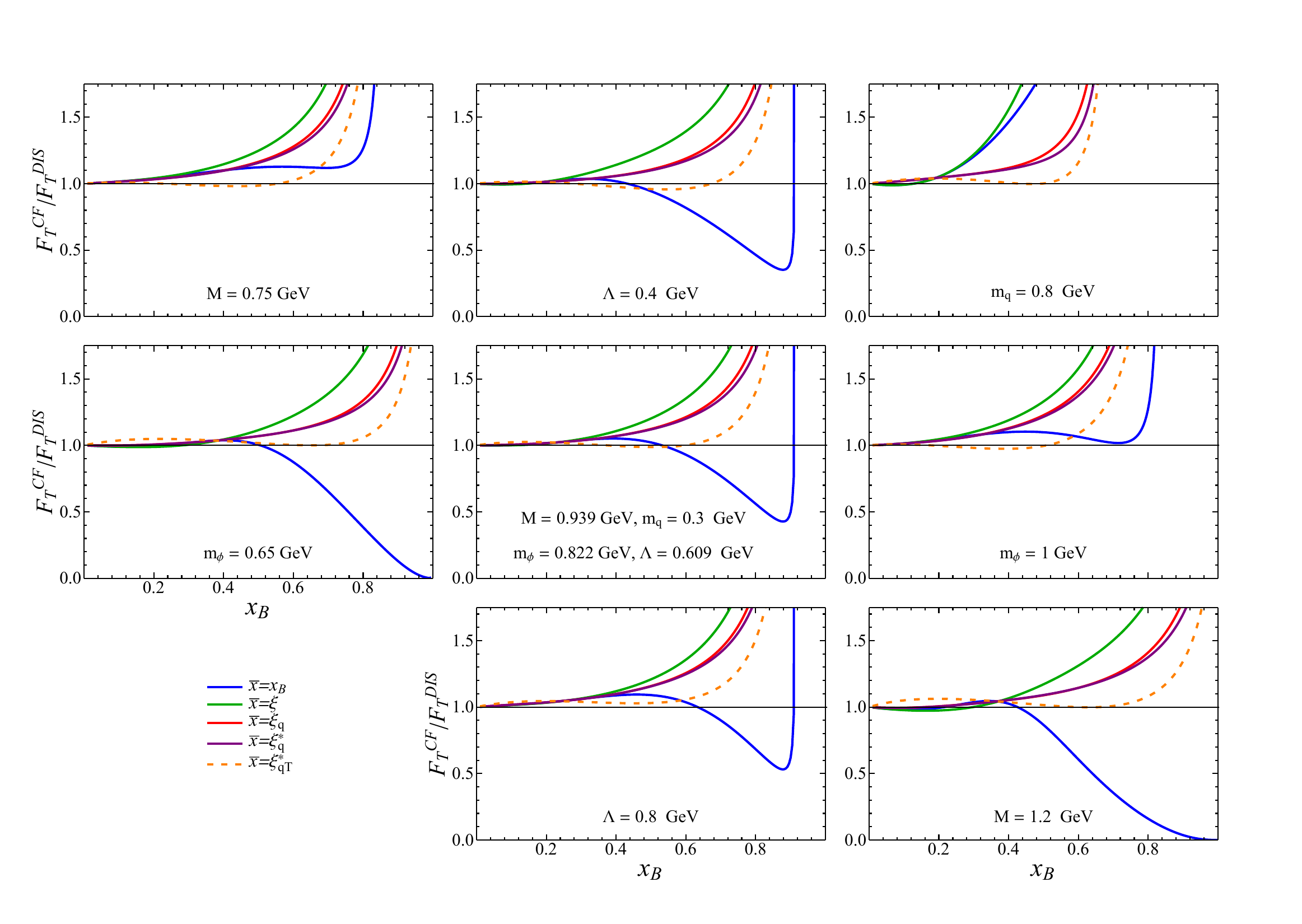}
	\caption{Ratio of the factorized to full DIS structure functions calculated in the model at $Q^2 = 4 \text{\ GeV}^2$. In the center panel plot, the default model parameters are used, and the other panels show the effect of varying one parameter at a time.
    }
	\label{fig:ratio_FT_syst}
\end{figure}

In Secs.~\ref{sec:testing_kin} and~\ref{sec:fact_limits} we tested the kinematic approximation and investigated the region of validity of collinear factorization for a selection of representative values of the target mass $M$ and the quark $m_q$, both ``external'' parameters of the model. In this Appendix, complete the analysis of the systematics of the model results, by also varying  
the ``internal'' parameters of the model, which mimic physics that can not be experimentally controlled in DIS: the confinement scale $\Lambda$, and the proton remnant mass modeled by the spectator mass $m_\phi \sim \langle m_X \rangle$.
The results are presented, in Figures~\ref{fig:avg_kT2_II}-\ref{fig:ratio_FT_syst}, that share the same structure: 
\begin{itemize}
\item 
the central panel corresponds to the default model parameters, determined by a fit of the model PDF to the quark PDFs of QCD phenomenologically fitted to experimental data as discussed in Section~\ref{sec:model}, and the parameters are varied one by one in the horizontal, vertical, and diagonal directions, respectively; 
\item 
the horizontal and vertical rows show the variation of the spectator mass $m_\phi$, and confinement scale $\Lambda$, respectively;
\item
the two diagonal rows show variations of the external target mass $M$ and quark mass $m_q$.
\end{itemize}

The plots in Figure~\ref{fig:avg_kT2_II} show the behavior of  $\avekTsq$ as a function of $x_B$.
As expected, $\avekTsq \approx O(\Lambda^2)$ is not \textit{a priori} negligible compared to the target and quark squared masses, $M^2$ and $m_q^2$.

In Figure~\ref{fig:avg_k2_kT2_II}, we present a comparison between the average light-cone virtuality in the DIS process compared to the $\vsqbar = 0$, $\vsqbar = \vsqbarnew$ [Eq.~\eqref{eq:vbarsq_new}], and $\vsqbar = \vsqbarnewT$ [Eq.~\eqref{eq:vbarsq_newT}] approximations. Claerly, the last two approximate $\langle v^2 \rangle$ at large $x$ better. However, the light-cone virtuality only contributes at $O(1/Q^4)$ to the determination of the struck quark momentum fraction, see for example Eq.~\eqref{eq:x_sol_fact_I}, and these non-zero choices for $\vsqbar$ have a minor impact on the calculation of the CF structure functions (see also below).

In Figure~\ref{fig:avg_x_II}, we compare the average light-cone fraction $\langle x \rangle$ to the collinear choices collected in Table~\ref{tab:xapprox}.
As already discussed in Section~\ref{sec:numerical_validation}, the choice $\bar x= x_B$ approximation provides an inaccurate description of the parton's longitudinal kinematics, while $x = \xi_q$ describes $\sim 95$\% of the light-cone fraction, with only minor improvements obtained when including non-zero virtuality $O(1/Q^4)$ corrections in $\bar x = \xiqnew$. Only keeping into account the partonic transverse dynamics through $\avekTsq/Q^2$ corrections in $\bar x = \xi_q$ can one obtain a fully accurate approximation.

Finally, in Figure~\ref{fig:ratio_FT_syst} we show the ratio of the collinear structure functions to the exact one. This plots illustrate that  the conclusions made from Figure~\ref{fig:ratio_FT} are not very sensitive to the variation of $\Lambda$ and $m_\phi$: a factorized calculation utilizing $\bar x = \xi_q$ provides nearly the best possible description of the full DIS structure function achievable only considering external variables, and the quality of this approximation is independent of the value of the internal (unobservable) model parameters. Taking into consideration the transverse momentum dynamics allows one to maximize the kinematic range of validity of the CF calculation before this unavoidably breaks down due to neglect of momentum conservation. 

In Table~\ref{tab:xbreak_II} we present for all cases discussed in this Appendix the $\xBbreak$ factorization breaking thresholds, and the corresponding invariant mass value. We also present the corresponding $\Rbreak$ relative contribution of the $\kTsq > \ktmaxsq$ tails to the factorized $F_T$ structure function. As a rough summary, factorization breaks down at $W^2 \lesssim 4$~GeV$^2$, which is when 20\% or more of the CF structure functions originates from $\kTsq$ values beyond the kinematic $\ktmaxsq$ threshold imposed by four momentum conservation.

\begin{table}[bth]
\renewcommand{\arraystretch}{1.4}
\begin{tabular}{ c c c | c c c | c c c }
    $\xBbreak$ & $\Wsqbreak$ [GeV$^2$] & $R_\text{break}$ &
    $\xBbreak$ & $\Wsqbreak$ [GeV$^2$] & $R_\text{break}$ &
    $\xBbreak$ & $\Wsqbreak$ [GeV$^2$] & $R_\text{break}$ \\
    \hline\hline             
    \multicolumn{3}{c|}{$M=0.75$ GeV} 
    & \multicolumn{3}{c|}{$\Lambda=0.45$ GeV} 
    & \multicolumn{3}{c}{$m_q=0.8$ GeV} \\
     0.67 &  (2.5) &  0.34 &  0.75 &  (2.2) &  0.31 &  0.59 &  (3.7) &  0.24 \\
    \hline
    \multicolumn{3}{c|}{$m_\phi=0.65$ GeV} 
    & \multicolumn{3}{c|}{Default parameters} 
    & \multicolumn{3}{c}{$m_\phi=1$ GeV} \\
     0.83 &  (1.7) &  0.24 &  0.72 &  (2.4) &  0.29 &  0.61 &  (3.4) &  0.34 \\
    \hline
    \multicolumn{3}{c|}{} 
    & \multicolumn{3}{c|}{$\Lambda=0.8$ GeV} 
    & \multicolumn{3}{c}{$M=1.2$ GeV} \\
          &        &       &  0.66 &  (2.9) &  0.25  &  0.82 &  (2.3) &  0.21\\ 
\hline\hline
\end{tabular}
\caption{
Factorization breaking thresholds $\xBbreak$ calculated for $Q^2 = 4 {\rm\, GeV^2}$ collisions, accompanied by the corresponding values of the invariant mass $W^2_\text{break}$ (in parentheses) and by the relative contribution $\Rbreak$ of the large $\kpTsq > \ktmaxsq$ tail to the factorized $\bm{k_T}$-integrated $F_T$ structure function. The table is organized in blocks that correspond to the parameter choices of the panels in Figs.~\ref{fig:avg_kT2_II}-\ref{fig:ratio_FT_syst}. The central block reports values calculated with the default model parameters, and the other blocks modify the value of one parameter at a time as indicated.}
\label{tab:xbreak_II}
\end{table}

\end{appendix}

\newpage
\bibliography{CFwithmasses} 

\end{document}